%% file: glowy.tex
\documentclass[useAMS,usenatbib,usegraphicx,usenatbib]{mn2e}
\usepackage{amssymb}
\usepackage{amsfonts}
\usepackage{amsmath}
\usepackage{wasysym}
\usepackage{multirow}
\usepackage[usenames]{color}
\DeclareSymbolFont{boldss}{OT1}{cmss}{bx}{n}
\DeclareSymbolFontAlphabet\mathbsf{boldss}

\input{aas_macros}

\def\kpc{{\rm kpc}}

\def\GRADvec{\boldsymbol\nabla}

\def\nvec{\boldsymbol n}
\def\vvec{\boldsymbol v}
\def\Bvec{\boldsymbol B}
\def\Ivec{\boldsymbol I}
\def\Jvec{\boldsymbol J}

\def\Mcal{\mathcal M}

\title[Time-dependent radio emission from evolving jets]{
   Time-dependent radio emission from evolving jets}
\author[C.~J.~Saxton et al.]
{Curtis~J.~Saxton$^{1}$,  Kinwah~Wu$^1$, Svetlana Korunoska$^1$, Khee-Gan Lee$^2$,   \and 
  Kai-Yan Lee$^{1,3}$ and Nicola Beddows$^{1}$     \vspace*{0.25cm}  \\ 
$^1$ Mullard Space Science Laboratory, University College London, 
   Holmbury St Mary, Dorking, Surrey RH5 6NT  \\ 
$^2$ Department of Astrophysical Sciences, Princeton University, Princeton, NJ 08544, USA \\ 
$^3$ Department of Physics, University of Hong Kong, Pokfulam Road, Hong Kong, Hong Kong SAR  
}

\date{Received: }

\begin{document}
\twocolumn 

\maketitle

\begin{abstract} 
{\color{Black}
\color{Black}
We investigated the time-dependent radiative and dynamical properties 
   of light supersonic jets launched into an external medium,
   using hydrodynamic simulations
   and numerical radiative transfer calculations. 
These involved
   various structural models for the ambient media,  
   with density profiles
   appropriate for galactic and extragalactic systems. 
The radiative transfer formulation took full account of
   emission, absorption, re-emission,
   Faraday rotation and Faraday conversion explicitly. 
High time-resolution intensity maps were generated, frame-by-frame,
   to track the spatial hydrodynamical
   and radiative properties of the evolving jets.
Intensity light curves were computed
   via integrating spatially over the emission maps. 
We apply the models to jets in active galactic nuclei (AGN). 
From the jet simulations and the time-dependent emission calculations 
   we derived empirical relations for the emission intensity and size for jets 
   at various evolutionary stages.     
The temporal properties of jet emission  
   are not solely consequences of intrinsic variations
   in the hydrodynamics and thermal properties of the jet.  
They also depend on the interaction between the jet and the ambient medium. 
The interpretation of radio jet morphology
   therefore needs to take account of environmental factors. 
Our calculations have also shown that
   the environmental interactions can affect specific emitting features,
   such as internal shocks and hotspots.  
Quantification of the temporal evolution
   and spatial distribution of these bright features, 
   together with the derived relations between
   jet size and emission,
   would enable us to set constraints
   on the hydrodynamics of AGN
   and the structure of the ambient medium.
}
\end{abstract}


\section{Introduction}

Jets are present in astrophysical systems  
  from active galaxies
  \citep[see][]{begelman1984}
  to stellar systems, such as X-ray binaries 
  \citep{hjellming1981,mirabel1999,fender2006}
  and symbiotic stars
  \citep{taylor1986,sokoloski2004}.
They are supersonic. 
The jets from active galactic nuclei (AGN) are also relativistic,  
   and they show a large variety of morphologies and properties.  

Although the jet plasmas have densities orders of magnitude smaller 
   than the density of the ambient media, 
   the large momentum flux carried in the plasmas   
   enable the jets to propagate great distances, even up to Mpc scales.        
These jets impart momentum to the ambient gas. 
A termination shock (which appears as a hotspot) 
   is often formed at the jet front, 
   but in some cases, a series of lesser shocks along its length.
Shocked jet plasmas fill low-density central cocoons,
   comprising eddies, backflows as well as transient shocks.
Internal shocks may stand or propagate along the jet itself,
   corresponding to intrinsic pinching modes
   \citep[e.g.][]{hardee1979,cohn1983,norman1982,hardee1983,hardee1987,falle1987,koessl1988}
   or stochastic disturbances from motions of the cocoon plasma.
Studies have show that the termination shock may be stable or unstable, 
   depending on the system conditions. 
There are cases that the termination shock
   is susceptible to vortex shedding, throbbing and disintegration
   \citep[e.g.][]{norman1982,norman1988,chakrabarti1988,
		koessl1988,falle1991,oniell2005}.  
The disrupted hotspots and shocks in the backflow 
   may be the cause of the bright rings or filaments
   seen in some radio galaxies.

The energetic particles in jets emit strong synchrotron radio waves, and in some cases, synchrotron X-rays 
   \citep[e.g.\ M87,][]{marshall2002,perlman2005}.   
In FR2 radio galaxies, jets tend to end in termination shocks,
   where the jet plasma decelerates abruptly
   as it encounters a dense external medium.    
The shocks compress the magnetic filed
   and convert the kinetic energy in the bulk flow 
   into microscopic kinetic energy. 
In shock regions, charged particles are accelerated to relativistic speeds,
   giving rise to synchrotron emission.  
The energetic particles in the jet also upscatter the low-energy photons (from the jet itself or from the ambient radiation field) 
   to the X-ray energy ranges. 
Radio synchrotron emissions from jets are linearly polarized
   \citep[see e.g.][]{mantovani1997}.  
Some AGN also show substantial circular polarization,
   in their core radio emission and in the jet knots
   \citep{hodge1977,weiler1983,komesaroff1984,homan1999,rayner2000,aller2003,
    homan2004,homan2006,vitrishchak2008,cenacchi2009}. 
{\color{Black}%
Theoretical studies  \citep[see][]{mcnamara2009} 
   have shown that Compton scattered X-rays from jets
   can also be strongly polarized. 
}

Jets are dynamical: 
   their structures change over time.
The jet evolution is manifested 
   in morphological changes in their radio image and 
   in their emission lightcurves.  
Studies on the emission properties of jets tend to adopt stationary structures. 
Simultaneous time-dependent radiative transfer calculations  
  along with hydrodynamical simulations of jet evolution are rare
  as they are usually computationally demanding. 
However, only such time-dependent calculations/simulations 
   could allow us to generate appropriate multi-wavelength lightcurve templates
   so to have a holistic pictures of the time-evolution of various jets 
   (in particular, those in micro-quasars)
   and how they interact with the ambient media.
    
Here we present high time-resolution simultaneous  
   radiative transfer calculations and hydrodynamic simulations of  
   evolving jets in a variety of ambient media. 
The radiative transfer calculations are carried out frame by frame  
   following the evolution of the jet density, velocity and thermal structures 
   in the hydrodynamic simulations. 
Time-sequences of emission images and light curves of jets are computed. 
Our study will shed light on the time-dependent hydrodynamics
   and radiative properties of
   radio jets.
    
The paper is organised as follows.  
An introduction is given in \S1; 
  the methodology of the hydrodynamic simulations
  and radiative transfer calculations are laid out in \S2. 
The emission properties of simulated jets,
  their interactions with ambient media, 
  and the astrophysical implications are discussed in \S3.
We conclude in \S4.

\section{CALCULATIONS}

\subsection{hydrodynamic simulations} 

In this study, we consider supersonic jets with low density. 
Without losing generality, we consider axially symmetric jets. 
(Note that we have carried out full 3D hydrodynamic simulations 
  for a few test cases without imposing the axi-symmetric condition.  
We  have found that the results are quantitatively similar 
  to the corresponding axi-symmetric cases, 
  but the computational time for each run does not let us 
  conduct a comprehensive survey of various jet conditions 
  and jet-interaction with the ambient media.)  
There are substantial works on the properties of such jets, 
  \citep[e.g][]{norman1982,deyoung1986,smith1985,rosen1999,carvalho2002a,
	saxton2002a,saxton2002b,krause2003,krause2005}.

Our hydrodynamic simulations adopt
   the piecewise parabolic method (PPM) algorithm:
   an explicitly conservative, shock-capturing, grid-based numerical scheme
   \citep{colella1984,blondin1993}.  
We the {\sc ppmlr} code
   \citep{sutherland2003a,saxton2001,saxton2005}
   a descendant of the standard
   {\em VH-1} University of Virginia
   hydrocode,\footnote{{\tt http://wonka.physics.ncsu.edu/pub/VH-1/}}
   modified to improve numerical stability and parallelisation 
   and to consider better tracer variables to distinguish fluids of differing origins.
The computational grid consists of $600\times300$ cells
   (in longitudinal and radial directions respectively). 
A reflecting condition applies on the cylindrical axis.
Numerically, this border copies the nearby cells into corresponding ghost cells beyond the edge
   (with vector quantities reversed to ensure balanced restoring forces).
The right boundary and the outer cylindrical boundary are open to outflow.  
The jet nozzle spans 15 cells radially at the left boundary.
Unless otherwise stated, a constant inflow condition is applied.    
Outside the nozzle, the rest of the left boundary has 
   either an ``open'' condition or a reflecting condition (``closed boundary'').  
For ``open'' boundary condition, we consider selective reflection to prevent inflow (if $v>0$ at a border cell).
For edge outflow ($v<0$ at the edge),  
   zero gradients (and zero force) is enforced, 
   through copying the single boundary cell into all ghost cells. 
No explicit treatment of radiative cooling is in the hydrodynamics formulation.  
We also omit self-gravity effects.  
These omissions would not cause significant effects
   and our results generally hold,    
   provided that the free-fall timescales are longer than the crossing times 
   associated with supersonic motions.

In our formulation, the simulated jets are characterized by two parameters: 
   the jet Mach number $\Mcal\equiv~v_{\rm j}/c_{\rm s,j}$
   and the density contrast
   $\eta\equiv\rho_{\rm j}/\rho_{\rm icm}$. 
Here $c_{\rm s,j}$ is the adiabatic sound speed in the jet plasma, 
   and the subscript ``icm'' denotes the ambient medium. 
For adiabatic jets, variables such as mass, length and time can be rescaled,
   but the contrast variables $\eta$ and $\Mcal)$ are fixed parameters. 
Generally, the thermal velocity dispersion in the ambient medium 
  may be used to set the velocity unit of the simulations, 
In this study, we consider an astrophysical realization appropriate for giant radio galaxies.  
The cylindrical region in the simulations spans $150\kpc\times75\kpc$. 
The jet radius is about $3.75\kpc$.   
If we assume that the jet propagates in an intra-cluster medium (ICM)  
  with a temperature of $10^7$~K 
  and a density of $\sim2\times10^{-27}~{\rm g}~{\rm cm}^{-3}$ 
  (corresponding to a particle density of $n_{\rm icm}=10^{-3}~{\rm cm}^{-3}$), 
  then the velocity unit $U_v=365.5~{\rm km}~{\rm s}^{-1}$   and the time unit $U_t=10.03$~Myr.

\begin{figure*}
\begin{center} 
\begin{tabular}{cc}
 \includegraphics[width=9cm]{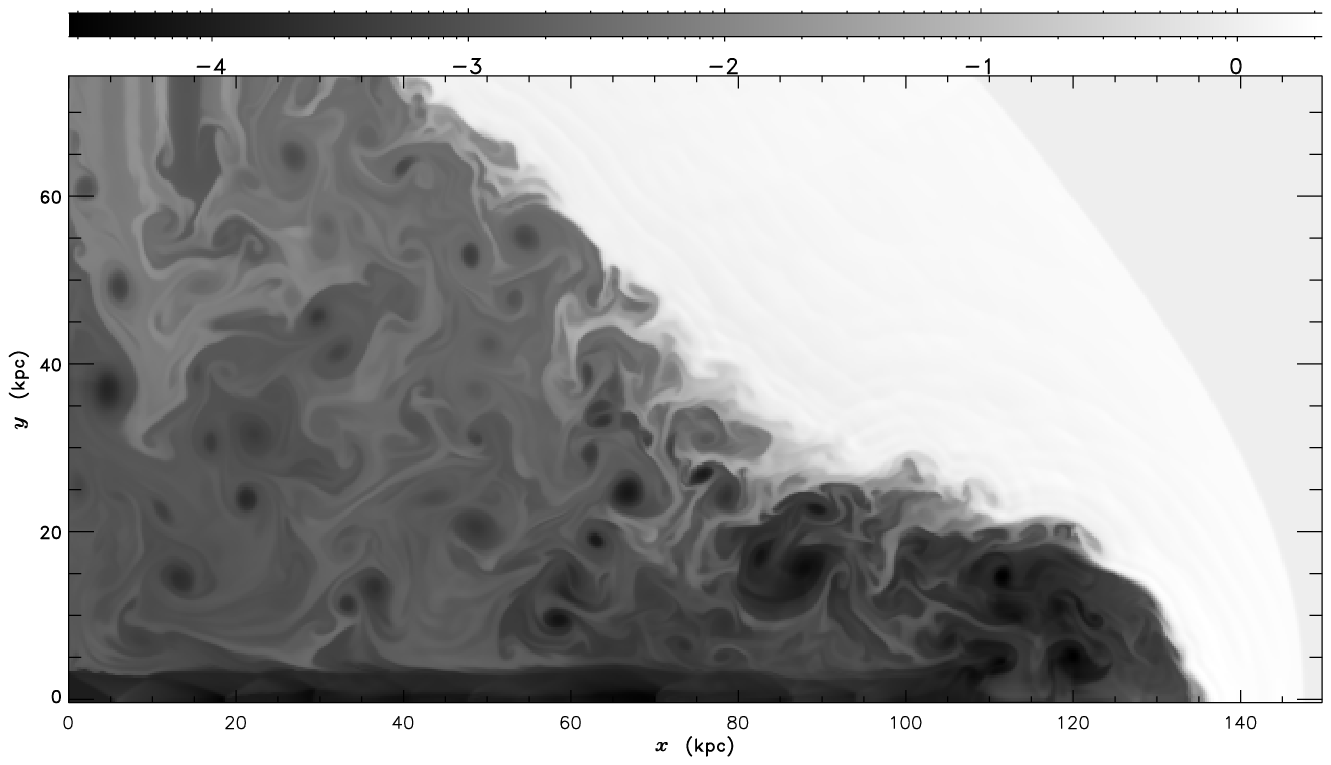}  
&\includegraphics[width=9cm]{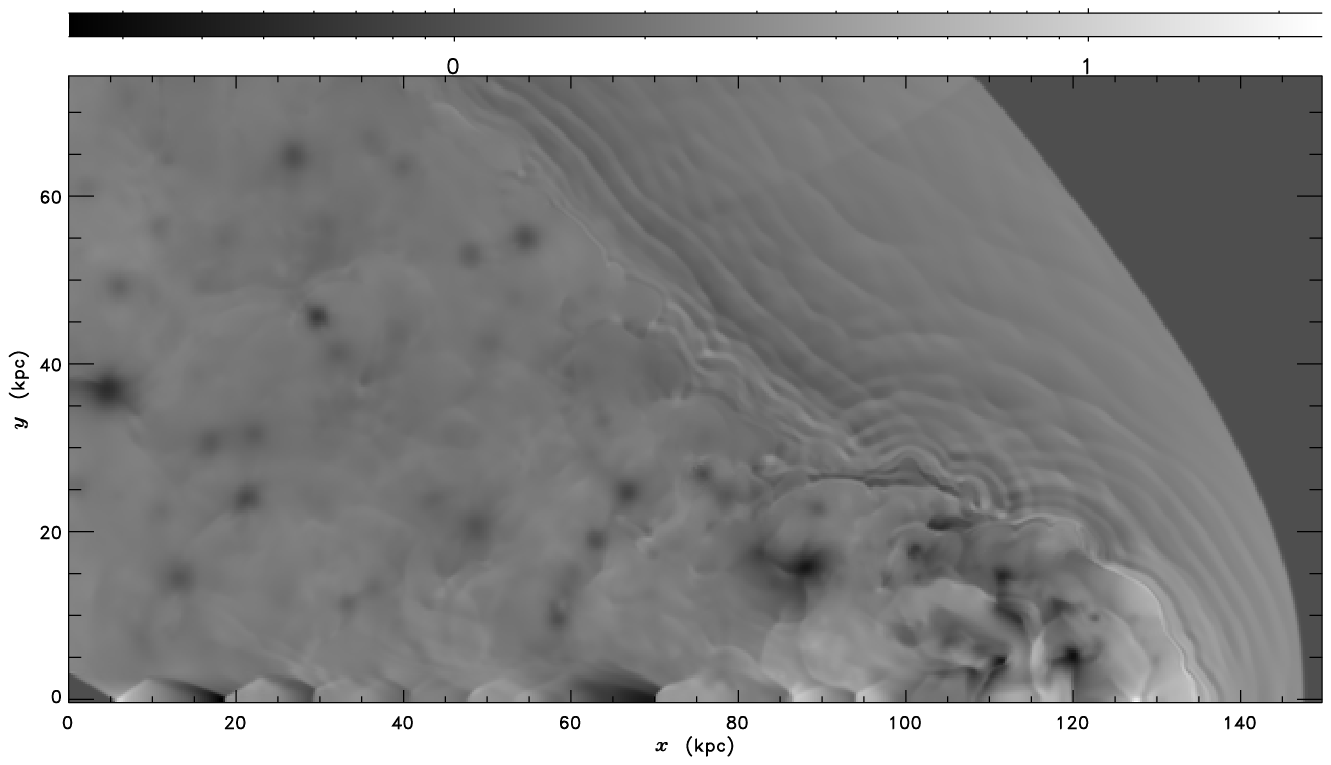}  
\\
 \includegraphics[width=9cm]{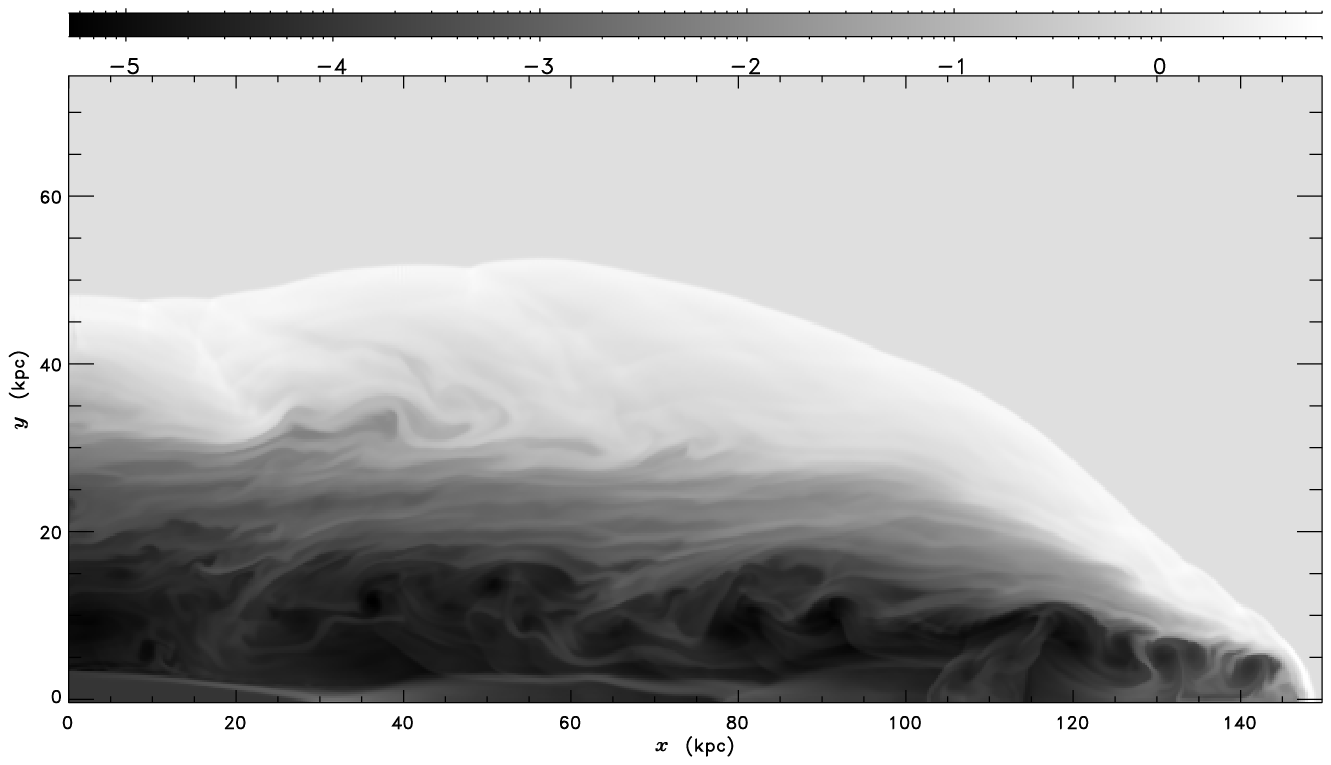}  
&\includegraphics[width=9cm]{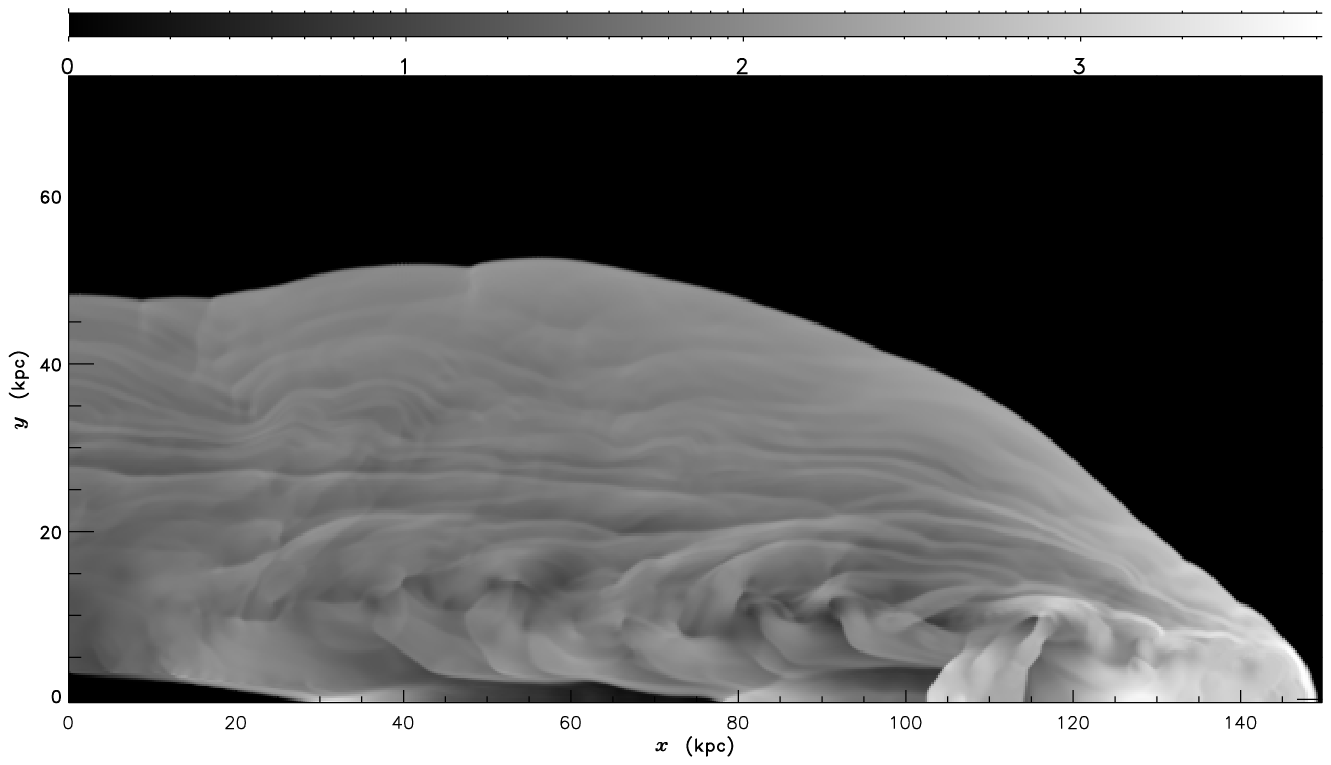}  
\\
 \includegraphics[width=9cm]{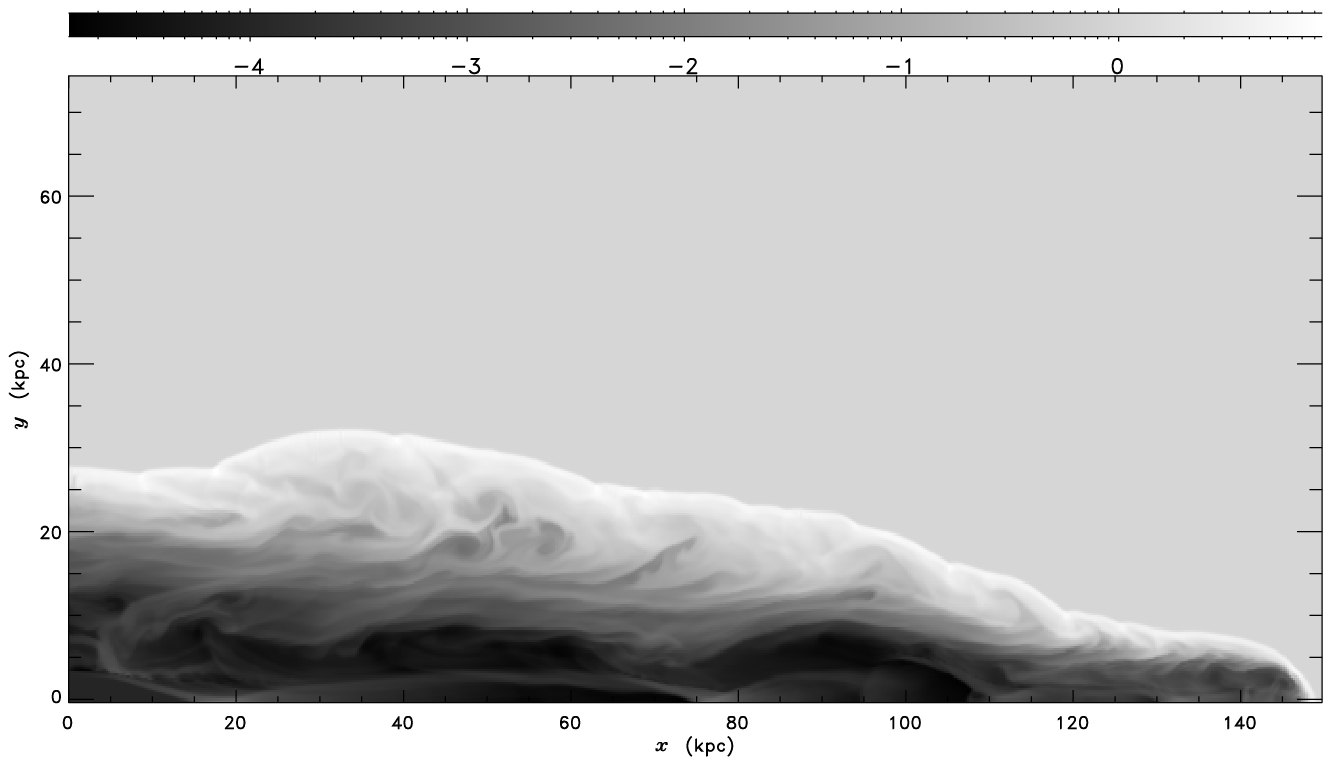}
&\includegraphics[width=9cm]{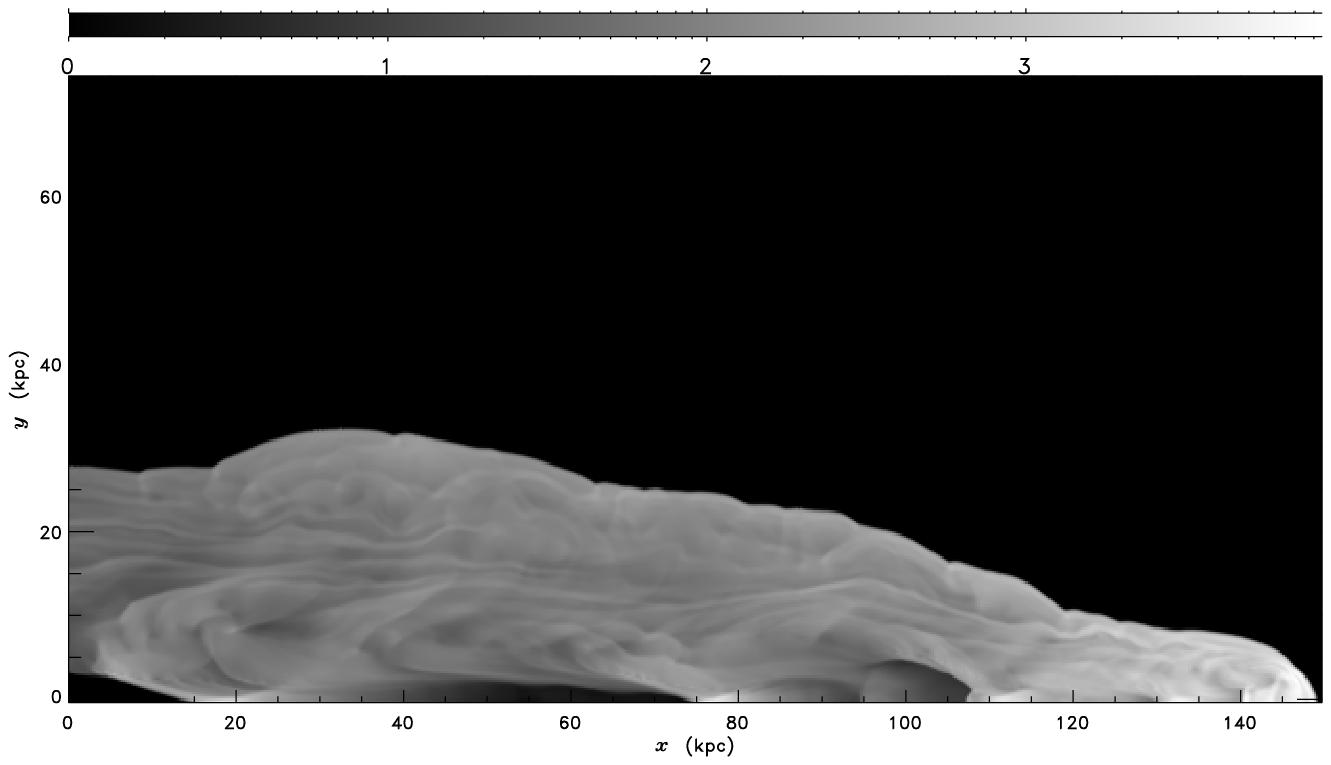}
\end{tabular}
\end{center} 
\caption{ 
Sections of jet through the cylindrical structure at final frames of three simulations. 
The grey values and scale bar are decimal logarithmic. 
The jet axis runs along the bottom of each frame. 
The radial direction runs upwards.
The left panels show the density; the right panels show the pressure (a proxy for $B^2/8\pi$).  
The {\sc slow} jet is on the upper, and  the {\sc fast1o} jet is on the middle row.  
The jet plasma is a polytropic gas with $\gamma=5/3$.
The {\sc fast1ro} jet is on the bottom row. 
The simulation has the parameters and set ups  same as those of the {\sc fast1o} jet, 
  except that the jet plasma is relativistic with $\gamma=4/3$.
}
\label{fig.hydro}
\end{figure*}

As a basis of comparison, we first carry out simulations 
   of jets launched into uniform media.  
This simple setting avoids 
   the unnecessary complications caused by the structures in the ambient medium 
   and enables us to identify the dynamics
   and emission properties intrinsic to the jets.  
We consider two jet models. 
In the first simulation ({\sc fast1o})
   we set $\Mcal=50$ and $\eta=10^{-4}$.  
The corresponding particle column density through the jet is
   $\approx2.3\times10^{15}~{\rm cm}^{-2}$.       
In this case, the jets are expected to produce hotspots and other shocks
   with high brightness contrast relative to their surroundings
   \citep[e.g.\ the radio galaxy Pictor~A,][]{saxton2002a}.
In the second simulation ({\sc slow})
   we consider a lower Mach number ($\Mcal=3$) for the jet.  
The density of the ambient medium is the same.  
These jets tend to give lower luminous contrast for their induced features  
   \citep[e.g.\  the radio galaxy Hercules~A][]{saxton2002b}.
The upper and middle rows of Figure~\ref{fig.hydro}
   show density and pressure slices of {\sc slow} and {\sc fast1o}
   at late times.

Next we run simulations for jets propagating
   into media with various density structures. 
In the {\sc fast2} and {\sc fast2o} simulations, 
   we assume a decreasing medium density following a flat-cored model
   \citep{cavaliere1976}
\begin{equation}
	\rho = {{\rho_0}\over{
			\left[{1+(r/r_0)^2}\right]^{3\xi/2} \ . 	
		}}
\end{equation}
We set the parameter $\xi=0.5$ and the scale radius $r_0=15~\kpc$. 
Cases with a reflecting and an open left boundary are investigated.     
(Open cases are named with a suffix ``{\sc o}''.)
In the {\sc fast3} simulations, 
   we consider a model where the density of the ambient medium
   increases with radius. 
We set $\xi=-1$ and $r_0=15~\kpc$.
In the {\sc fast4} simulations, the density of the ambient medium varies sinusoidally,
   with $\rho=\rho_0\ \exp[-a\sin(kr)]$, 
   where the wavenumber $k = 2\pi/\lambda_0$. 
We set the amplitude $a=\log_e 10$. 
This gives an effective wavelength of the variation $\lambda_0=30~\kpc$. 
In the {\sc fast1r} and {\sc fast1ro} simulations, 
   we investigate the effects of the equation state.  
The jet plasma now has an adiabatic index $\gamma=4/3$ 
  (characteristics of relativistic gas) instead of $5/3$.
Table~\ref{table.setup}
   lists the principal parameters of the simulations
   and the temporal sequence of output frames.

\begin{table}
\caption{
Settings and parameters of the simulations.
For those marked $\CIRCLE$ the left boundary is closed/reflecting;
for those marked $\Circle$ it is open.
The jet density contrast and Mach number are
$\eta=\rho_{\rm j}/\rho_{\rm ism}$
and $\Mcal$.
The duration of each simulation is $t_{\rm end}$
and the interval between frames is $\Delta t$
(both in units of $U_t\approx10.03$~Myr).
Runs with nonuniform ambient media have index $\xi$ and scale radius $r_0$.
(The grid size is $40\times20$ in the same units.)
$N$ counts the frames.
}
\begin{center}
$\begin{array}{lcrrr@{.}lrrrrrrrrrrrrrrrrrrrrrrrrrrrr}
\mbox{run}
&\mbox{BC}
&\eta
&{\mathcal M}
&\multicolumn{2}{c}{t_{\rm end}}
&\Delta t
&\xi
&r_0,
&N
\\
&&&&\multicolumn{2}{c}{( U_t )}
&( U_t )
&&\lambda_0
\\
\hline
\\
\multicolumn{3}{l}{\gamma=5/3}
\\
{\sc slow}
&\CIRCLE
&10^{-4}
&3
&15&0
&0.0024
&0
&\infty
&5500
\\
{\sc fast1}
&\CIRCLE
&10^{-4}
&50
&0&800
&0.0004
&0
&\infty
&2000
\\
{\sc fast1o}
&\Circle
&10^{-4}
&50
&0&172
&0.0004
&0
&\infty
&430
\\
{\sc fast2}
&\CIRCLE
&10^{-4}
&50
&0&328
&0.0004
&0.5
&4
&820
\\
{\sc fast2o}
&\Circle
&10^{-4}
&50
&0&184
&0.0004
&0.5
&4
&460
\\
{\sc fast3}
&\CIRCLE
&10^{-4}
&50
&1&44
&0.0004
&-1
&4
&3600
\\
{\sc fast4}
&\CIRCLE
&10^{-4}
&50
&1&28
&0.0004
&\sim
&8
&3200
\\
\\
\multicolumn{3}{l}{\gamma=4/3}
\\
{\sc fast1r}
&\CIRCLE
&10^{-4}
&50
&0&816
&0.0004
&0
&\infty
&2040
\\
{\sc fast1ro}
&\Circle
&10^{-4}
&50
&0&3864
&0.0004
&0
&\infty
&966
\\
\\
\hline
\end{array}$
\end{center}
\label{table.setup}
\end{table} 

\subsection{numerical radiative transfer}

In the radiative transfer calculations,  we use a 3D Cartesian grid,   
   consisting of  $500\times500\times500$ cells. 
The values of relevant variables on the radiative transfer grid  
  are computed from the 2D data of axially symmetric jets 
  obtained from the hydrodynamic simulations.
For each Cartesian cell centered at $(x,y,z)$,
   we find the four neighbouring points of the cylindrical mesh
   that have integer coordinates.
We use interpolation weights $w_i\propto1/(r_i+\delta)$
   where $r_i$ is the distance neighbouring cylindrical vertex $i$
   and $\delta=10^{-6}$ is a softenning term in case
   the projected position lies near a cylindrical cell boundary or corner.

The jets in this study are not magnetically dominated,   
  and our simulations did not involve
  an explicit treatment of the magnetic field.   
However, the calculation of synchrotron radiation requires 
   an explicit specification for the magnetic field.  
We therefore  use a parametric prescription for the local magnetic field in the jet plasma. 
The field strength is taken as $B=\sqrt{8 \pi P/\beta}$, 
   where $P$ is the pressure and $\beta\ (\geq1)$ is the strength parameter.
The orientation of the field vector assumes
    an ordered component and a random component. 
The random component  has no particular correlated length scales,  
   i.e.\ with a white power density spectrum.  

The polarized radiative transfer is in the 4-Stokes formulation,   
   with the Stokes parameters $I$, $Q$, $U$ and $V$. 
The linear polarization component is given by 
   $L = \sqrt{U^2 + Q^2}$, and the circular polarization is given by $V$.  
The degrees of linear and circular polarization are 
   $\Pi_{\rm L} = L/I =  \sqrt{U^2 + Q^2}/I$ and 
   $\Pi_{\rm C} = V/I$ respectively, where $I$ is the total intensity. 
The Stokes parameters  satisfy the time-dependent local polarized radiative transfer equations:  
\begin{equation} 
  \left( \hat{\mathbsf D} + {\mathbsf K}  \right) 
    \left[{\begin{array}{c}
       I\\Q\\U\\V
   \end{array}}\right] 
   =  
     \left[{\begin{array}{c}
	\varepsilon_{_I} \\ 
	\varepsilon_{_Q} \\ 
	\varepsilon_{_U}\\ 
	\varepsilon_{_V}
	\end{array}}\right] \ , 
\label{eq.transfer}
\end{equation} 
 where $\varepsilon_{_I}$, $\varepsilon_{_Q}$, $\varepsilon_{_U}$ and  $\varepsilon_{_V}$ 
 are the corresponding emission coefficients of the Stokes parameters. 
The propagation operator is defined as 
\begin{equation} 
 \hat{\mathbsf D} \equiv {\mathbsf I} \partial_{s}   
   =  {\mathbsf I} \left({c^{-1}}{\partial_t} +\nvec\cdot\GRADvec \right) \ , 
 \end{equation}   
   where $c$ is the speed of light, $\nvec$
   is the normal vector of the ray propagation, 
   ${\mathbsf I}$ is the identity matrix, $\partial_{s}$ is the projected differentiation operator.      
The transfer matrix is given by 
\begin{equation} 
  {\mathbsf K} = 
       \left[{\begin{array}{rrrr}
		\kappa&q&0&v\\
		q&\kappa&f&0\\
		0&-f&\kappa&h\\
		v&0&-h&\kappa\\
	\end{array}}\right] \ ,  
\label{eq.terms}
\end{equation}  
 where $\kappa$ is the absorption coefficient,  
 $q$ and $v$ are coefficients describing absorption effects;
while $f$ and $h$ describe the Faraday propagation effects. 
The functional form of the emission, absorption, conversion
  and rotation coefficients  
  are given in \citet[]{jones1977}.
Several studies have directly integrated the transfer equations
   (\ref{eq.transfer})--(\ref{eq.terms})
   including circular polarisation,
   in parametric models of shocks traversing cylindrical or conical jets
   \citep[e.g][]{jones1988,hughes1989a,hughes1989b}.
We apply a similar treatment to hydrodynamically ``live'' simulated jets,
   but with a mathematical alteration
   to improve computational speed
   and robustness in inhomogeneous structures of any opacity.

To solve the transfer equation we use the following method. 
Consider the transformation:      
\begin{equation} 
  \left(\hat{\mathbsf D} + {\mathbsf K}\right) {\Ivec}  =   \Jvec
   \rightarrow   
    \left[ {\mathbsf R}^{-1} (  \hat{\mathbsf D} + {\mathbsf K} ) {\mathbsf R}  \right]~ {\mathbsf R}^{-1}  \Ivec
   =  {\mathbsf R}^{-1} \Jvec \  ,  
\end{equation}  
where  ${\Ivec}=[I~Q~U~V]^{\rm T}$ 
 and ${\Jvec}=[\varepsilon_{_I}~\varepsilon_{_Q}~\varepsilon_{_U}~\varepsilon_{_V}]^{\rm T}$. 
Choose the transformation matrix ${\mathbsf R}$
   such that it is translationally invariant in space and time.   
The commutation ensures that 
  the propagation operator $\hat {\mathbsf D}$ 
  remains diagonalised and is hence is invariant under the transformation.   
It follows that 
\begin{equation} 
   \left(\hat{\mathbsf D} + \tilde{\mathbsf K}\right) \tilde{\Ivec} 
   =   \tilde{\Jvec}  \ ,  
\label{eq.ray.diagonal}
\end{equation} 
 where $\tilde{\Ivec}  = {\mathbsf R}^{-1}  {\Ivec}$, 
  $\tilde{\Jvec} =   {\mathbsf R}^{-1} {\Jvec}$, and 
  $\tilde{\mathbsf K}= {\mathbsf R}^{-1}  {\mathbsf K}  {\mathbsf R}$.   
When $\tilde{\mathbsf K}$ is diagonalised, 
  the components in the polarized transfer equation are decoupled. 
Once the initial condition (the Stokes parameters of the incident ray) is specified, 
  the equations can be integrated independently. 
This method is more computationally efficient and numerically stable 
  than the direct integration of the coupled polarized radiative transfer equation.  
  
Following \citet{wegg2003},   
   we use the algebraic package LAPACK  \citep{anderson1990,barker2001}
   for diagonalisation, 
   which determines the eigen-matrix $\tilde{\mathbsf K}$
   and the transformation matrix ${\mathbsf R}$. 
Direct integration of Equation~(\ref{eq.ray.diagonal})
   yields $\tilde{\Ivec}$, 
   of which an inverse transform under ${\mathbsf R}$
   gives the Stokes vector ${\Ivec}$.    
We consider a forward ray-tracing scheme, 
   in which the transfer equation is solved sequentially
   along the ray through the cells.  
The collection of rays through the jet
   generates the time sequence of Stokes images.    

\section{RESULTS AND DISCUSSIONS}

\subsection{jet morphology and emission images}

\subsubsection{standard cases}

\begin{figure*}
\begin{center} 
\begin{tabular}{cccc}
 \includegraphics[width=6cm]{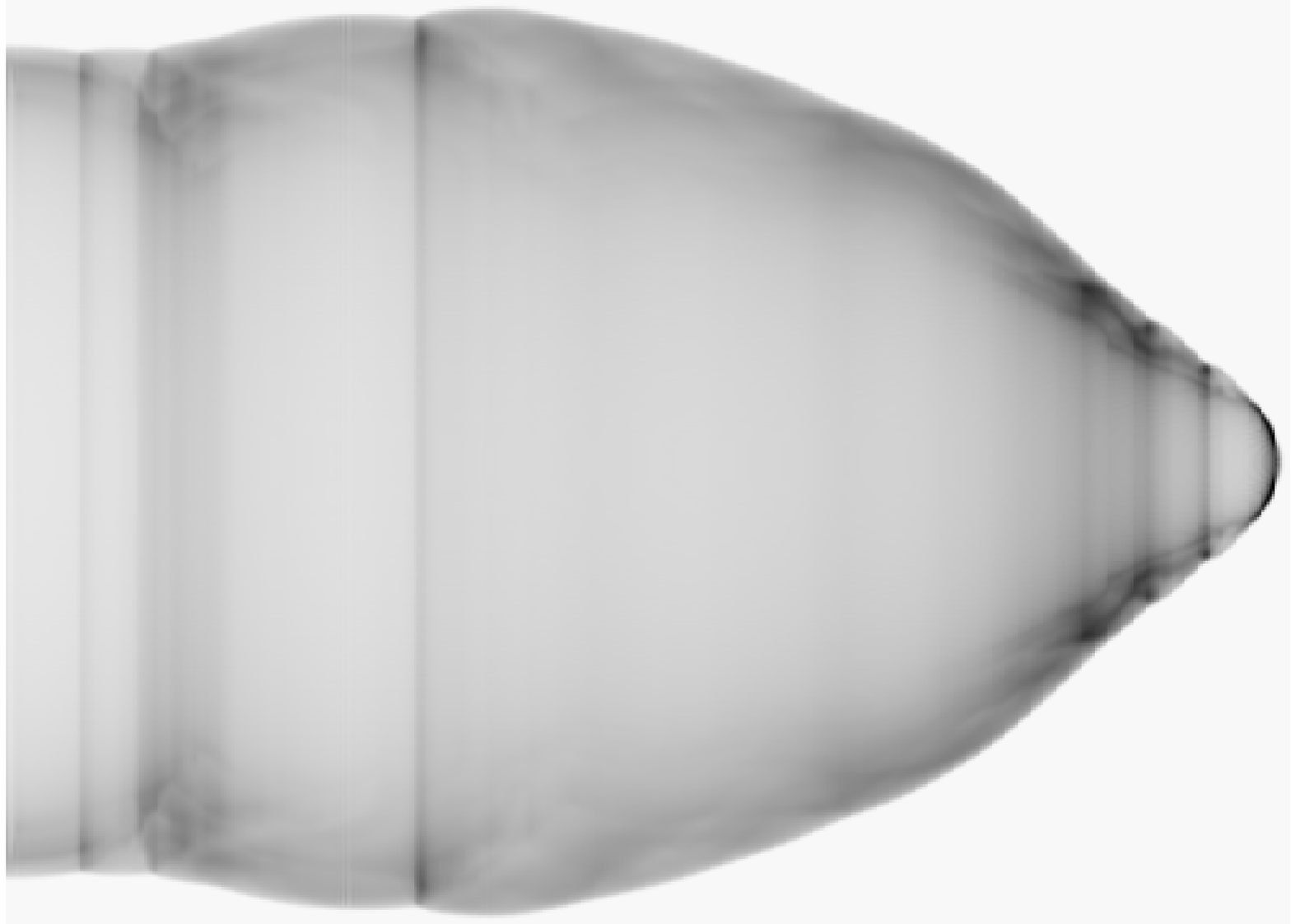}  
&\includegraphics[width=6cm]{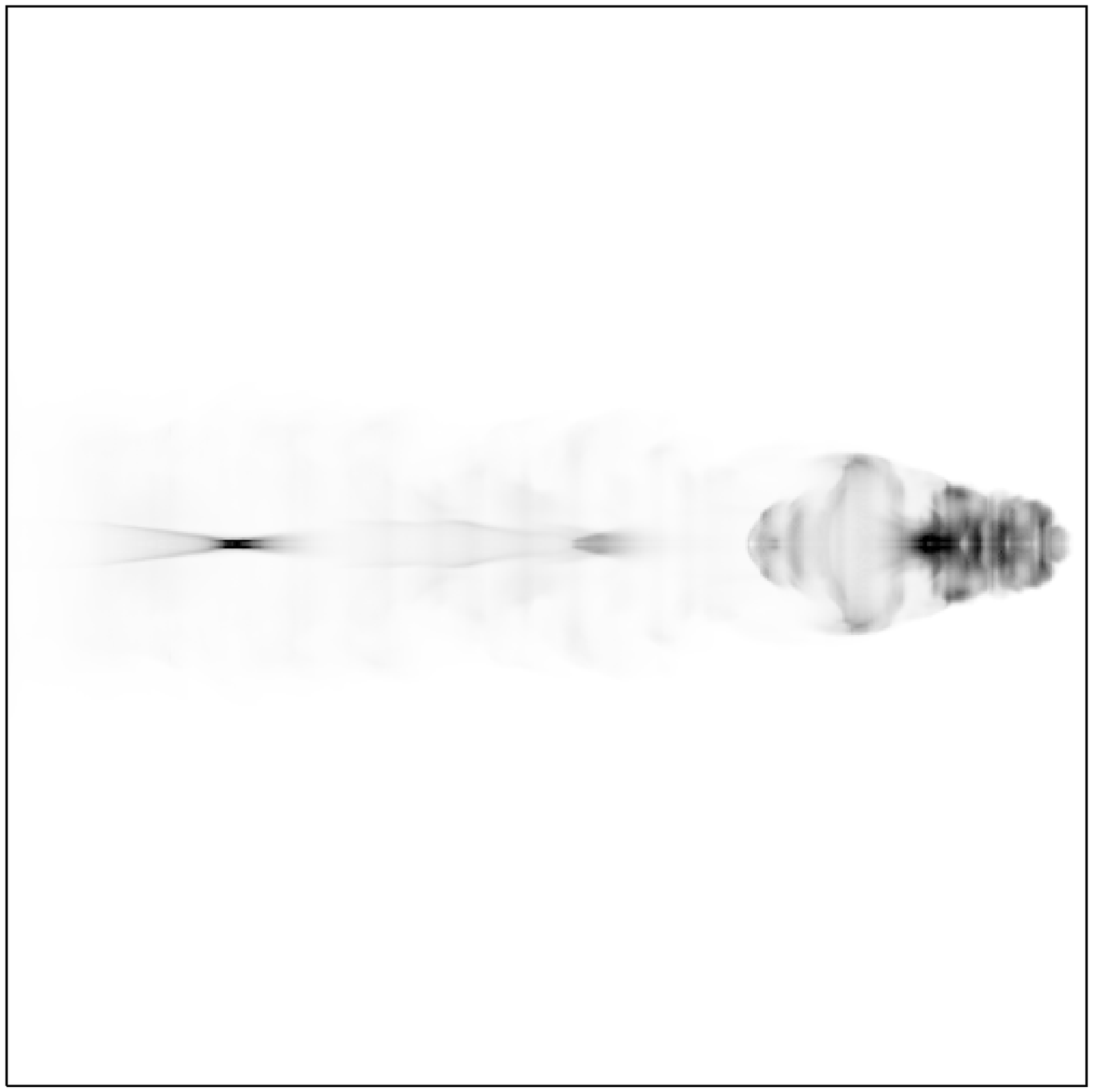}  
\\
 \includegraphics[width=6cm]{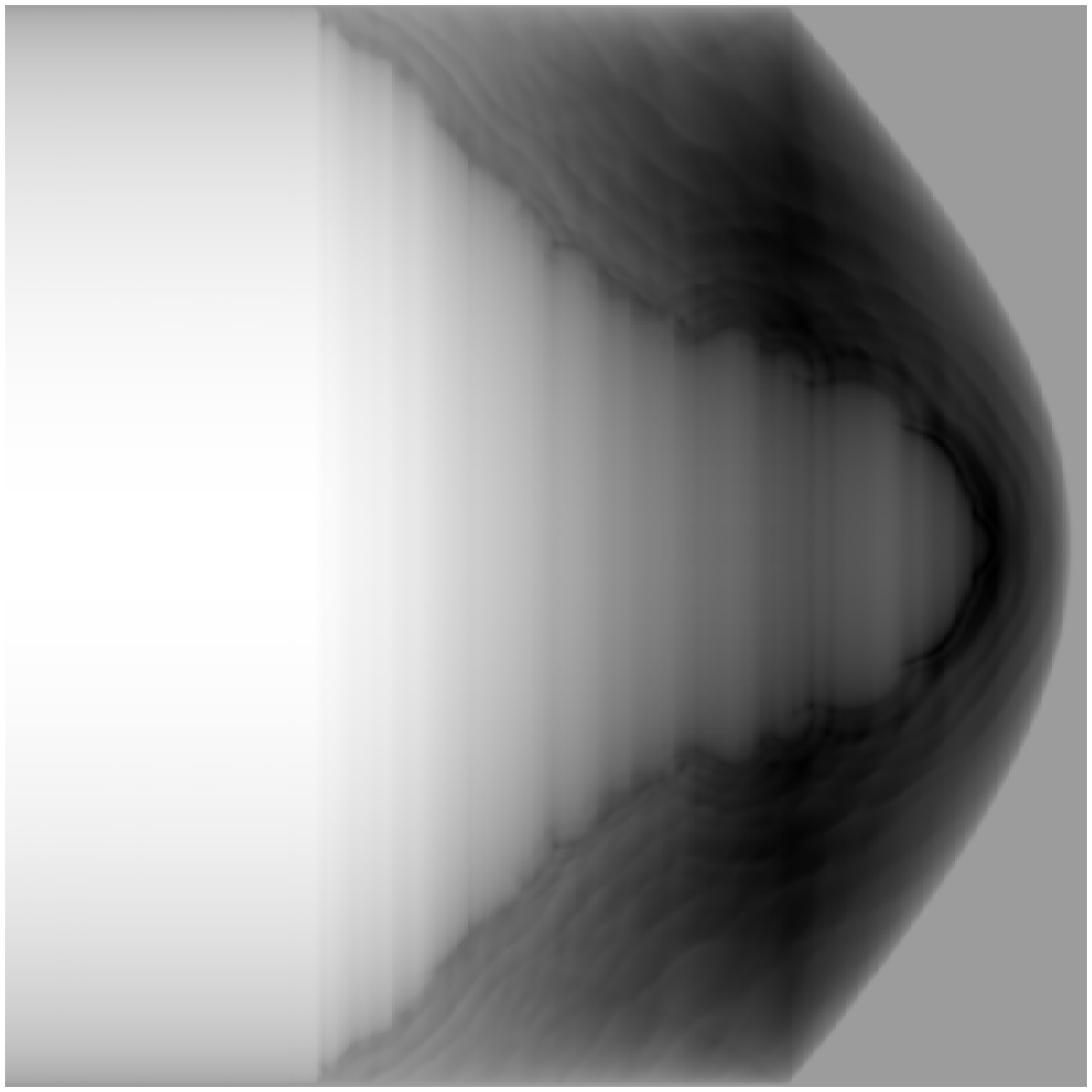}  
&\includegraphics[width=6cm]{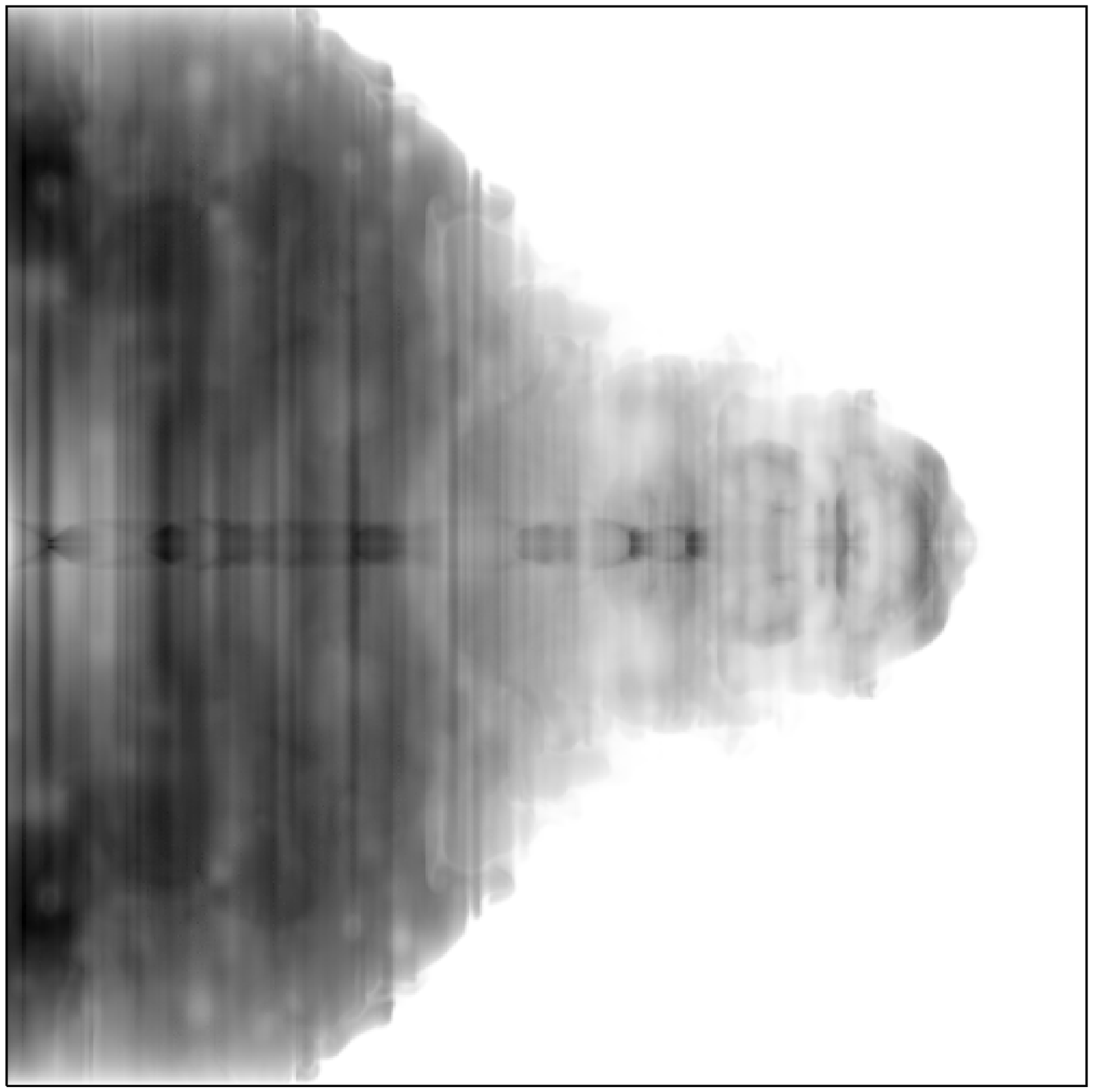}  
\end{tabular}
\end{center} 
\caption{ 
Emission images of the {\sc fast1o} jet (top row) and the {\sc slow} jet (bottom) 
  at their late evolutionary stage.    
The image sizes are $150~\kpc\times150~\kpc$.  
In both cases,  $\beta=10$
   and the field is quasi-poloidal ($\Bvec\parallel\vvec$).
Columns from left to right show: 
  thermal bremsstrahlung X-rays, 
  and synchrotron radio intensity ($I$).
The emissions are linearly scaled.
The synchrotron intensity is evaluated at 6~cm. 
The jet morphology in the non-thermal synchrotron X-ray images (unshown)
  is similar to that in the synchrotron radio emission image. 
}
\label{fig.advanced}
\end{figure*}

Figure~\ref{fig.advanced}
   shows the emission images of the {\sc fast1o} and {\sc slow} jets, 
   calculated during advanced evolutionary stages. 
These two model jets are light and unobstructed. 
Bremsstrahlung maps are calculated by line-of-sight integration
   of the bolometric emissivity $\propto\rho^2T^{1/2}$.
In both simulations,
  most of the bremsstrahlung X-rays
  are emitted from the compressed ambient medium within the bow-shock;  
  the contribution of the jet itself to the total
  bremsstrahlung emission is insignificant. 
The radio emission from the jets is synchrotron emission, 
   mostly from the hotspot, the jet's recollimation, pinch mode shocks,  
   and the weaker shocks in the turbulent cocoon.   
In producing 6~cm radio synchrotron emission maps of these jets,  
   we adopt a relativistic electron energy spectral index $p=2.2$.
By default we have assumed $\beta=10$
   (i.e. hydrostatic pressure dominates magnetic pressure).
In some extra comparisons,
   we also calculate the maximal field case ($\beta=1$)
   or a weaker field case ($\beta=100$).

\begin{figure*}
\begin{center} 
\begin{tabular}{ccccc}
 \includegraphics[width=4cm]{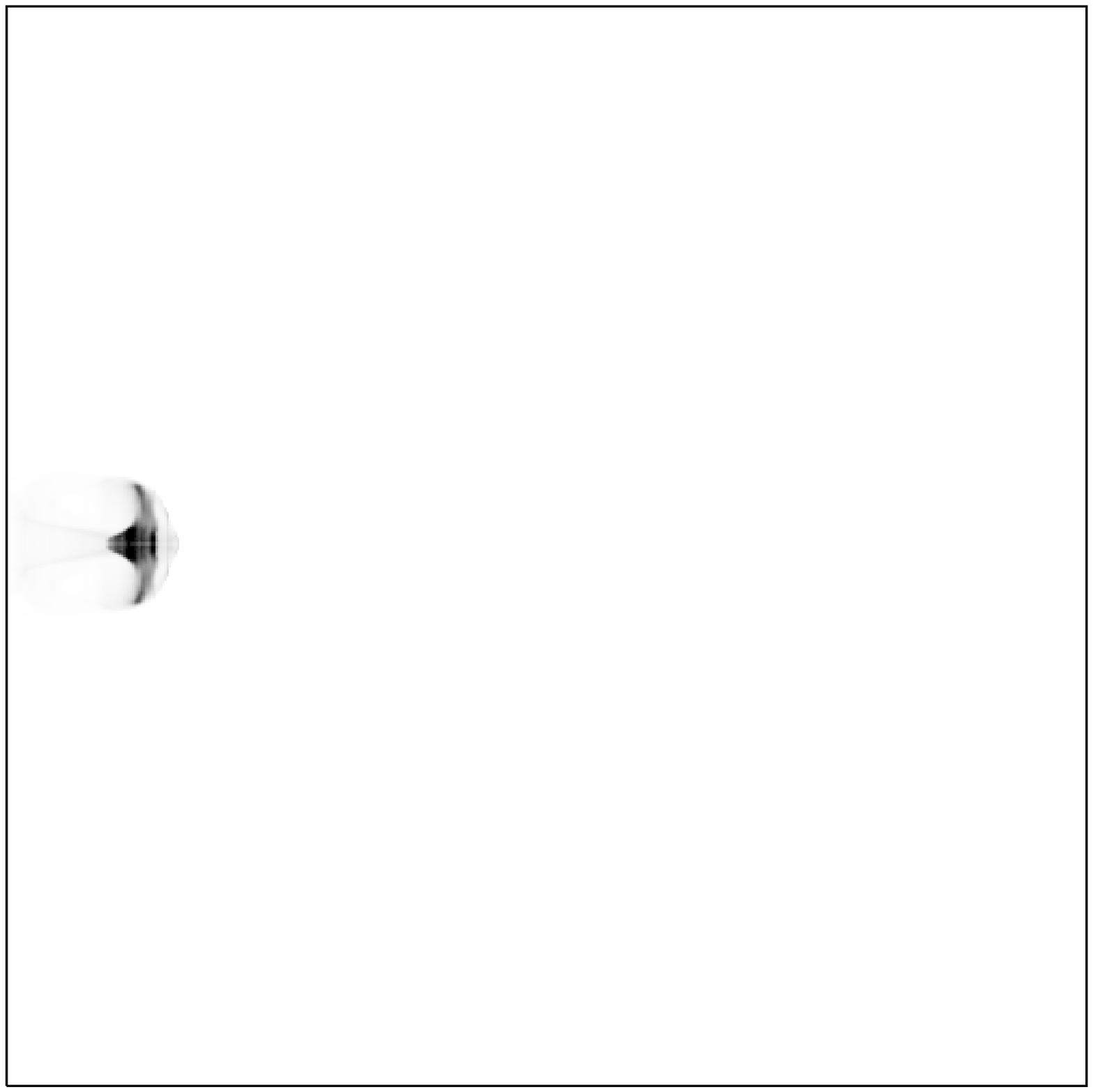}  
&\includegraphics[width=4cm]{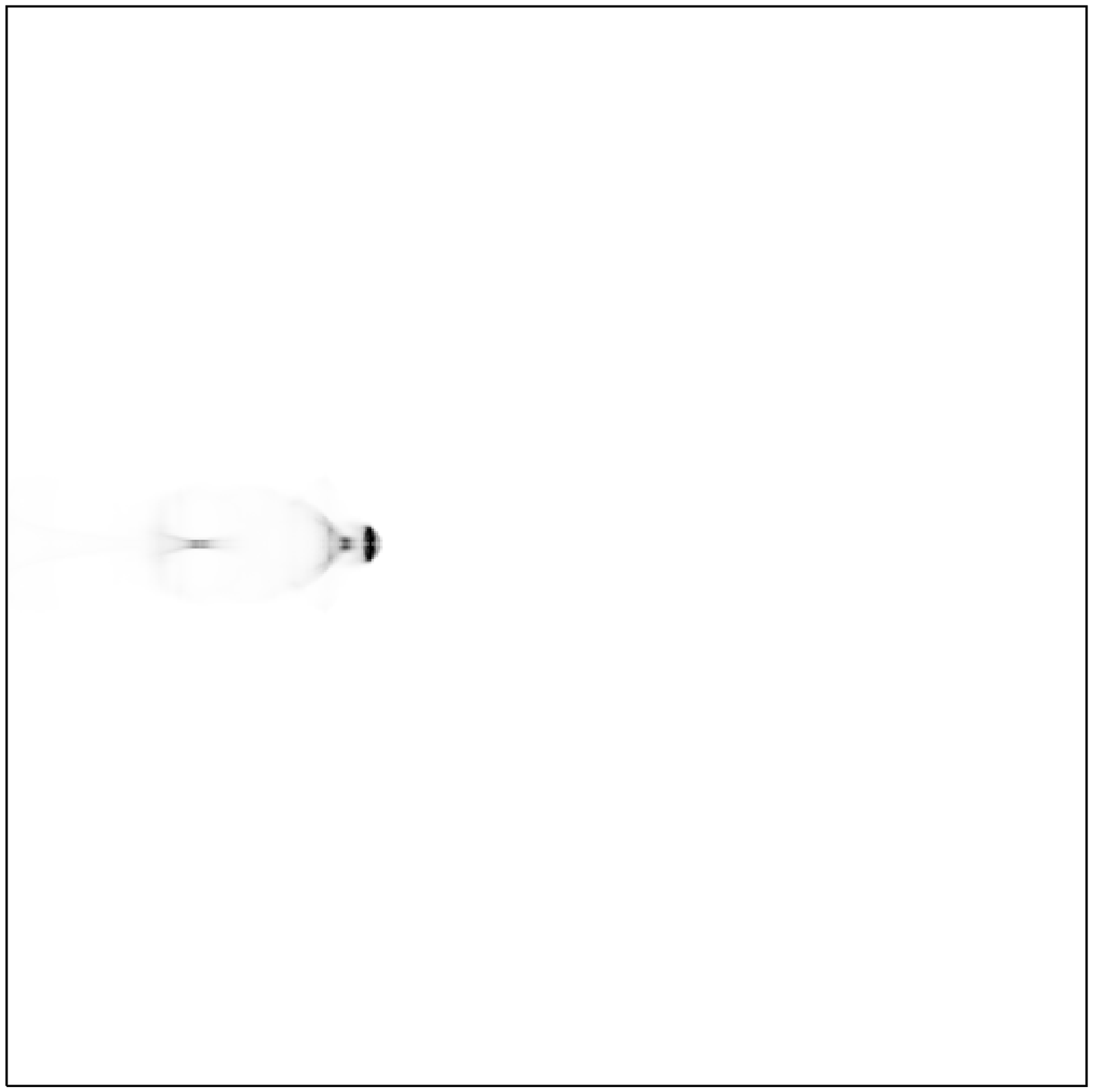}  
&\includegraphics[width=4cm]{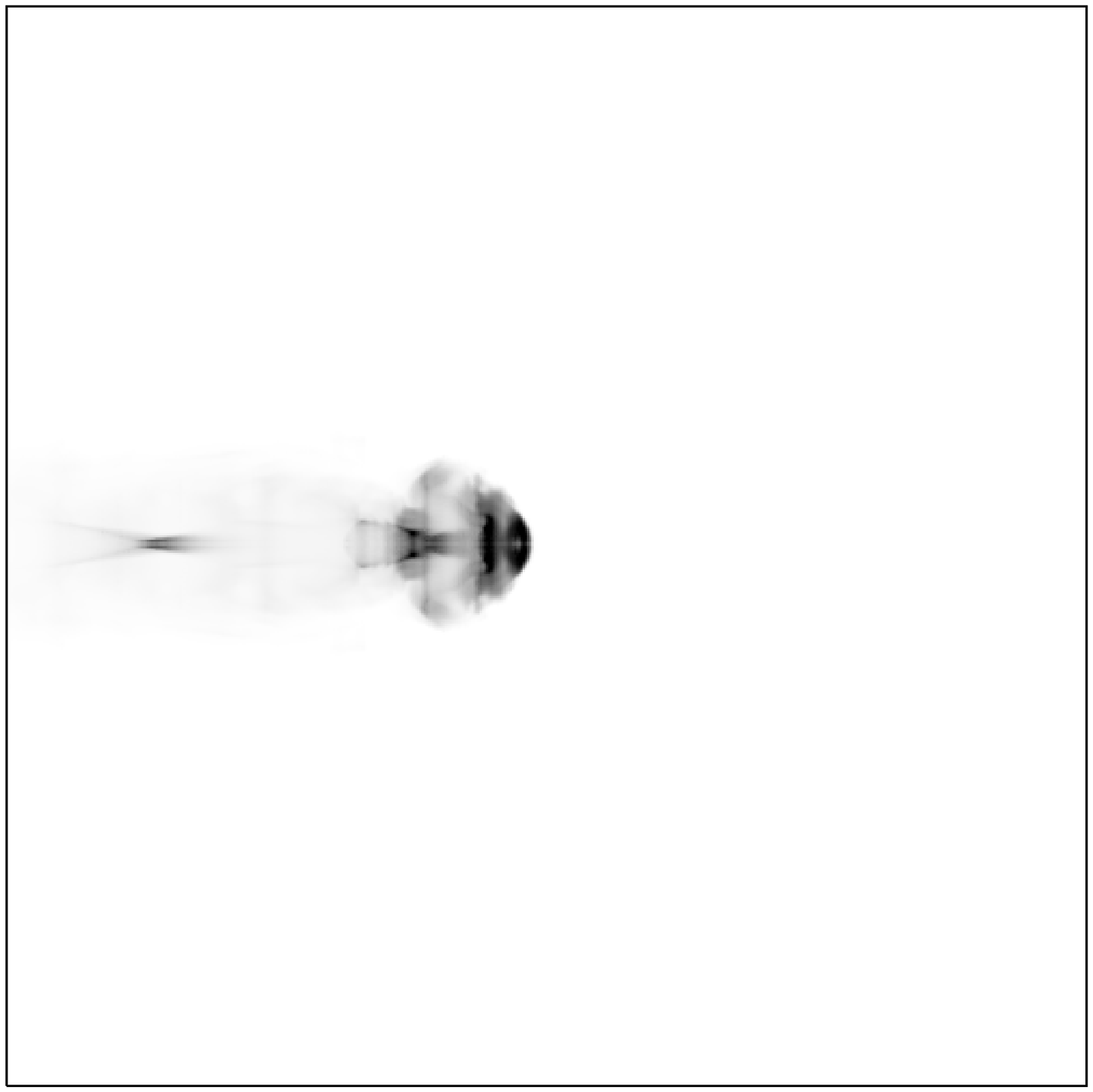}  
&\includegraphics[width=4cm]{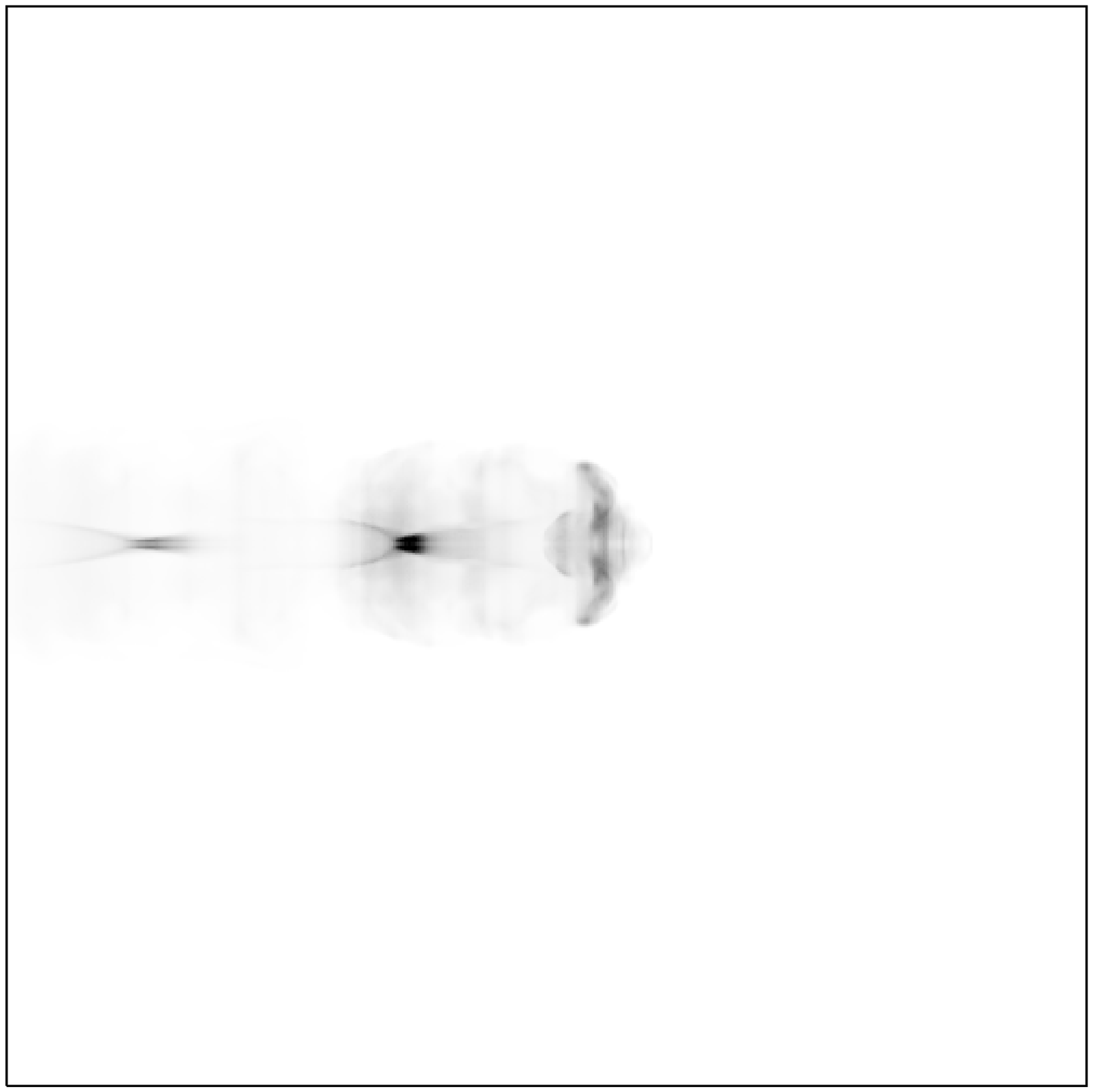}  
\\
 \includegraphics[width=4cm]{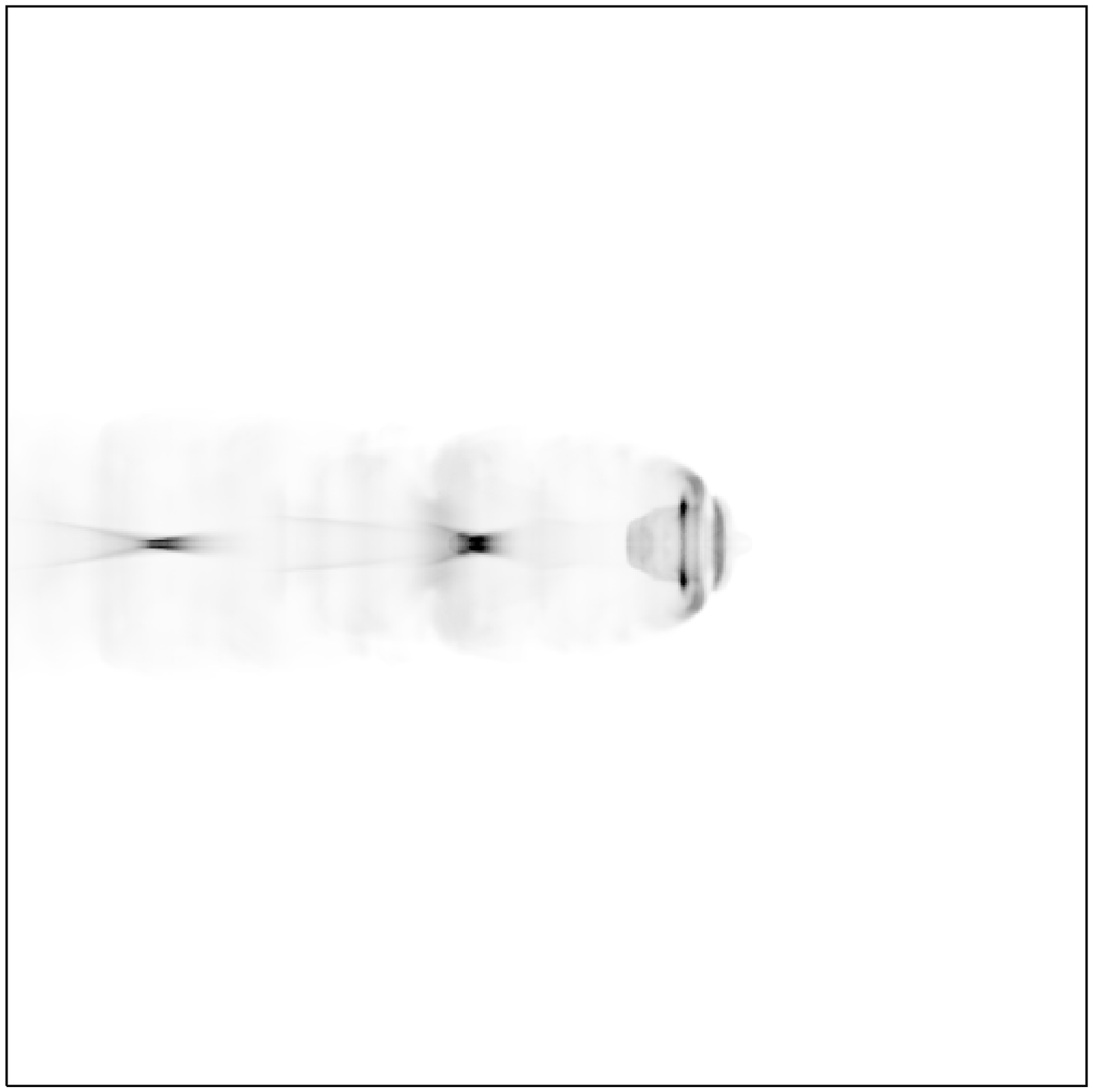}  
&\includegraphics[width=4cm]{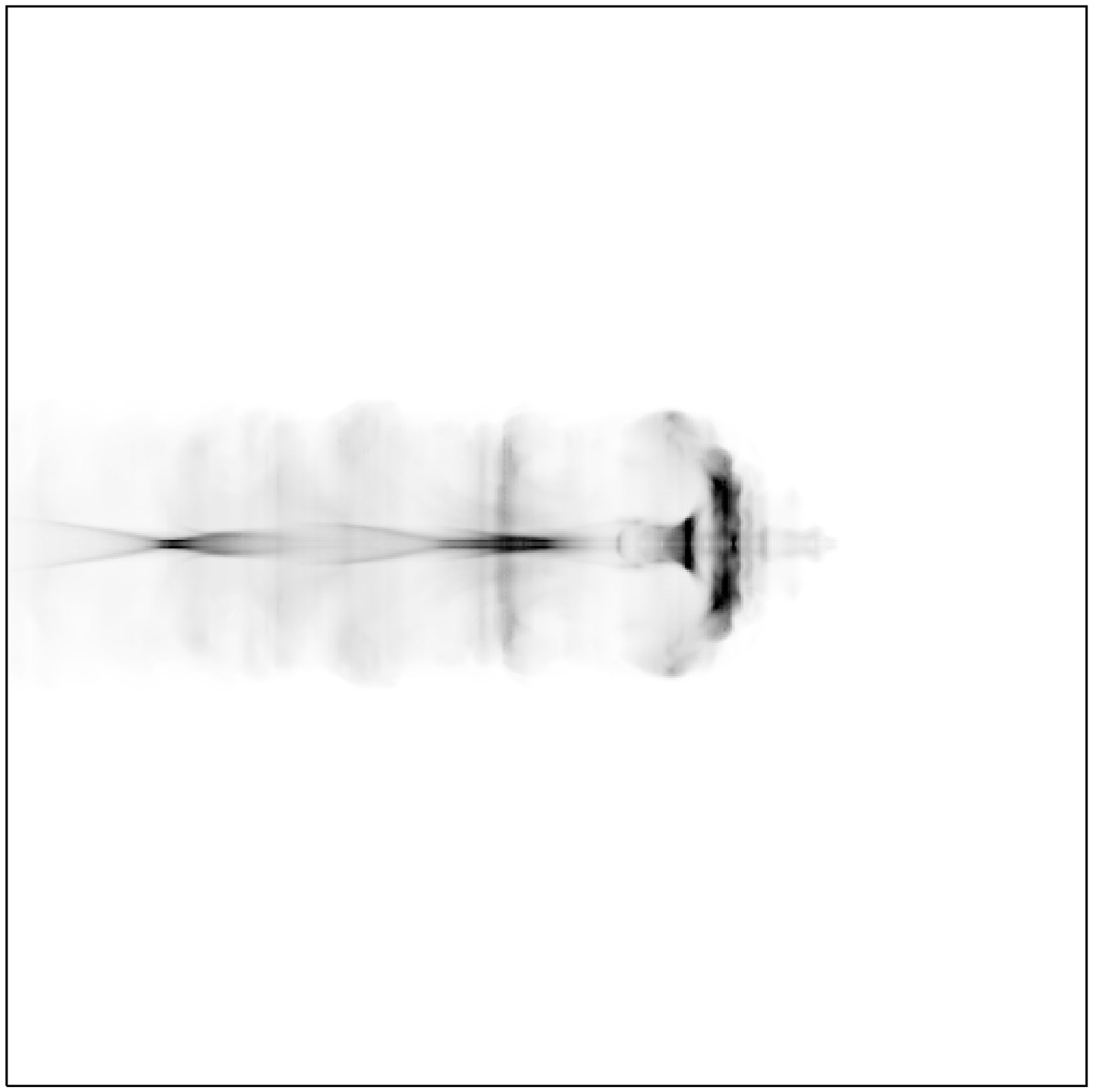}  
&\includegraphics[width=4cm]{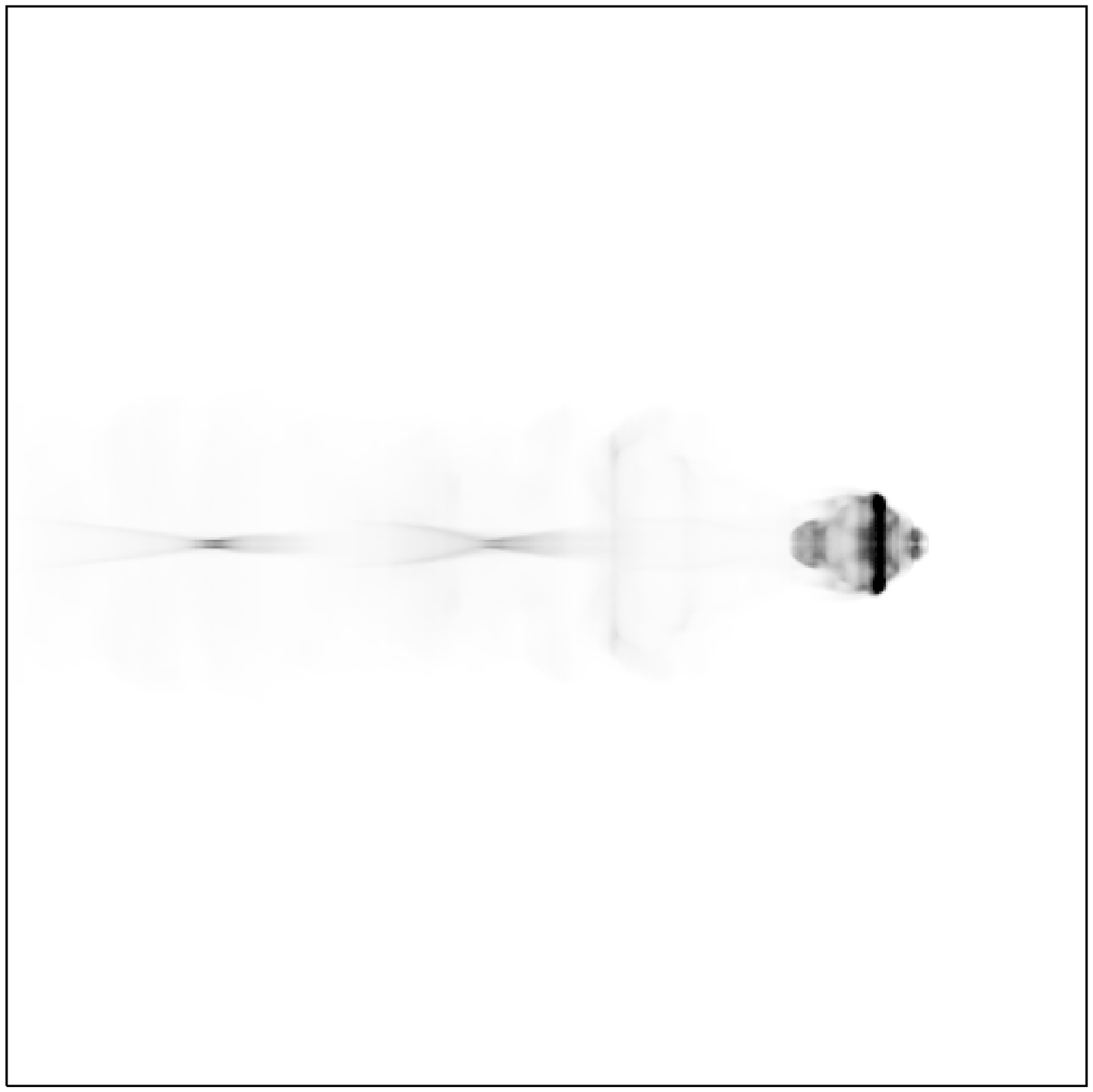}  
&\includegraphics[width=4cm]{int_fast_0430.ps}  
\end{tabular}
\end{center} 
\caption{ 
A time sequence of synchrotron radio intensity images 
  of the {\sc fast1o} jet at various evolutionary stages 
 (frames 54, 108, 161, 215, 269, 323, 376 and 430, 
  corresponding to times
  $t\approx0.217, 0.433, 0.646, 0.863, 1.08, 1.30, 1.51, 1.73$~Myr
  after the initial launch of the jet).  
In this calculation, $\beta=10$
   and the field is ordered (quasi-poloidally $\Bvec\parallel\vvec$).
}
\label{fig.sequence}
\end{figure*}

For jet parameters equivalent to {\sc fast1o},
   \cite{saxton2002a}
   noted axisymmetric shocks spawned off the throbbing hotspot,  
   and transverse filaments emerged,
   resembling structures seen in the western lobe of the radio galaxy Pictor~A.
Here we reproduce comparable models as a benchmark. 
{\color{Black}
   In addition, we have carried out explicit numerical polarized radiative transfer calculations 
   frame by frame to show the emission and structural evolution of the jets.}
Figure~\ref{fig.sequence} shows a sample sequence of image frames 
   of the radio intensities of the evolving {\sc fast1o} jet. 
The morphology is relatively simple at the early times, 
   with the emission dominated by the hotspot. 
As the jet evolves and propagates, 
   internal recollimation shocks begin to appear,
   and filamentary shocks are thrown off the jet's advancing head.
At certain epochs, the filamentary shocks appear ahead of the hotspot, 
   but at other times they are coincident or behind
   \citep[resembling structures in Pictor~A,][]{roesser1987,perley1997,
		wilson2001,tingay2008}.
The hotspot and recollimation shocks appear sharp in contrast to the cocoon. 
Reassuringly, the synchrotron morphologies generated by full radiative transfer
   resemble the approximate imaging by
   \cite{saxton2002a}. 

Radio intensity maps of the {\sc slow} jets
   show more internal shocks for a given length of jet.
Globally the images are less contrasty than for the fast jets.
There is a conspicuous, fuzzy bulge is located near the jet base.
It is turbulent jet-derived plasma in the backflow around the jet.  
When the boundary is open, allowing outflow off the grid
   (e.g. compare the {\sc fast2o} jets to the {\sc fast2} jets), 
   the fuzzy feature diminishes.
We suspect that it might diminish further in the 3D simulations  
   and when the dimming of aged plasma were taken into account. 
The hotspot's local behaviours
   are effectively independent of the left boundary condition   
   when the jet has progressed far enough,
   in the late evolutionary epochs of each simulation. 

Our radiative transfer calculations 
   show that the jet emission structure 
   depends on the assumed model of the magnetic-field configurations.  
If we assign the local magnetic-field orientations
   parallel to the velocity vectors,
   the intensity distribution in the jet image will be quite smooth 
   (after the subtraction of bright artefacts 
   that occasionally appear in the cocoon). 
If a random field orientation is used
   (with directions independent between adjacent cells)
   then the jet image will appear speckly. 
In \S\ref{s.recipes}
   we consider the effect of some alternative configurations
   of the ordered magnetic field.

In some simulations,  
   large eddies with high concentrations of jet plasma
   are spun into large rings (by axisymmetry).
Projections of these structures gives rise to patches with intense emission.  
The similarity of the light-curves in the ``random'' and ``ordered'' cases
   of Figures~\ref{fig.temporal.slow}--\ref{fig.temporal.fast4}
   demonstrates that 
   these twinkles cause only local fluctuations.   
Their effects on spatially integrated intensity and Stokes parameters
   are usually negligible.

For synchrotron emission from relativistic electrons
   with a power-law energy distribution of an index $p$
   gyrating in an orderly oriented magnetic field, 
   the intensity spectral index $\alpha$
   and the linear polarization $\Pi_{\rm L}$ 
   are given by $\alpha=(p-1)/2$
   and $\Pi_{\rm L}=(p+1)/(p+{\frac73})$ respectively.  
Assuming $p=2.2$ in the calculations  
   gives an index $\alpha = 0.6$ of the synchrotron spectra  
   and a maximum linear polarization $\Pi_{\rm L}\la0.7$.   
The peak local values of $\Pi_{\rm L}$ in the simulated jets 
   generally approach the theoretical maximum value of 0.7 
   regardless of the assigned magnetic-field configurations, 
   as expected.  
Synchrotron emission has little intrinsic circular polarization
   \citep{legg1968}.
However in a plasma, some circular polarization can be produced 
   by Faraday conversion of the linear polarization components of radiation.  
{\color{Black}
(We discuss an aspect of polarization evolution briefly in \S\ref{s.stages},
   and in more detail in a forthcoming paper.)
}

Jet pinch shocks show a simple luminous knot
   where the biconical shock occurs.
Hot-spots are composed of more complex and transient shocks,
   resulting in variable, closely spaced luminous patches.
Resembling FR1 radio galaxies,
   the {\sc slow} jets do not have a clear hotspot,
   and so the cocoon dominates the radio emission. 
For random fields, the jet is inconspicuous.
There is little cocoon obscuration in the {\sc fast1o} jets, 
   and the jet shocks and the hotspot are distinct,
   like the morphology of FR2 sources.

\subsubsection{alternative environments}

We now comment on the characteristics
   of several variants of simulated fast jets:

{\sc fast1}: 
The jets propagate in a uniform background density, 
   as was the case for the {\sc fast1o} jets. 
However a closed left boundary is now imposed.
For {\sc fast1} jets, 
   the jet itself and the shocks are generally distinguishable,
   but cocoon emission is more apparent.
Many pinch shocks appear in the jet
  at the late stages of the evolution. 
The hotspot surges irregularly back and forth,
  typically traversing about a third the jet's length.

{\sc fast1ro}:
   The background is uniform and the left boundary is open to outflow,
   but the adiabatic index $\gamma=4/3$.
Compared to {\sc fast1o},
   the cocoon is narrower and the jet takes longer to cross the grid.

{\sc fast1r}:
   This is equivalent to {\sc fast1ro} except that the left boundary is closed.
Closure causes a wider cocoon to accumulate
   (as in {\sc fast1}), despite the choice of $\gamma=4/3$.

\begin{figure}
\begin{center} 
\begin{tabular}{cc}
 \includegraphics[width=6cm]{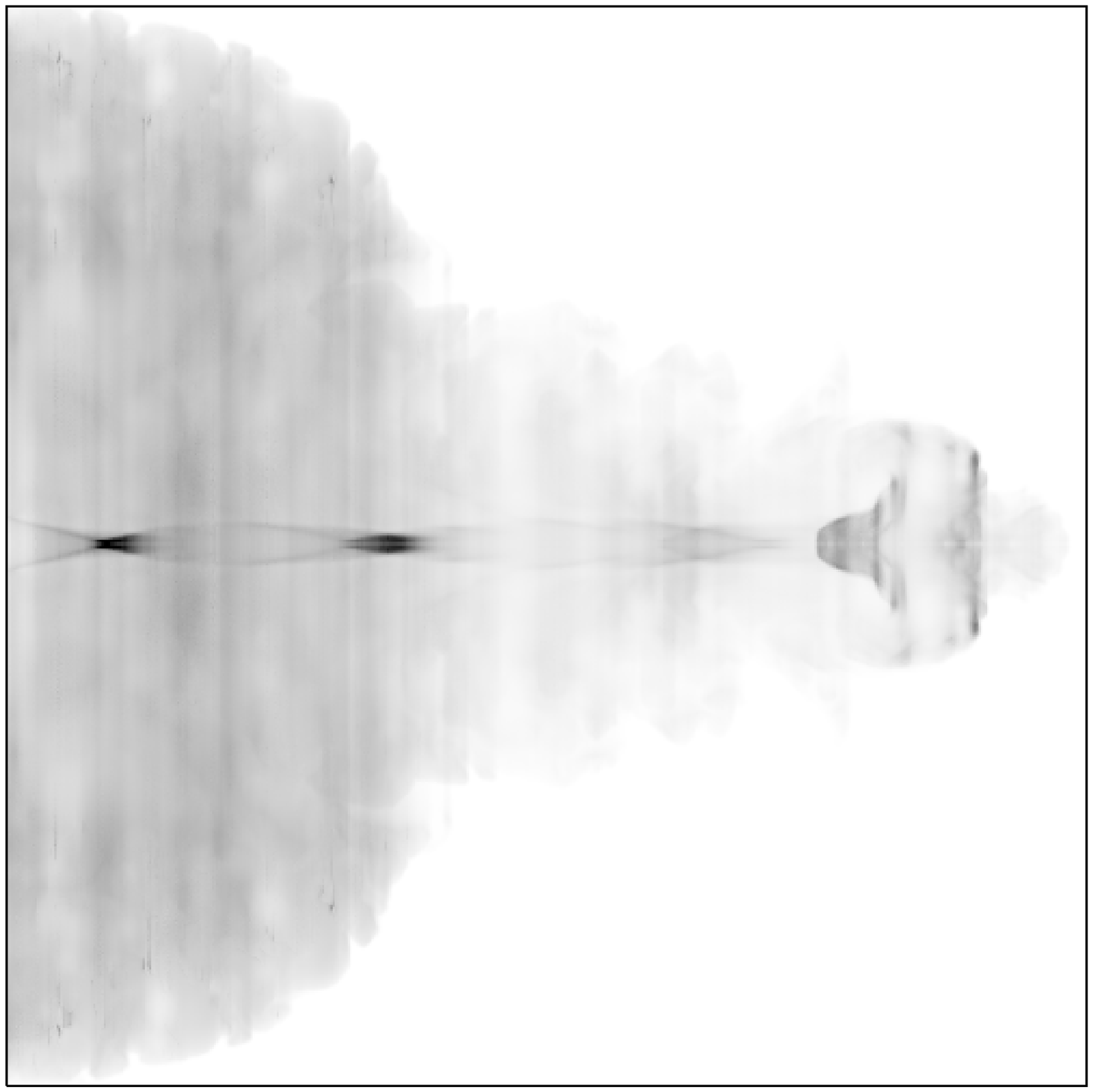}
\end{tabular}
\end{center} 
\caption{ 
The last frame of the {\sc fast2} jet simulation.  
In the simulation, $\beta=10$, 
  and the local magnetic field orientation is aligned with the flow velocity.
Note that the hotspot is developing into a funnel morphology.
}
\label{fig.pica2}
\end{figure}

{\sc fast2}: 
The jets propagate in a radially declining background density. 
The jet emission remains distinctly visible through a cloudy cocoon
   (e.g. Figure~\ref{fig.pica2}).  
There are fewer pinch shocks than the case with a uniform background density, 
  and the shocks are farther apart.
The first two pinch shocks tend to outshine the hotspot.  
The hotspot throws off rings or filaments,
   as in the case with a uniform background density.  

{\sc fast2o}: 
An open boundary reduces the emission intensity of the cocoon. 
Apart from the reduction of the emission of the cocoon, 
   the emission structure of the {\sc fast2o} jets
   resembles that of the {\sc fast2} jets. 
The jet and hotspot are clearer
   than those of the {\sc fast2} jets.

\begin{figure}
\begin{center} 
\begin{tabular}{cc}
 \includegraphics[width=6cm]{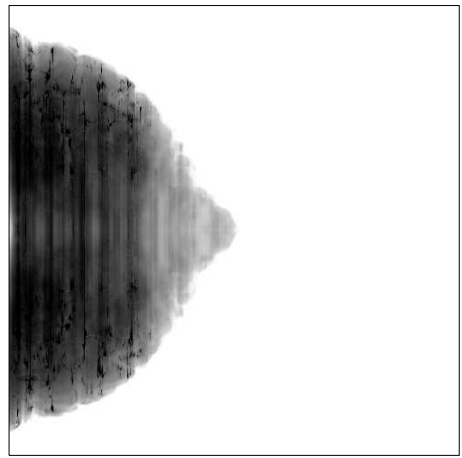}
\end{tabular}
\end{center} 
\caption{ 
The 6~cm synchrotron radio intensity ($I$) map
of the final frame of the {\sc fast3} jet simulation. 
In this simulation, the jet encounters a rising density profile. 
The magnetic field is quasi-poloidal ($\Bvec\parallel\vvec$)
and $\beta=10$.
}
\label{fig.pica3_Pic}
\end{figure}

{\sc fast3}:
A rising background density confines the expansion of the turbulent cocoon. 
The jet's advance and cocoon's growth decelerate
   (the front of the bow shock $x_{\rm s}\propto t^{0.44}$).
We stop the simulation before the jet traverses the grid.
With ordered fields and $\beta=10$,
   the late-time images only show a translucent cocoon
   with eddy induced substructures.
(The case with $\beta=1$ is unassessable,
   because the twinkle artefacts dominate.)
The jet is obscured and there is no apparent hotspot
    (see Figure~\ref{fig.pica3_Pic}).

{\sc fast4}: 
{\color{Black}%
The background medium has a ripply density variation.   
Naturally, the intensity of the jet emission varies
   as the jet propagates into these shells.}  
In spite of the density variation,  
  the jet morphologies are similar to those of the {\sc fast1} jets. 
The emission of the cocoon dominates in the intensity image, 
   as in all cases with a closed left boundary.
There is some subtle pinching of the cocoon outline
  where the jet reaches breakthrough points.   
At later evolutionary epochs, 
  the jet head surges stochastically as far as half the jet's length.

\subsubsection{magnetic field configuration}
\label{s.recipes}

\begin{figure}
\begin{center} 
\begin{tabular}{ccc}
 \includegraphics[width=6cm]{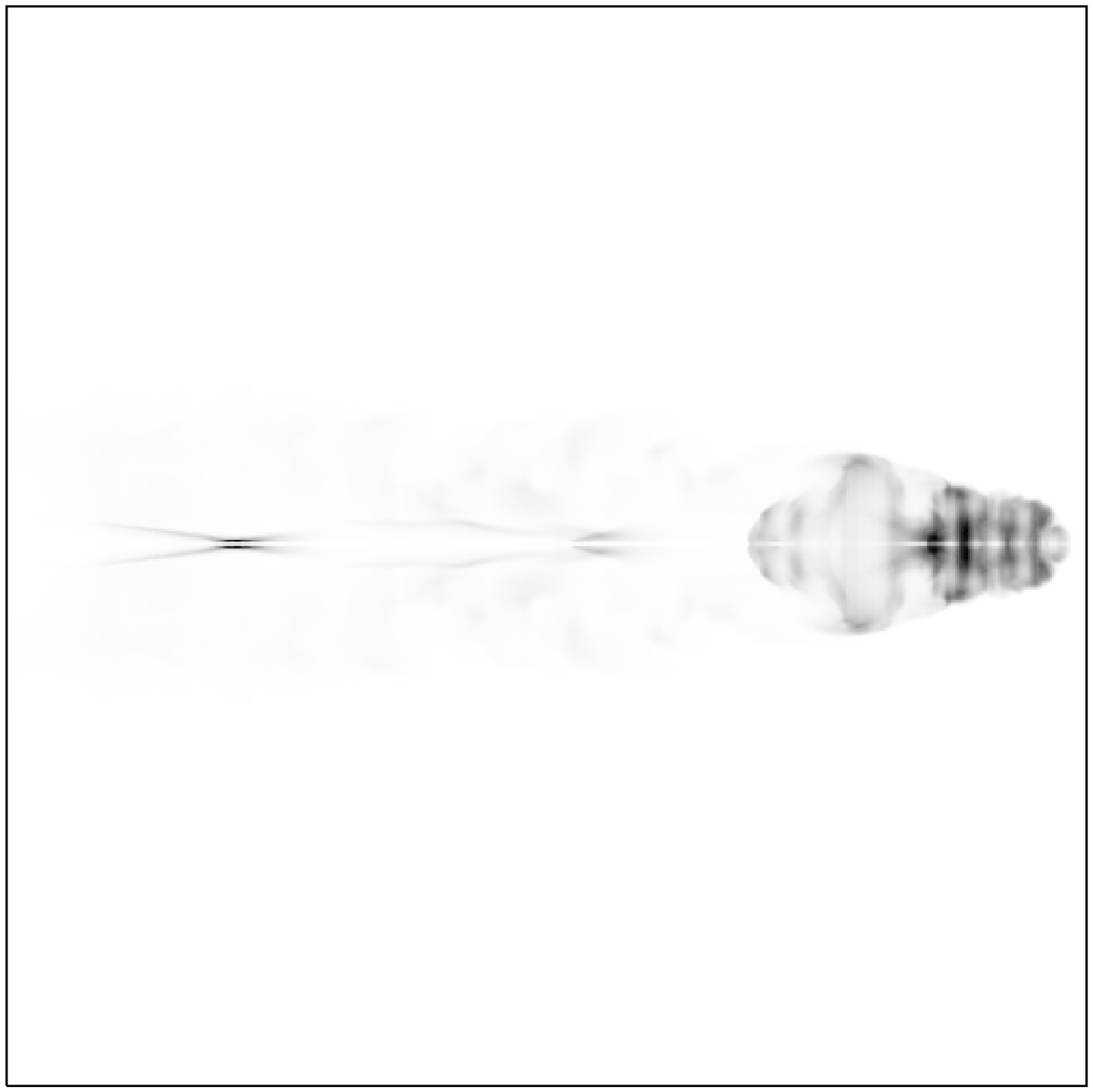}
\\
 \includegraphics[width=6cm]{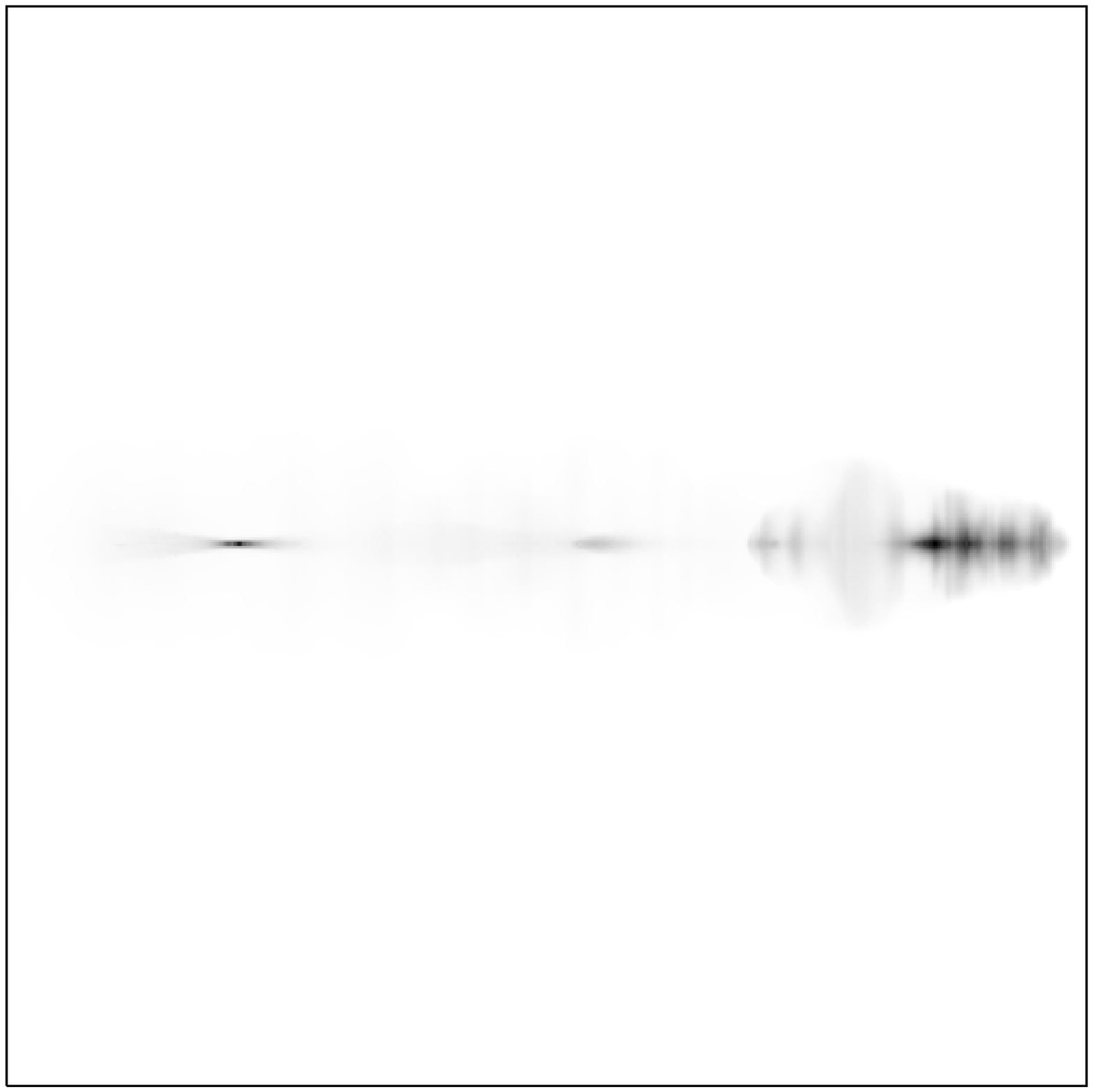}
\end{tabular}
\end{center} 
\caption{ 
Synchrotron radio intensity ($I$) maps (at 6~cm) 
  of the fast jet as in Figure~\ref{fig.advanced} ({\sc fast1o}),
  but with a different prescription for the magnetic field geometry.
Top panel shows 
  {\sc turn1},
  with a quasi-radial field ($\Bvec\perp\vvec$).
Bottom panel shows the result of a quasi-toroidal recipe ({\sc curl1},
   where $\Bvec\parallel\nabla\times\vvec$).
}
\label{fig.recipes}
\end{figure}

In our standard calculations, we have assumed that  
   the magnetic field is parallel to the flow, i.e.\  $\Bvec\parallel\vvec$. 
The field strength is parametrized,
   and its energy density scales with plasma thermal pressure.
This corresponds to a quasi-poloidal field in the jet,
   but with ``live'' deviations at transient features
   such as shocks and vortices. 
As radio synchrotron radiation is determined by the magnetic field, 
   we therefore investigate the jet emission
   for other magnetic field configurations. 
As a test of generality,
   we perform additional calculations with two alternative field configurations.
   (Figure~\ref{fig.recipes}).

In the sequence of the {\sc turn1} simulations,
   we rerun calculations for the {\sc fast1o} simulations, 
   with magnetic field directions perpendicular to local velocities.
The cylindrical components are
   $\Bvec=(B_z,B_r,B_\varphi)\parallel(v_r,-v_z,0)$.
Fields in the jet tend converge towards the axis
   (a quasi-radial configuration)
   which may be less physical than the usual $\Bvec$ recipe.
The emission morphology is essentially the same
   as that of the {\sc fast1o} jets 
   (top panels, Figure~\ref{fig.recipes}),
   but with a dark midline due to symmetry in $\Bvec$.

In a futher sequence of the {\sc curl1} simulations,
   we set $\Bvec\parallel\nabla\times\vvec$.
This prescription guarantees $\nabla\cdot\Bvec=0$,
   and yields quasi-toroidal fields around the jet.
In the intensity images, the brightness is concentrated closer to the axis,
   but knots and the hotspot still appear in the usual places
   (bottom panels, Figure~\ref{fig.recipes}).

The general appearance of the jet and local features
   do not vary significantly under drastic changes of the $\Bvec$ configuration.
The 6~cm radio emission morphologies are limited by the underlying 
   plasma density and pressure distributions.
Within the luminous features,
   there are cosmetic variations caused by the minor polarisation effects
   and magnetic configuration.
Mindful of these small differences,
   we proceed under the fiducial, quasi-poloidal assumption,
   and calculate radio maps and light-curves
   of fast jets penetrating various non-uniform media.

{\color{Black}
\subsection{Effects on jet evolution and emission variability}
}

\begin{figure*}
\begin{center}
\begin{tabular}{ccc}
 \includegraphics[width=5cm]{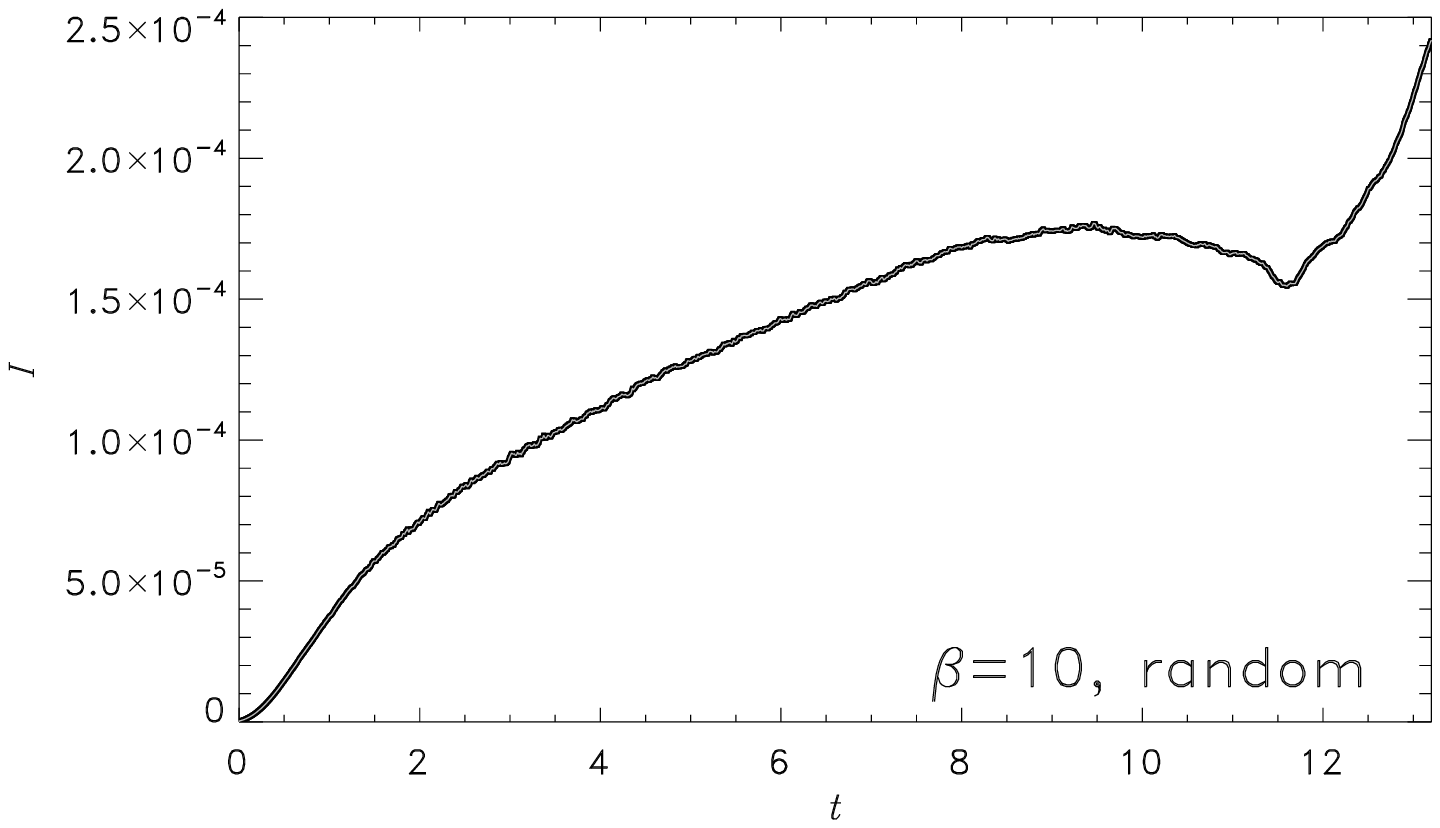}
&\includegraphics[width=5cm]{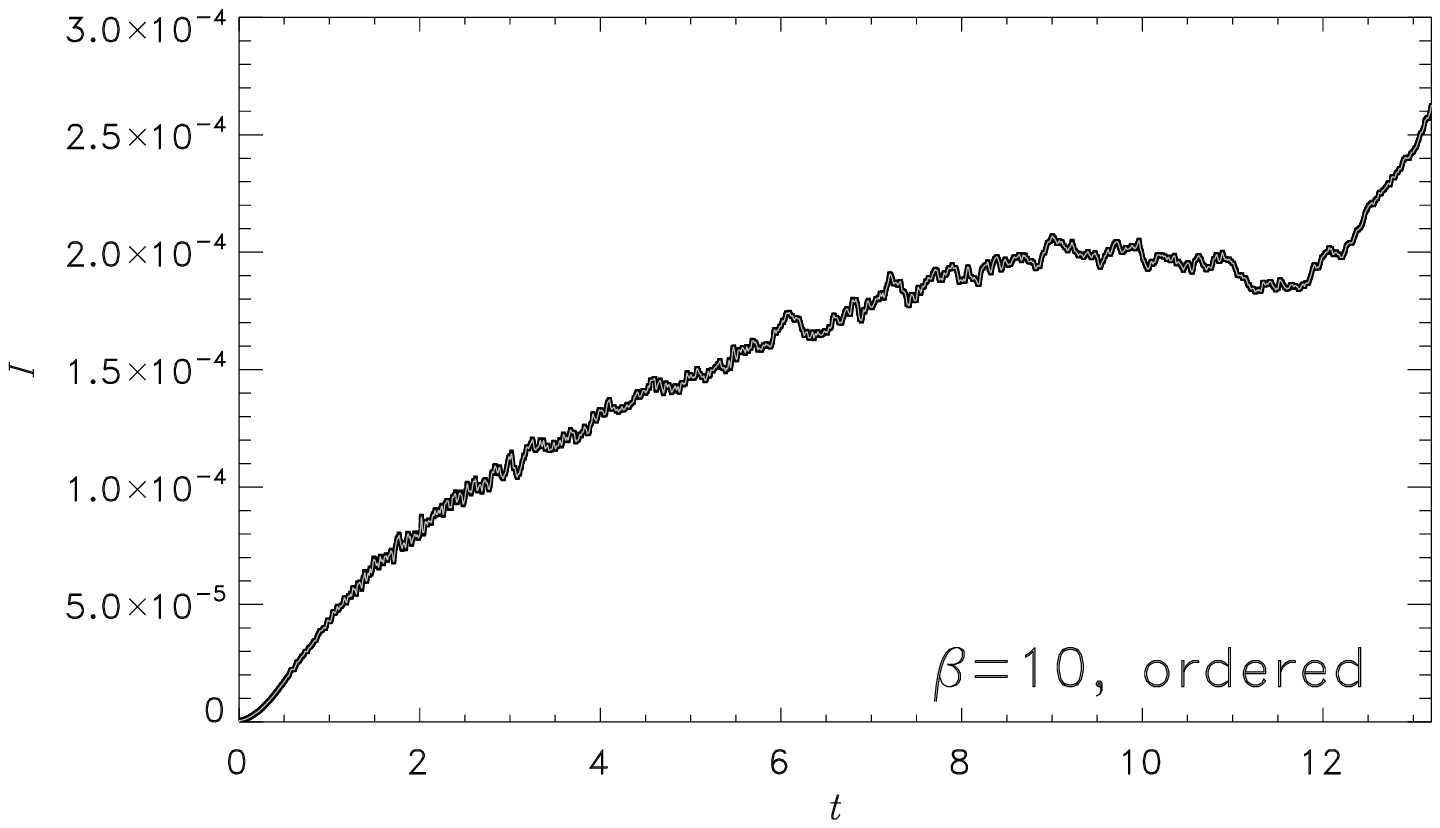}
&\includegraphics[width=5cm]{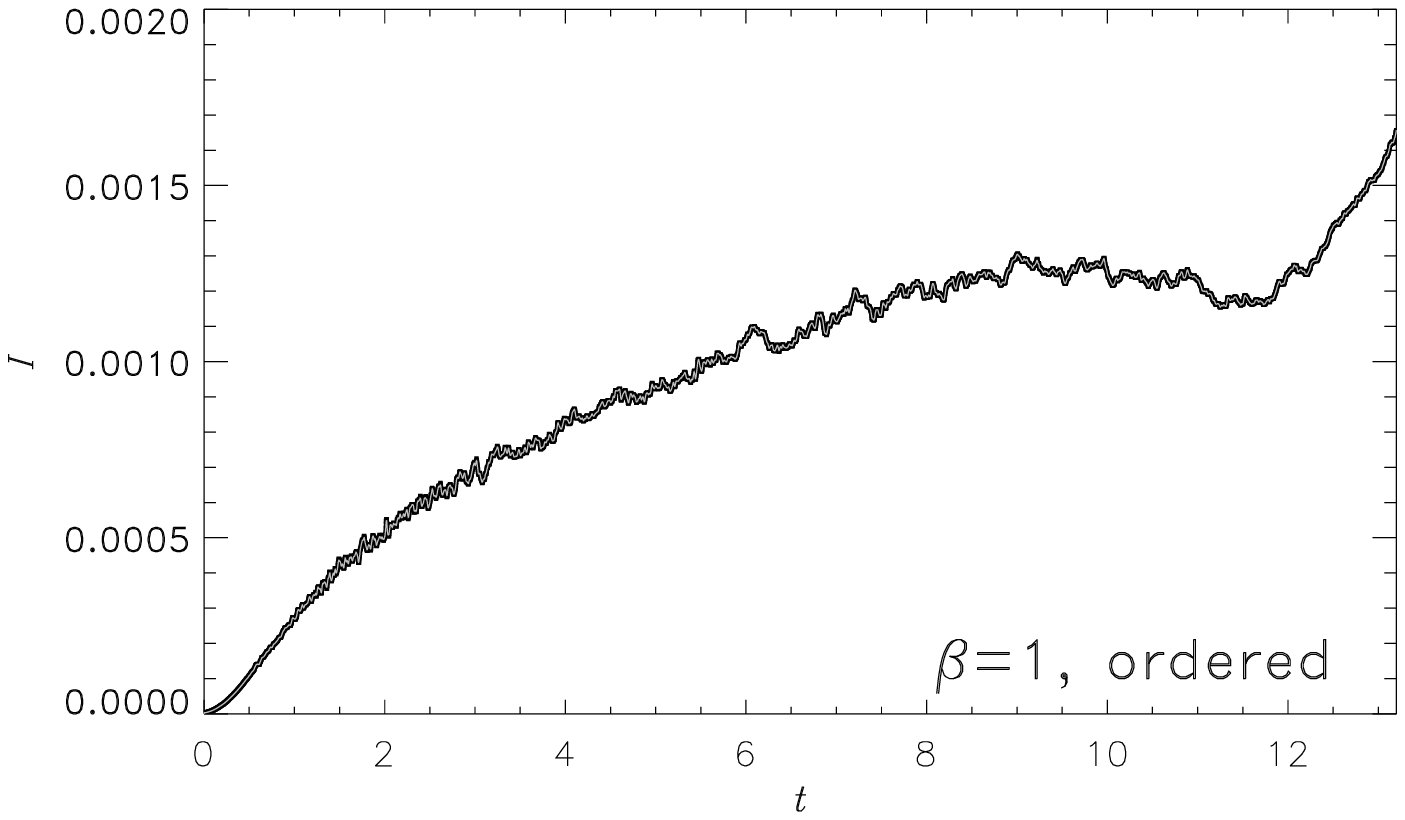}
\end{tabular}
\end{center}
\caption{
Light curves of the simulated {\sc slow} jet
   (uniform background; closed boundary).
The left column show the case with a random field orientation and $\beta=10$. 
The middle and right columns show cases
   with ordered local fields aligned with the matter flow velocities, 
   with $\beta=10, 1$ respectively.
The integrated 6~cm synchrotron radio intensity $I$ light curves
   are on the top panels. 
Note that the fuzzy cocoon dominates the emission,
   masking variability on short timescales.
}
\label{fig.temporal.slow}
\end{figure*}

\begin{figure*}
\begin{center}
\begin{tabular}{ccc}
 \includegraphics[width=5cm]{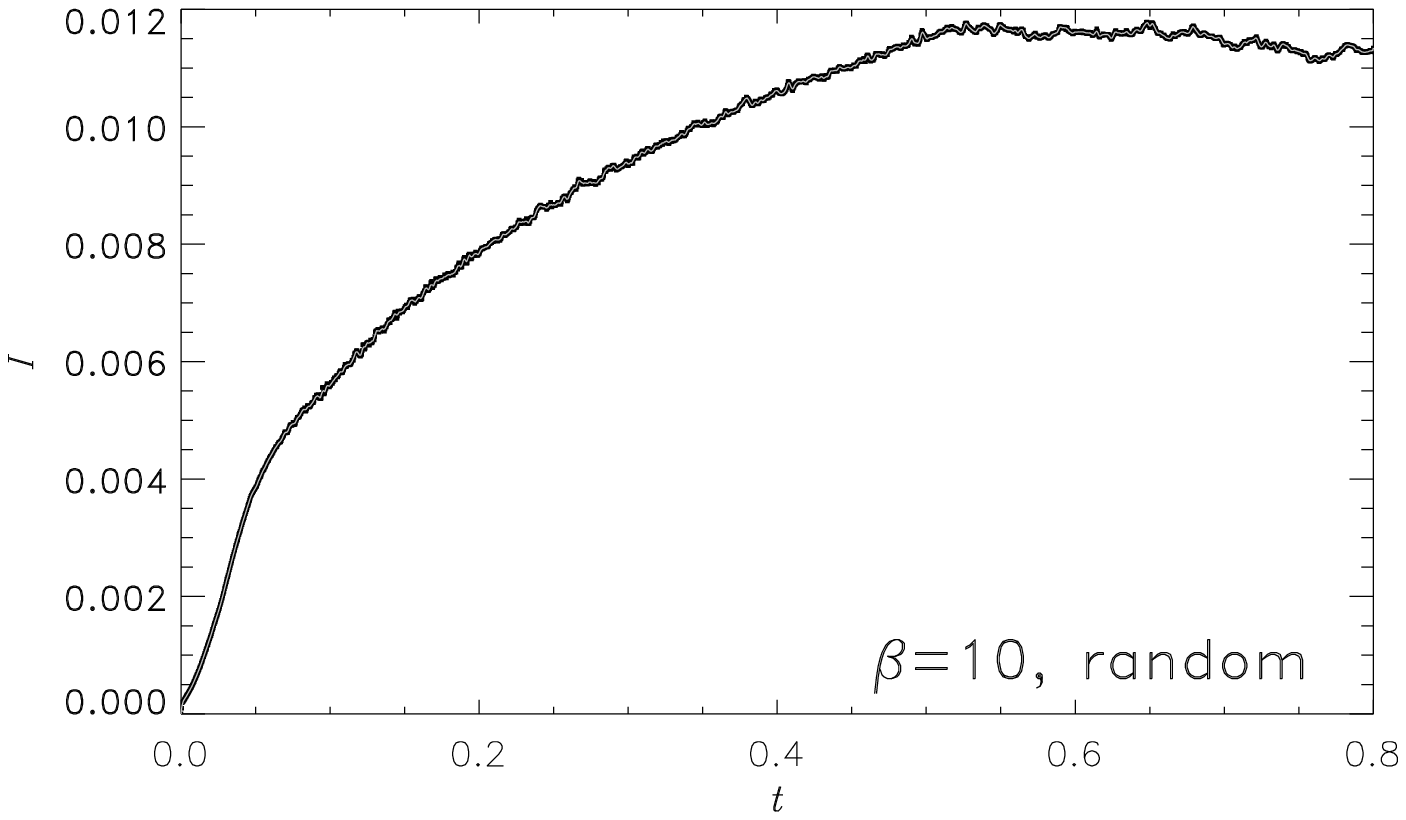}
&\includegraphics[width=5cm]{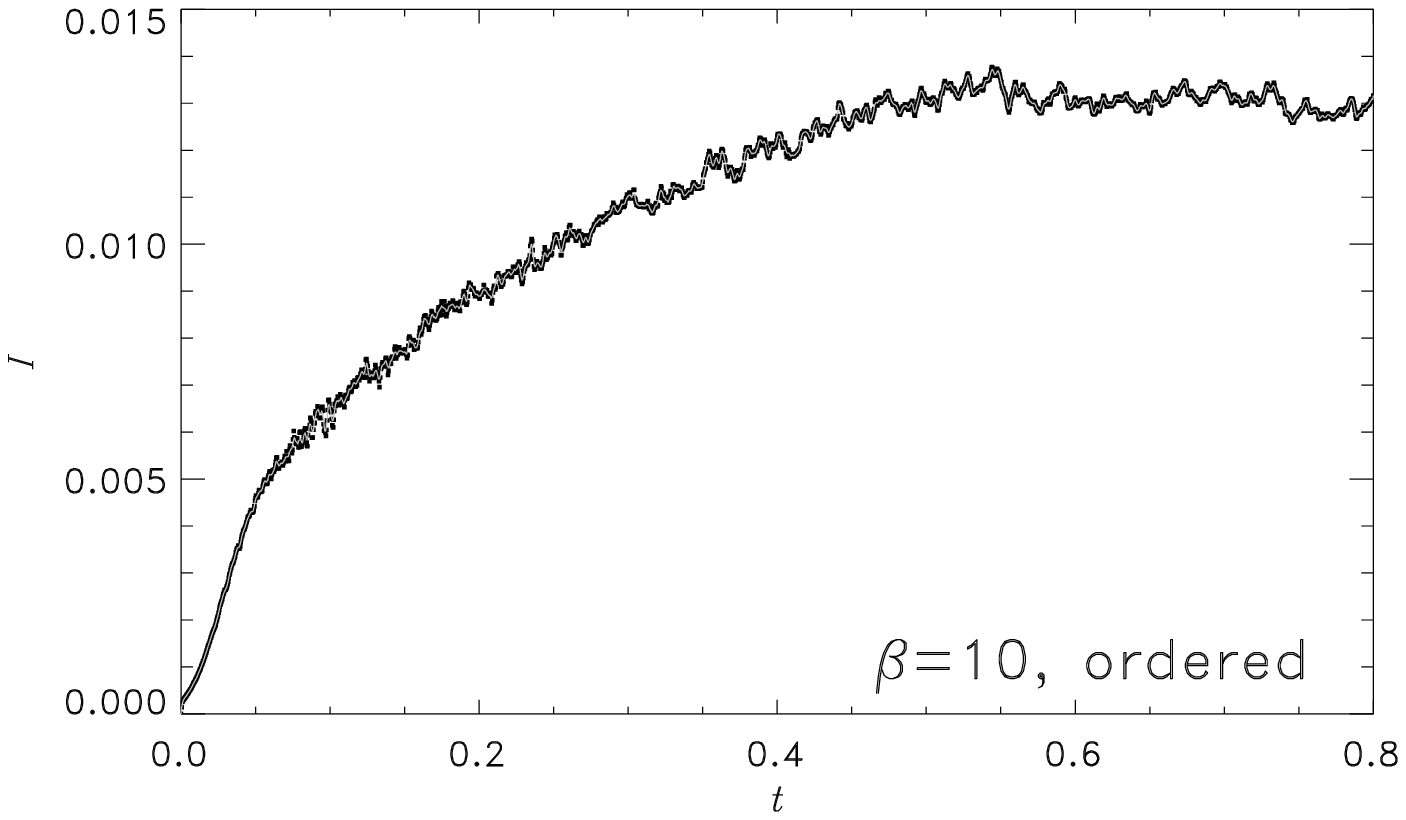}
&\includegraphics[width=5cm]{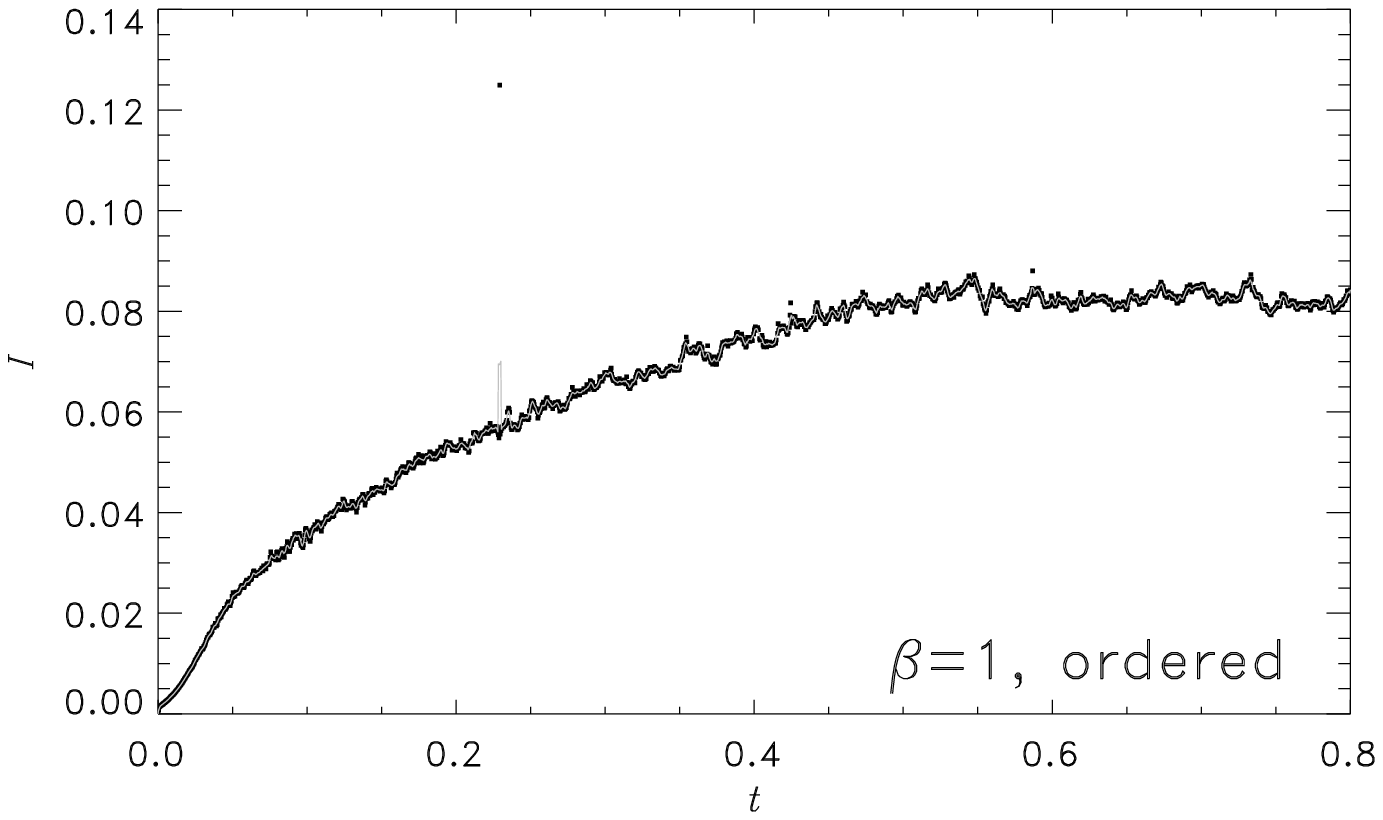}
\end{tabular}
\end{center}
\caption{Same as Figure~\ref{fig.temporal.slow}
   but for the {\sc fast1} jet simulations.  
In the simulations, a fast jet propagates in uniform background medium, 
   and a closed boundary condition is adopted.
Compared to {\sc fast1o},  there is more emission from the cocoon. 
The lightcurves thus track the expansion
   and emission fluctuations of the cocoon 
   rather than the variability in the jet shocks.
}
\label{fig.temporal.fast1c}
\end{figure*}

\begin{figure*}
\begin{center}
\begin{tabular}{ccc}
 \includegraphics[width=5cm]{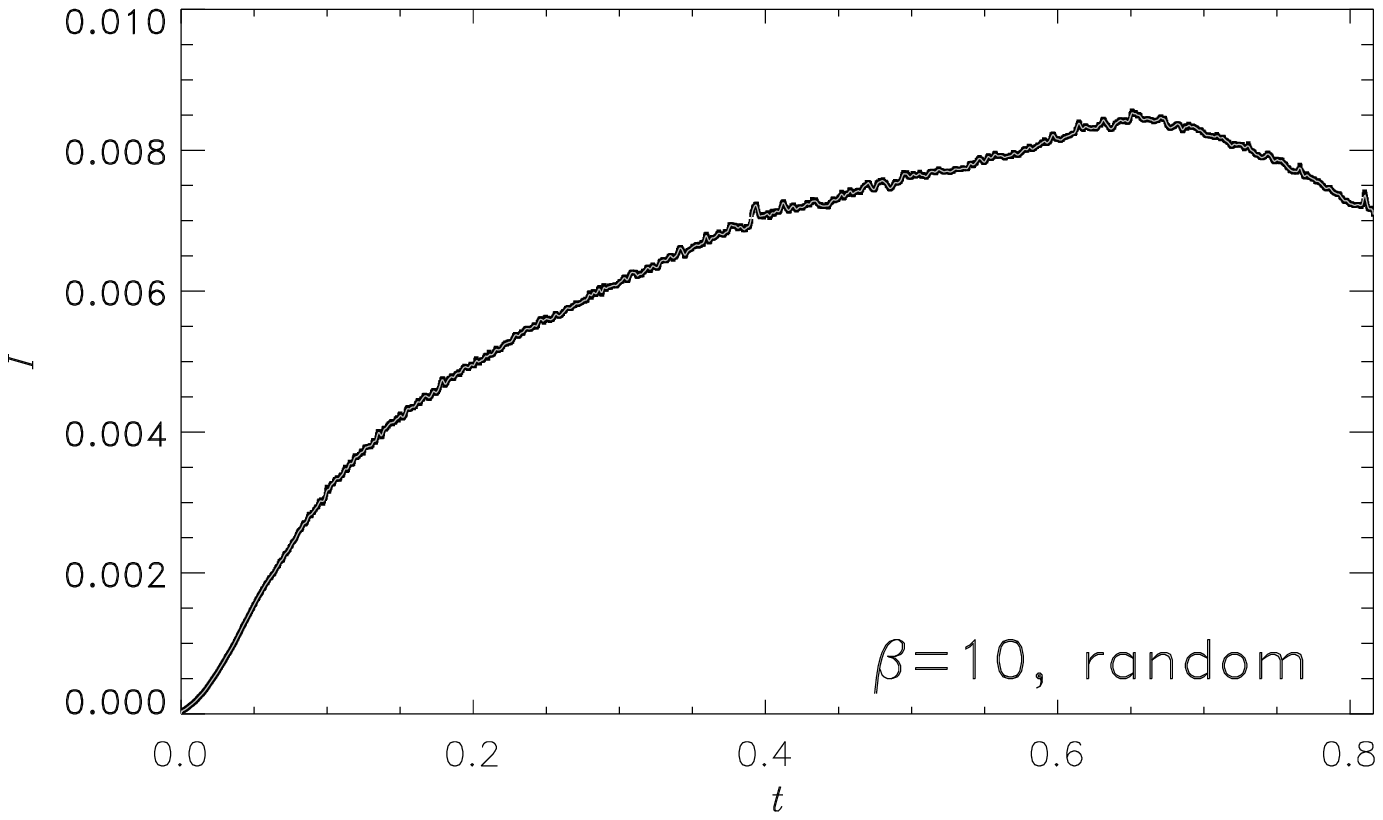}
&\includegraphics[width=5cm]{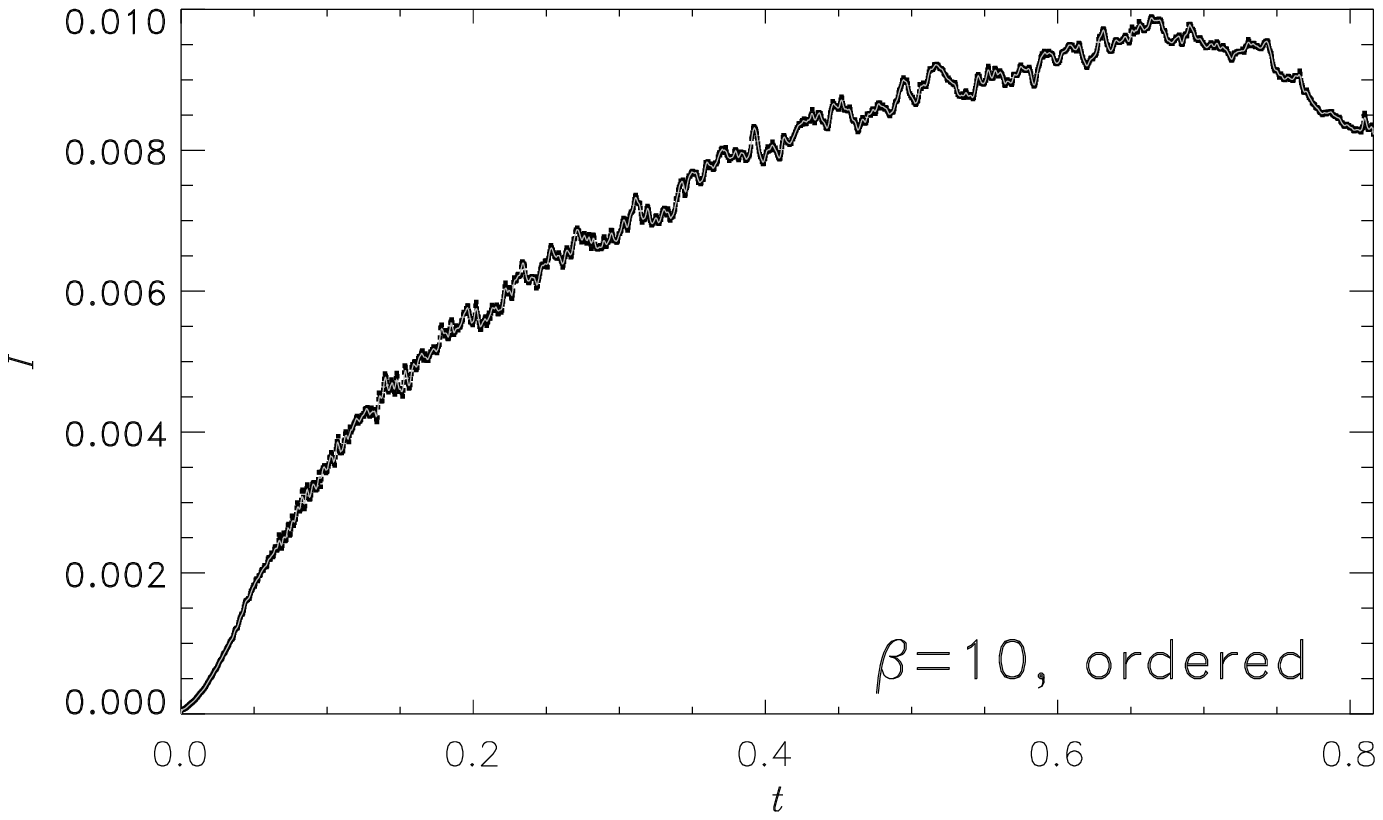}
&\includegraphics[width=5cm]{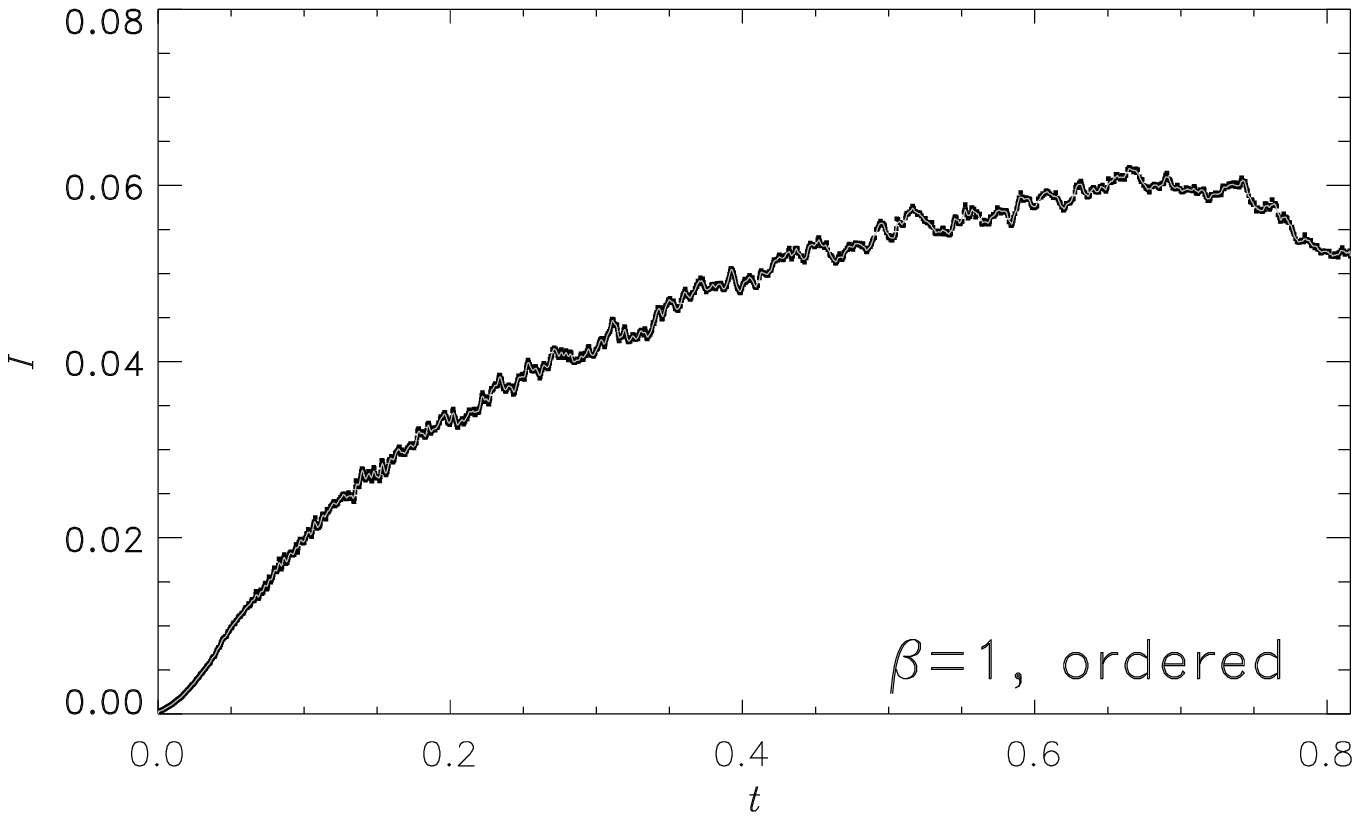}
\end{tabular}
\end{center}
\caption{
Same as Figure~\ref{fig.temporal.slow}, but for the {\sc fast1rc} jet simulations.  
In the simulations, a fast jet propagates in a uniform background medium, 
   and a closed boundary condition is adopted.  
The jet plasma is relativistic, with $\gamma=4/3$.
}
\label{fig.temporal.fast1rc}
\end{figure*}

\begin{figure*}
\begin{center}
\begin{tabular}{ccc}
 \includegraphics[width=5cm]{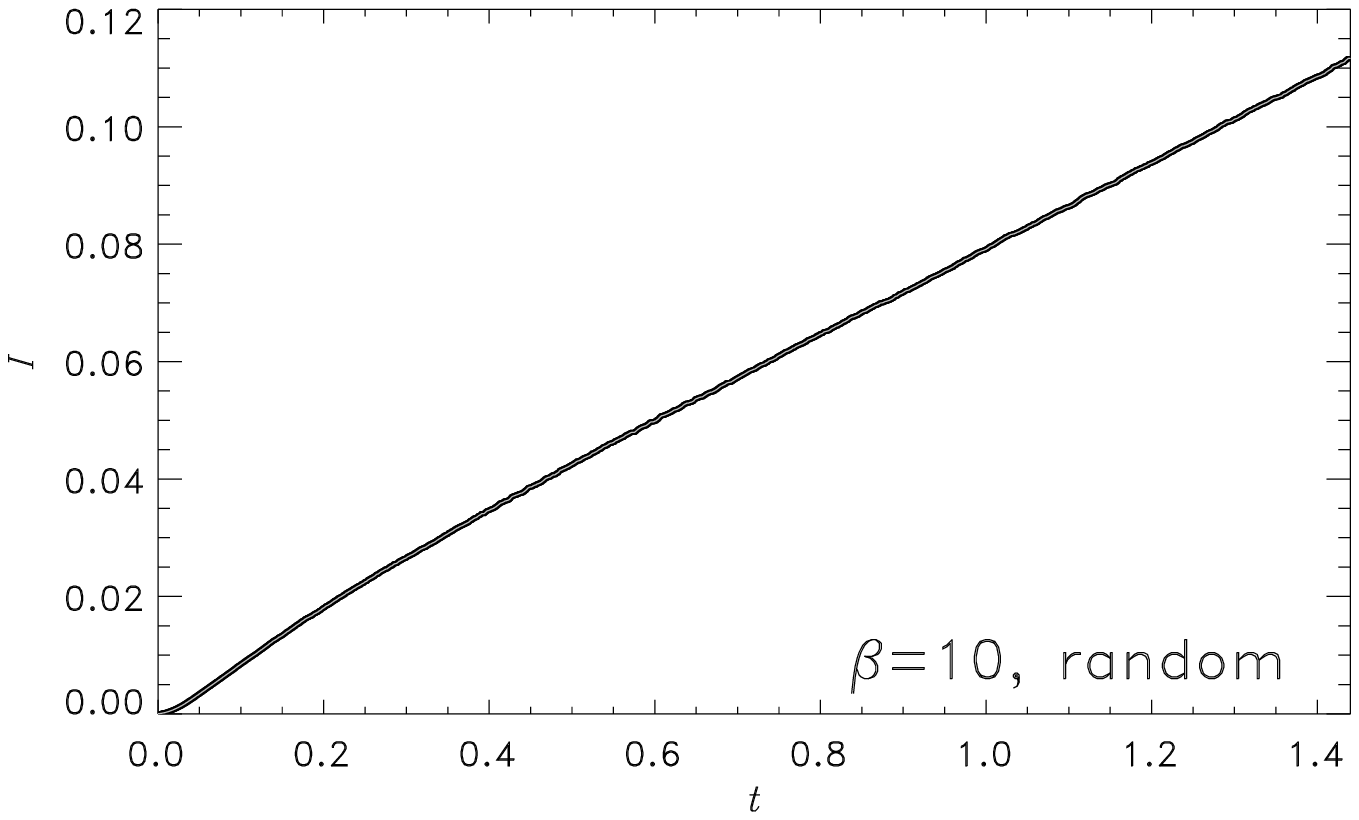}
&\includegraphics[width=5cm]{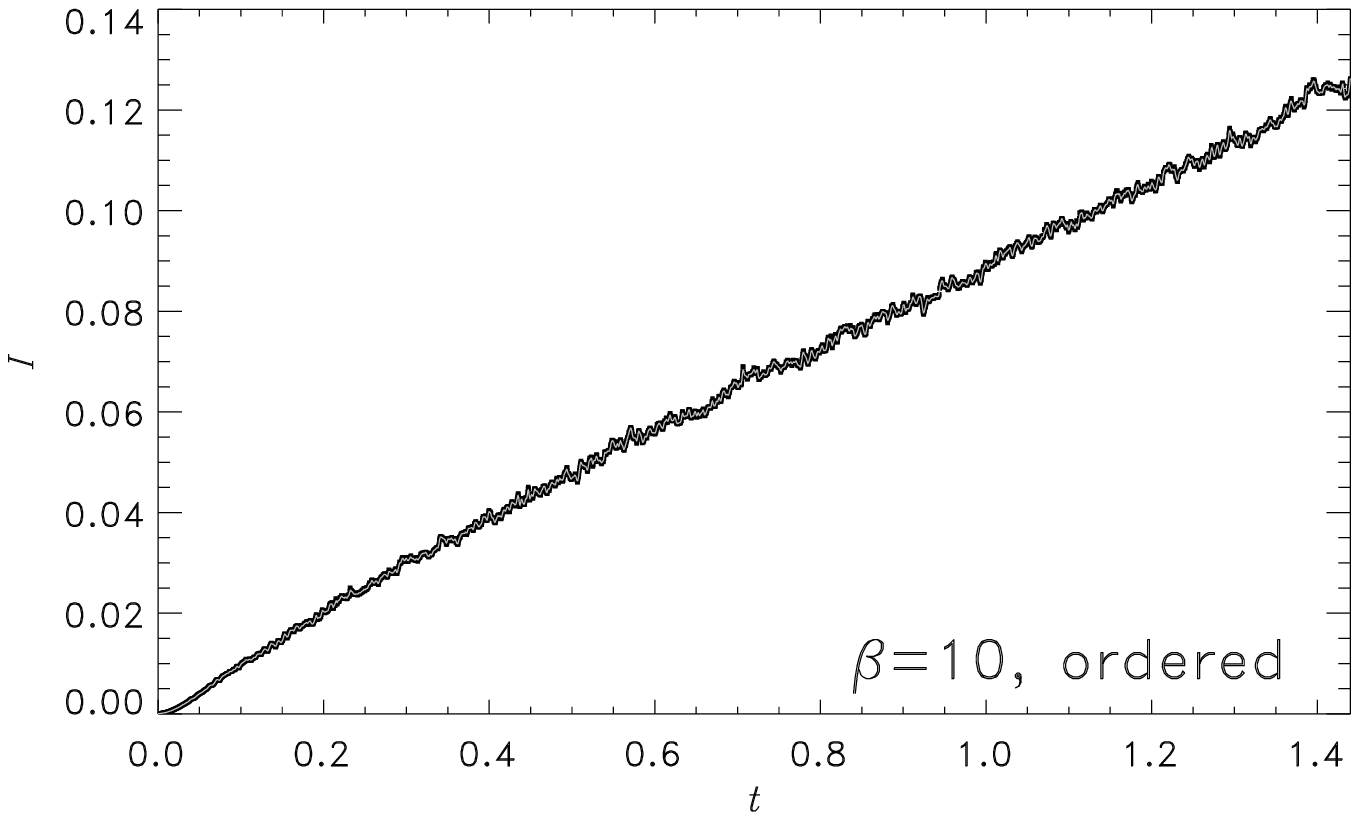}
\end{tabular}
\end{center}
\caption{
Synchrotron radio lightcurves (at 6~cm) of the {\sc fast3} jets. 
In the simulations, a fast jet propagates in a medium with a rising density. 
The jet generates an opaque fireball with no identifiable hotspot.
}
\label{fig.temporal.fast3}
\end{figure*}

\begin{figure*}
\begin{center}
\begin{tabular}{ccc}
 \includegraphics[width=5cm]{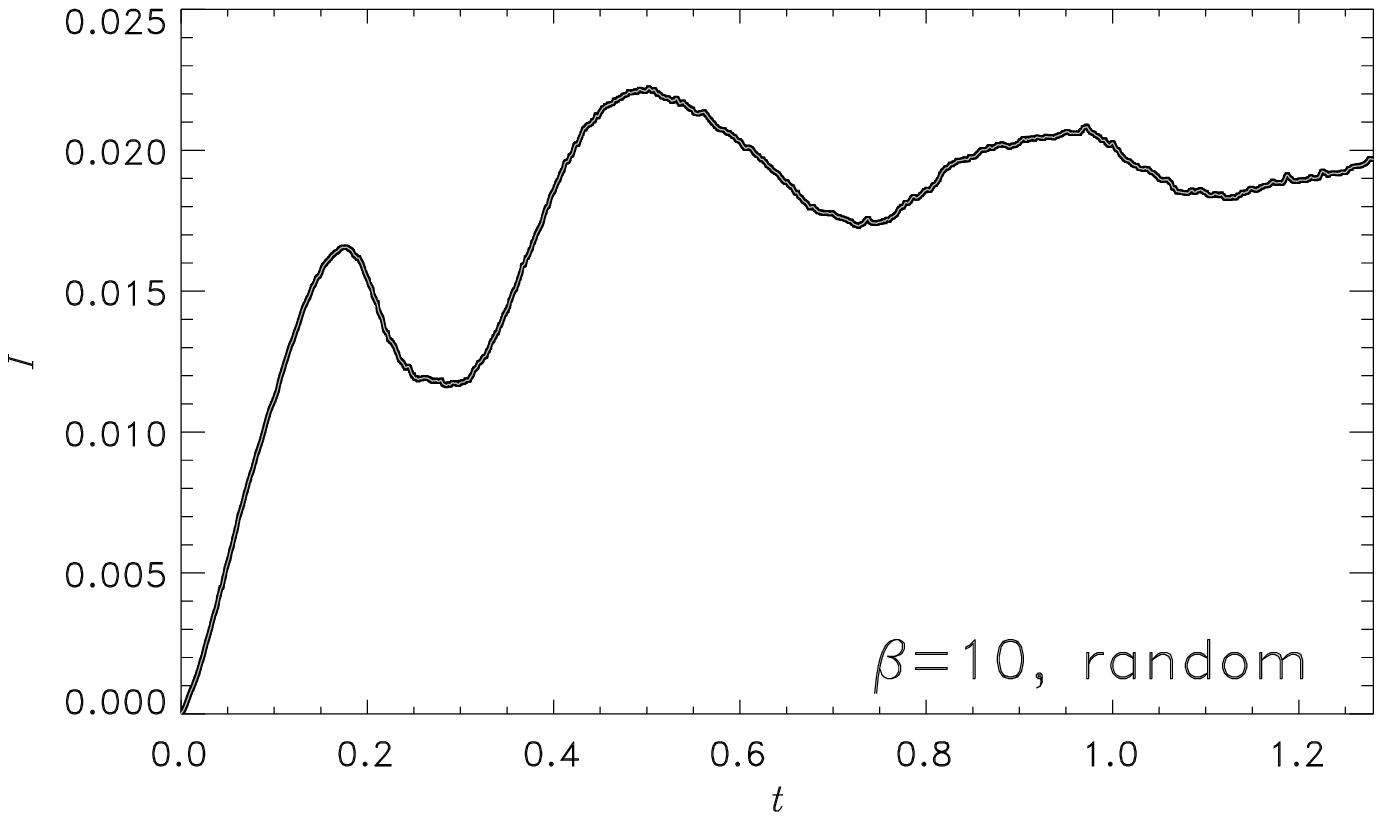}
&\includegraphics[width=5cm]{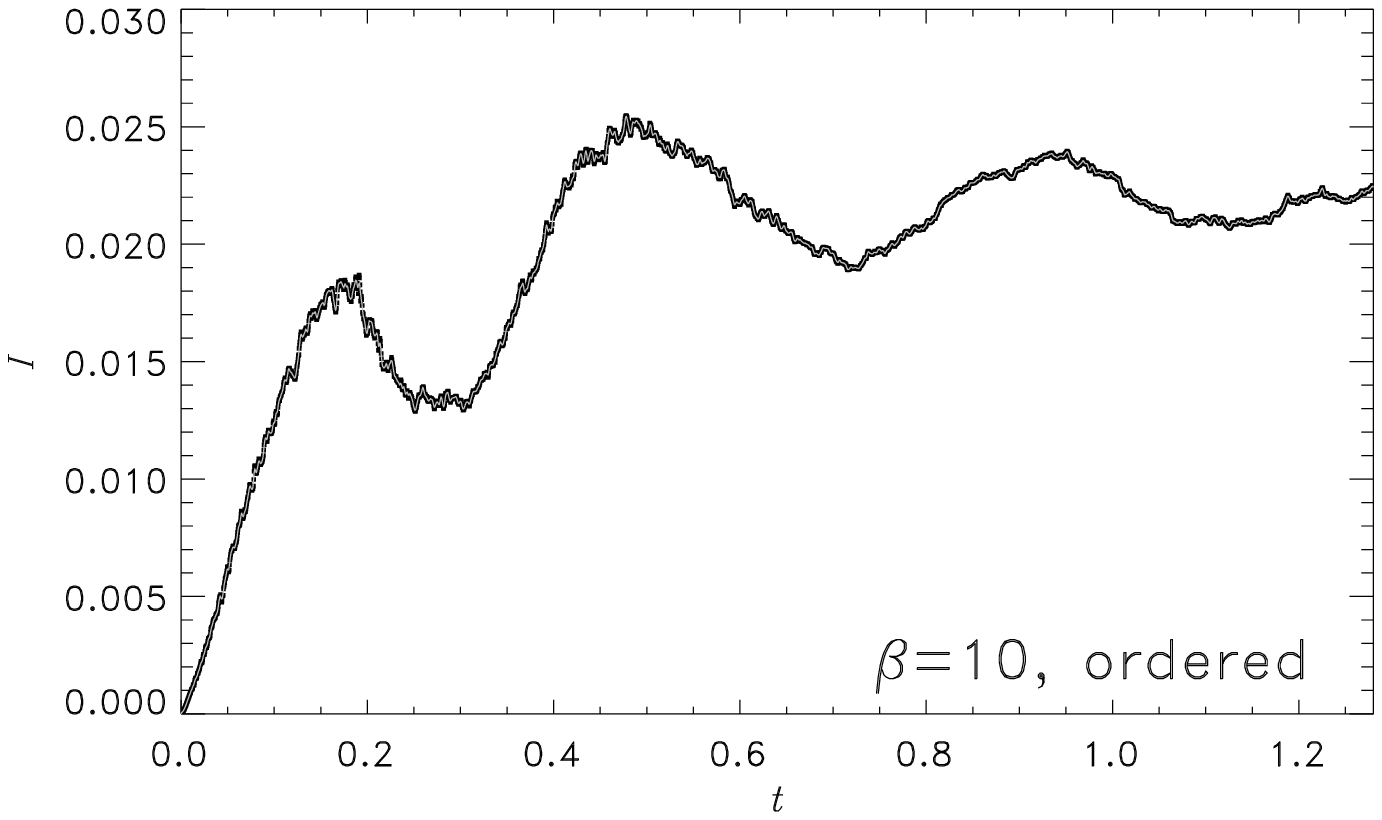}
&\includegraphics[width=5cm]{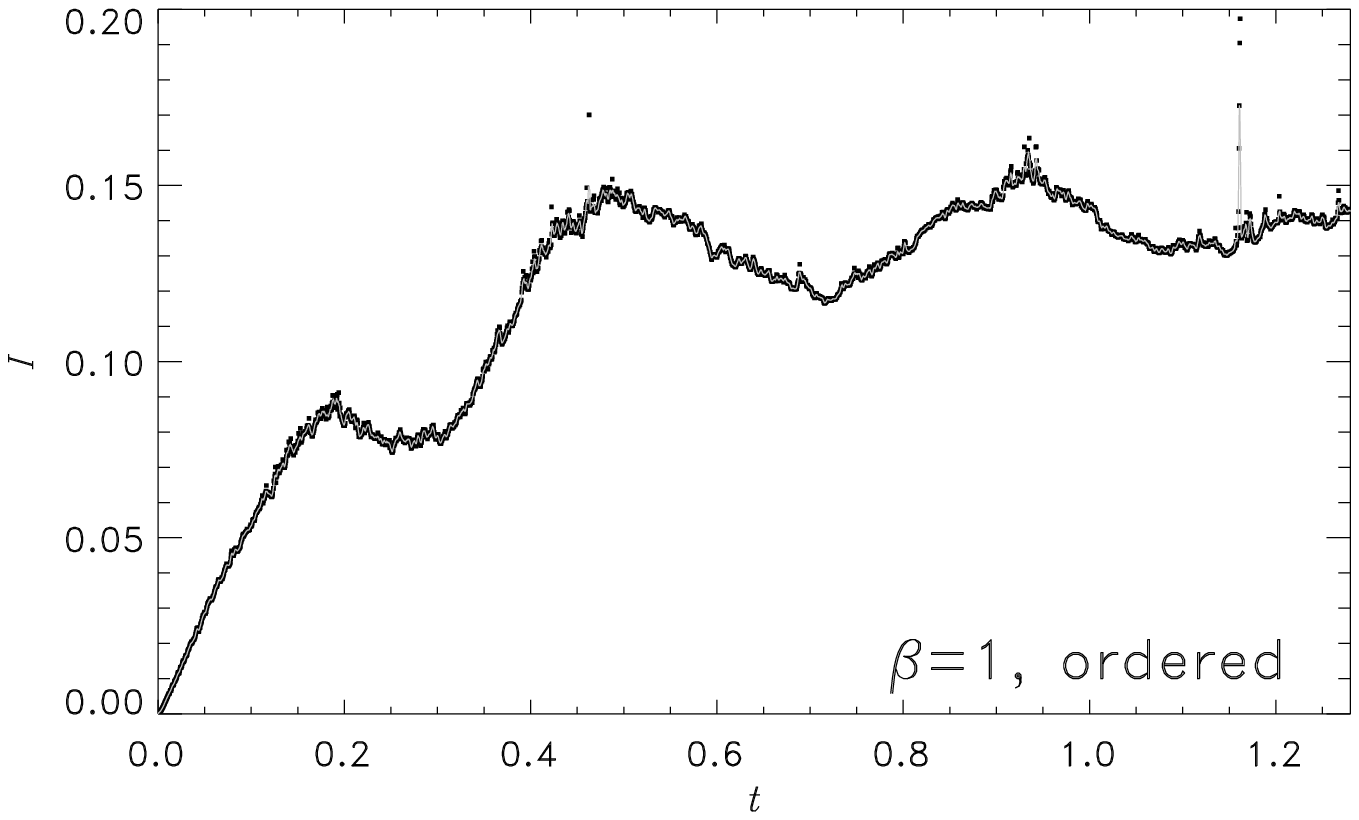}
\end{tabular}
\end{center}
\caption{
Synchrotron radio lightcurves of the {\sc fast4} jets,
    which penetrate a background density with ripply density variations.
}
\label{fig.temporal.fast4}
\end{figure*}

Figure~\ref{fig.temporal.slow}
   shows the evolution of the integrated intensity $I$,
   of the {\sc slow} jet.
Figures~\ref{fig.temporal.fast1o}--\ref{fig.temporal.fast4}
   are equivalent light-curves for the fast jets.
In simulations where the hotspot is unambiguous and unobscured by the cocoon,
   we also present a local light-curve of the hotspot ($I_{\rm hs}(t)$).
Here we define the instantaneous ``hotspot'' as the intensity maximum
   that is farthest right along the jet axis,
   and we integrate the emission within a projected radius of 20 pixels.

{\color{Black}    
\subsubsection{left boundary conditions} 
}

We have considered two types of left boundary conditions in the simulations.   
The open boundary condition mimics situations 
   where the jets have drilled far from the central engine. 
This is appropriate for modeling the jets in  giant radio galaxies.
The closed boundary takes account of the effects 
   of backflows and material accumulation.
This is appropriate for modeling compact jets
   that are yet to break away from the core region,
   where backflows from both jet and counter-jet can interact with each other.
Comparison of runs that differ at the left boundary 
   (i.e. {\sc fast1o} vs {\sc fast1} and {\sc fast2o} vs {\sc fast2}) 
   shows that the (spatially integrated) temporal fluctuations are more rapid
   and have larger amplitudes 
   for an open left boundary.  
This may be attributed to fact that the exit of plasma through the boundary 
   leaves the unsteady structures
   (such as the jet shocks and hotspot)
   brighter than the cocoon. 
Also, with an open boundary condition, 
   the cocoon contributes less to the total emission.

The cloudy cocoons of the $\gamma=5/3$ and $\gamma=4/3$ cases
   with closed boundaries
   have similar light curves
   (see Figures~\ref{fig.temporal.fast1c} and \ref{fig.temporal.fast1rc}).
The time taken to traverse the grid is not significantly affected
   by the alternative $\gamma$.
The intensity reaches marginally higher levels for $\gamma=5/3$.
When the left boundary is open
   (Figures~\ref{fig.temporal.fast1o} and \ref{fig.temporal.fast1ro})
   the effect of $\gamma$ is more conspicuous:
   $\gamma=4/3$ gives greater variability in the total intensity $I(t)$,
   as well as in bright features like the hotspot $I_{\rm hs}(t)$.
   
{\color{Black}
\subsubsection{magnetic field structure} 
}

The differences in the intensity light curves 
   between a random or an ordered field directions 
   are more obvious for jets with a closed boundary and conspicuous cocoon
   (e.g. comparing $\beta=10$ columns within
   Figures~\ref{fig.temporal.fast1c},
   \ref{fig.temporal.fast2},
   \ref{fig.temporal.fast3}
   and \ref{fig.temporal.fast4}).
Integrated intensity and polarization profiles tend to be smoother
   in the random case, 
   due to summation over incoherent zones of the image.
Ordered fields following the velocity structure
   produce larger coherent emssion patches
   that evolve on the same temporal and spatial scales
   as the principal shocks and eddies.
The hydrodynamic vacillation and surging of these structures
   is responsible for the high frequency quasi-periodicity
   apparent in the $I(t)$ curves.

\subsubsection{approximate homology of intensities}

A comparison of intensity light-curves calculated from the same simulation
   but with different $\beta$
   shows that they are similar to within a scale factor of $\approx0.16$
   for each factor of $10$ increase in $\beta$.
For cases with a closed boundary and a translucent cocoon
   ({\sc slow}, {\sc fast1}, {\sc fast2}, {\sc fast3}, {\sc fast4},
   {\sc fast1r})
   the homology holds at nearly all times.
For the cases with an open left boundary and prominent jet shocks,
   the homology holds at late times
   (say when the front of the bow shock is beyond $x_{\rm s}>30~\kpc$)
   but the initial behaviours differ in detail.
The scaling also breaks down temporarily
   around particularly bright and opaque features
   during the undulations of {\sc fast4b}
   and flashes of the jet in {\sc fast1ro}.
The extreme choice of $\beta=1$ (strong field)
   makes more substructures opaque,
   and the scaling fails more often for higher $\beta$.

{\color{Black} 
\subsection{Jet variability and astrophysical implications}
}

\begin{figure*}
\begin{center}
\begin{tabular}{ccc}
 \includegraphics[width=6cm]{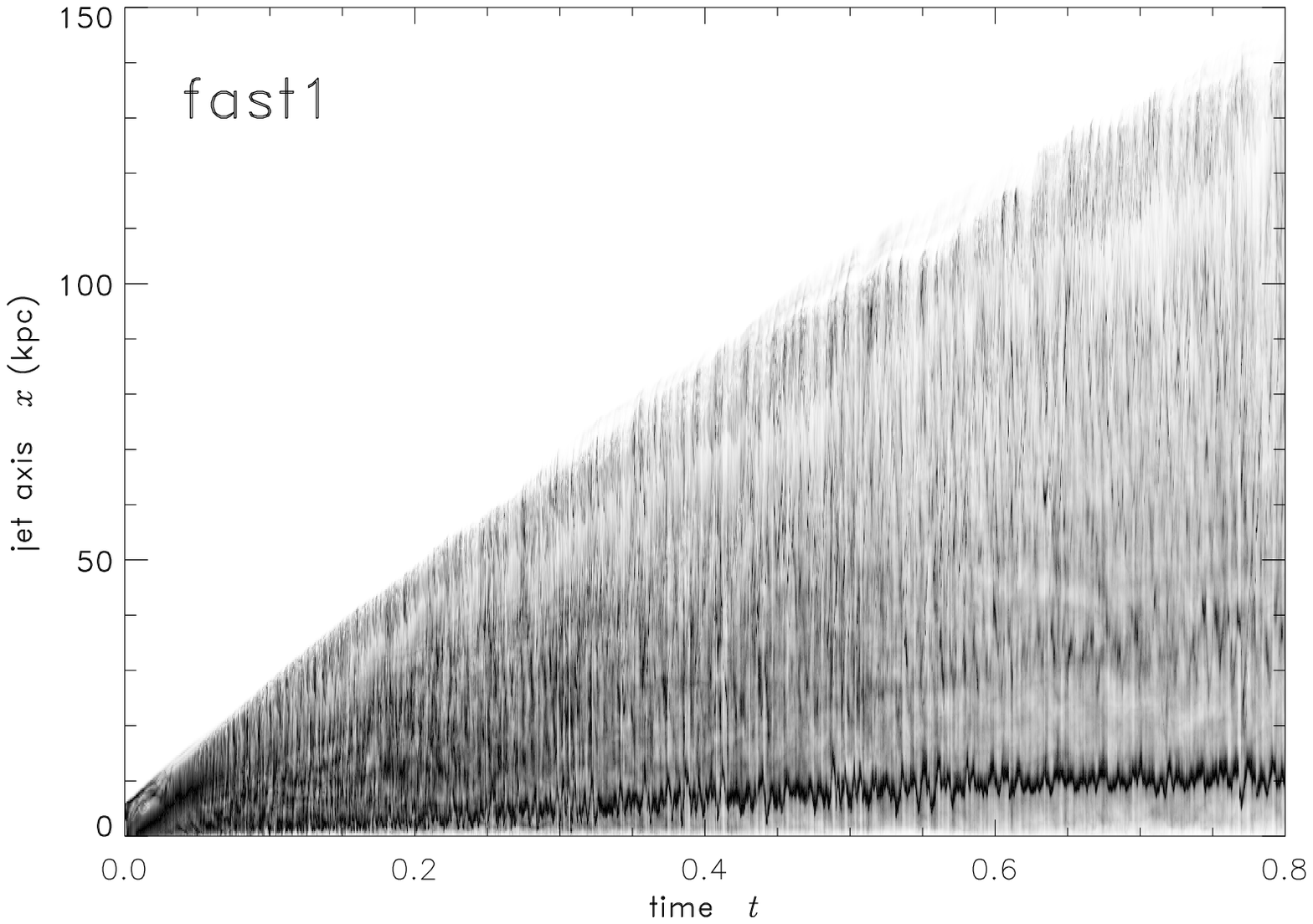}
&\includegraphics[width=6cm]{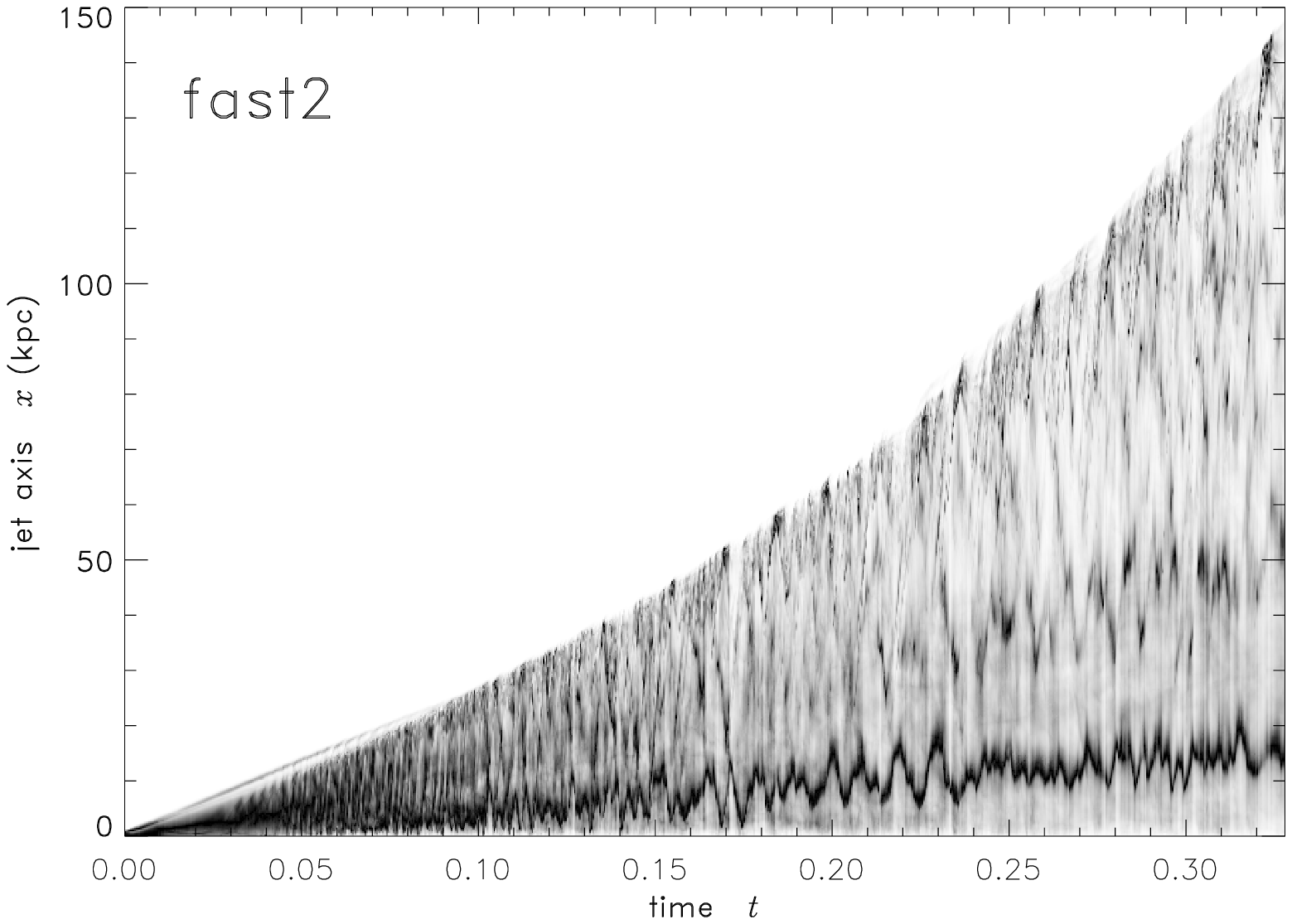}
&\includegraphics[width=6cm]{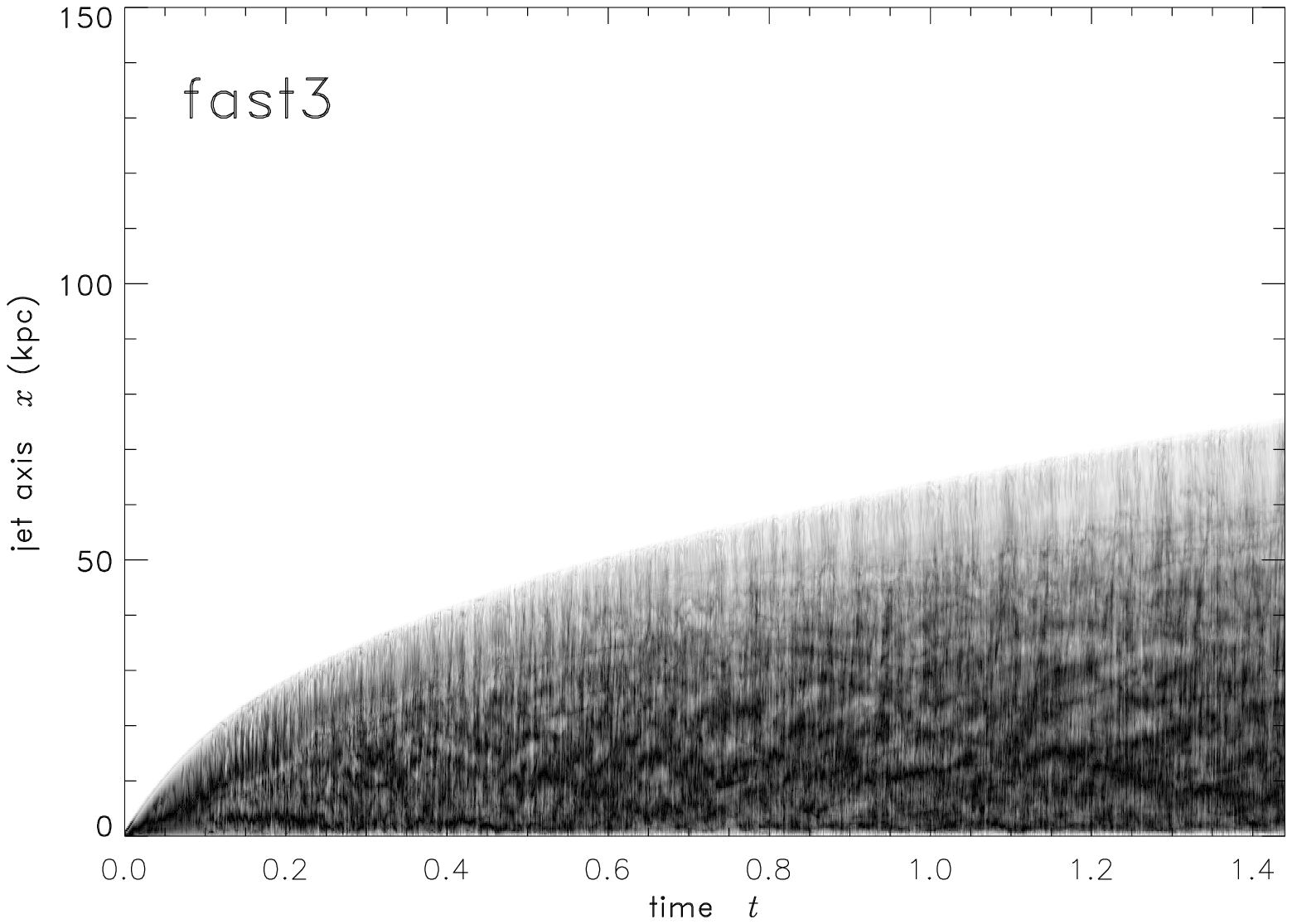}
\\
 \includegraphics[width=6cm]{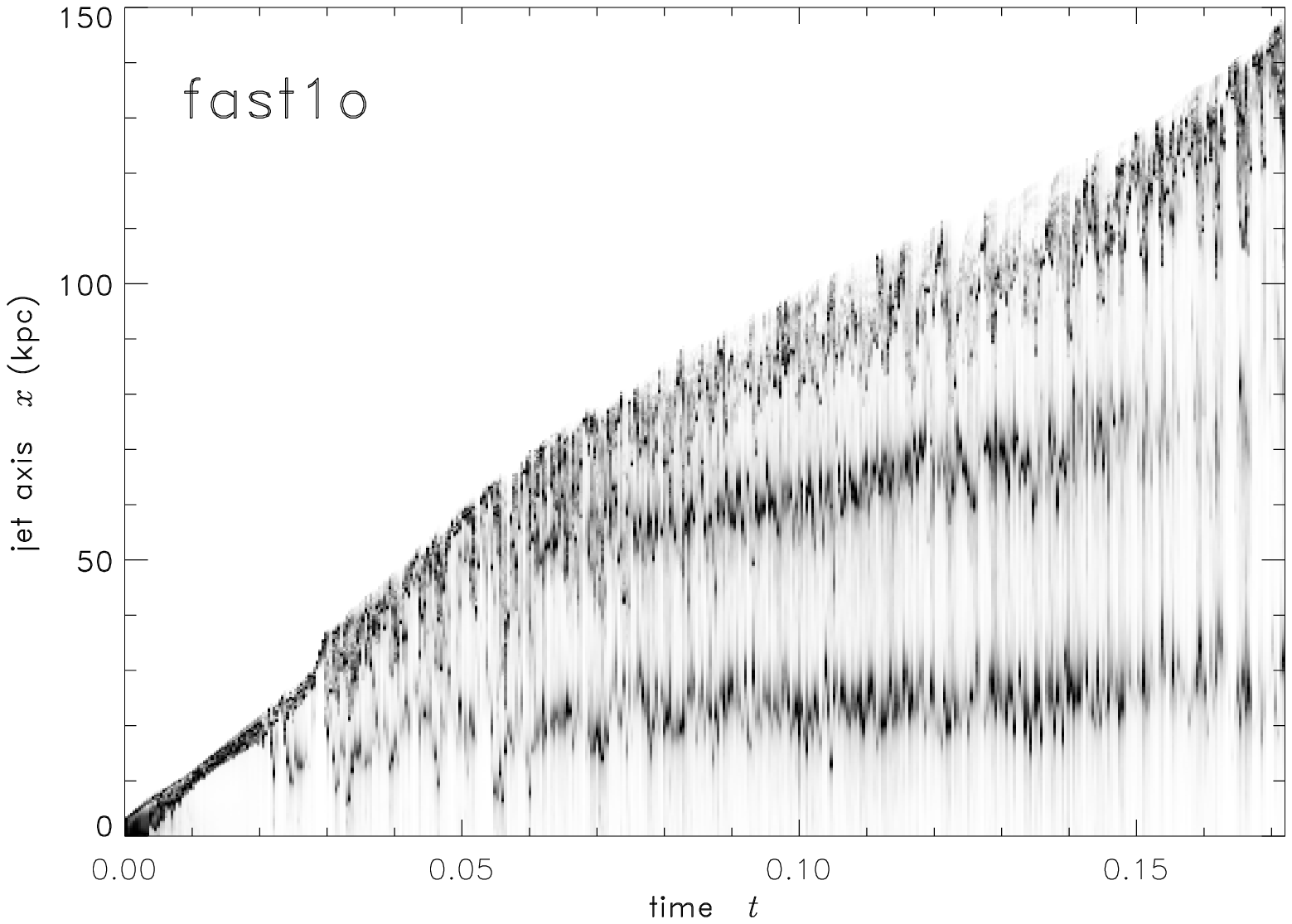}
&\includegraphics[width=6cm]{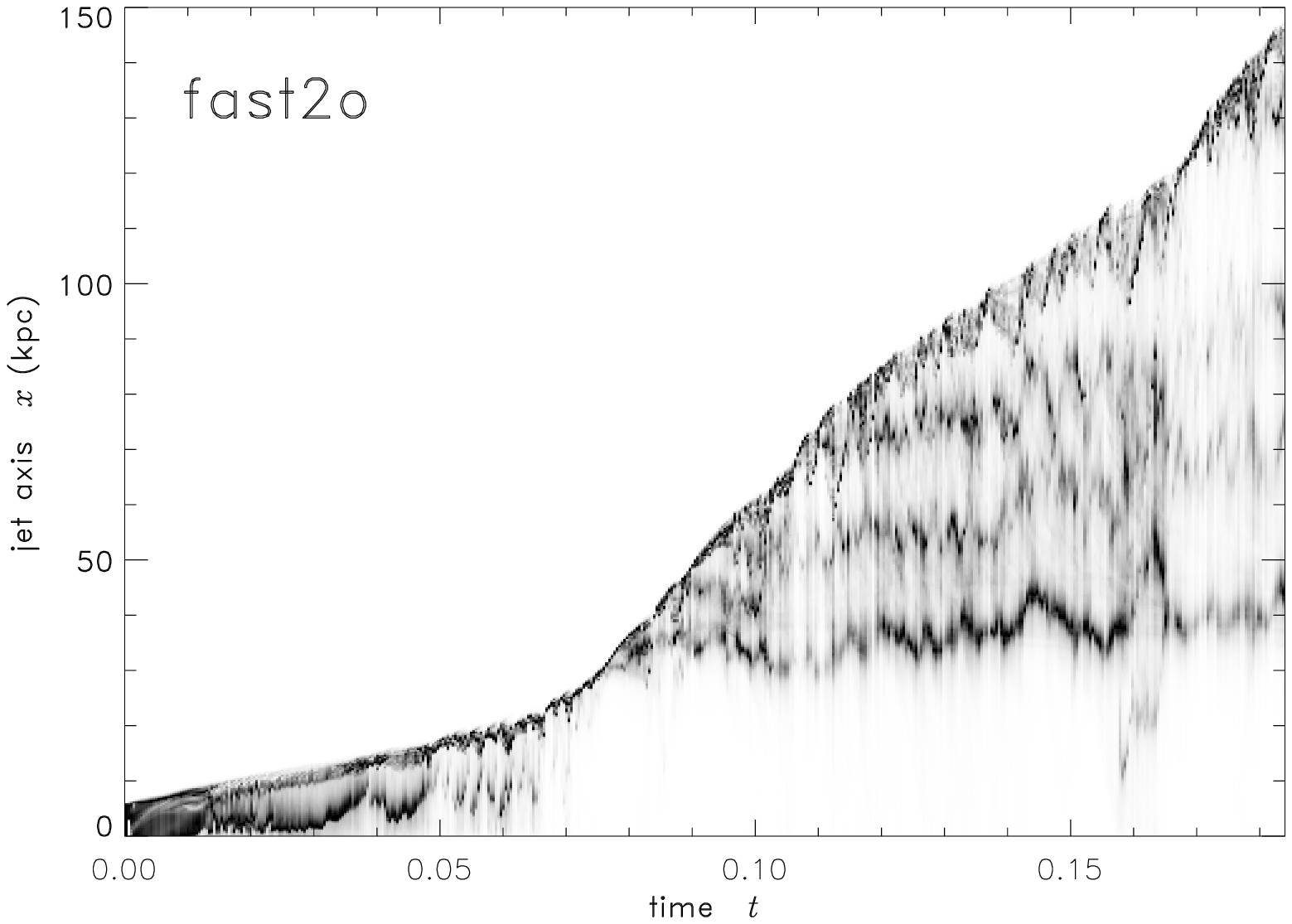}
&\includegraphics[width=6cm]{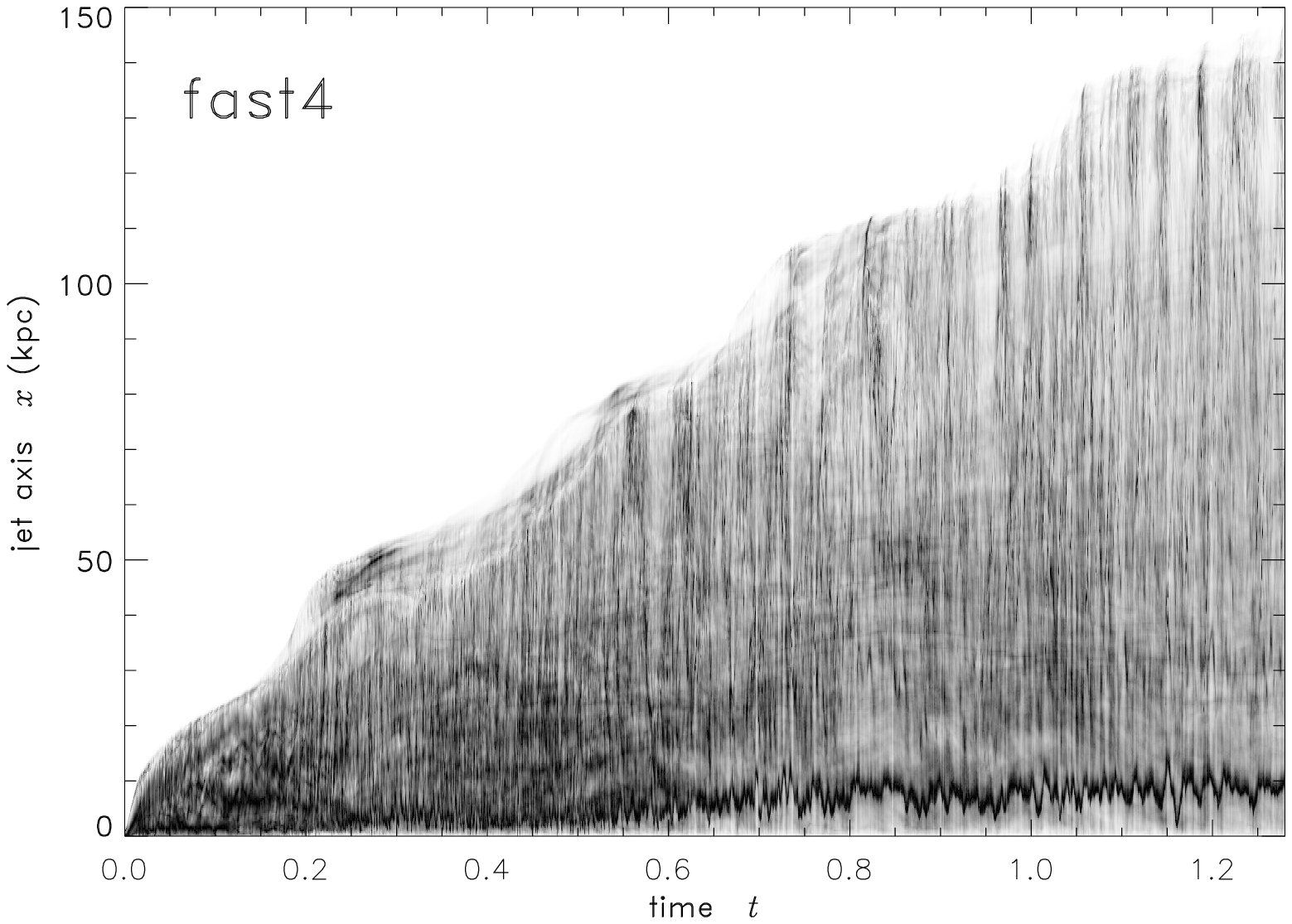}
\end{tabular}
\end{center}
\caption{
Evolution of intensity along the midline
   (projection of the jet axis)
   for the fast jet simulations (as annotated in respective panels,
   and assuming $\beta=10$).
In cases with an open boundary condition
   ({\sc fast1o}, {\sc fast2o})
   the jet shocks are brighter relative to the cocoon,
   and the hotspot is brighter relative to the pinch shocks.
The density ripples in the background medium make little difference:
   {\sc fast4} resembles {\sc fast1}. 
{\color{Black}
A radially declining density profile enables the jet's advance to accelerate
   ({\sc fast2}, {\sc fast2o})
   while a radially increasing profile restrains
   the expansion of a decelerating fireball ({\sc fast3}).} 
Except in {\sc fast3},
   the first jet shock is brightly visible,
   fluttering about its mean position.
These figures are derived from Stokes maps made with ordered fields,
   and each vertical stripe is separately normalised.
}
\label{fig.wedge}
\end{figure*}

{\color{Black}
\subsubsection{coruscating substructures}

Evolution of local substructures along the midline of the image
   is shown by intensity slices in Figure~\ref{fig.wedge}.  
The hotspot has rapid structural and morphological variabilities: 
  it throbs; it casts off vortex rings; it breaks and reforms. 
Identification of the hotspot is not always unambiguous.    
  }
{\color{Black} 
In one case ({\sc fast3}), 
  the hotspot and jet internal shocks are hidden within an opaque fireball.   
In general, the jet shocks  have comparable intensities 
  for cases with an open left boundary.   
However, when when left boundary is closed
   ({\sc fast1}, {\sc fast2}, {\sc fast4})
   the most persistently bright feature is the first jet shock. 
It outshines the jet hotspot,
   and would appear as  a ``pseudocore'' as in jet observations   
  \citep{jones1988,darcangelo2007}. 
At later evolutionary stages, a second pinching shock appears.  
It could be as bright as the hotspot, and it tends to migrate forwards
   (whereas the first shock hovers about a certain place).

Subtler diagonal streaks through the diagram
   represent more ephemeral shocks traversing the jet. 
The formation of bright emission knots in jets 
   is sensitive to the environment, 
   and the distribution of bright knots in the jets 
   during any particular epoch 
   could be a diagnostic of the density structure of
   the external medium.
Our simulations have shown that 
    the knots occur at nearly equal intervals 
    for a uniform background (e.g. {\sc fast1}, {\sc fast1o}). 
The knots can also appear evenly spaced, 
    for jets propagating through a medium
    with density that undulates about a mean ({\sc fast4}). 
For a medium with a radially declining density ({\sc fast2o}),
   the knots are closer together in the outskirts,
   beyond a long gap in the denser inner  region. 
}

{\color{Black}   
\subsubsection{intensity and jet advance}

As shown by each wedge's outline in Figure~\ref{fig.wedge}, 
   the radio-emitting region advances as time elapses, 
   and the front of the bow shock (denoted as $x_{\rm s}$, not illustrated)
   evolves similarly. 
Their evolutions are dependent on the external density profile.
In a uniform medium,
   the jet advances and the cocoon expands with a gentle deceleration.
The deceleration is more severe for media with a radially rising density
   ({\sc fast3}). 
However, acceleration would occur 
   if the density of the ambient medium decreases radially
   ({\sc fast2}, {\sc fast2o}).

In nature, the age of a jet activity episode is not directly observable, 
   and neither are the density profiles of the medium in the vicinity 
   (say, within a few tens kpc) of an AGN core.
In contrast, the size of the system and the radio luminosities 
   can be estimated from observational data. 
Thus, deriving a relation between the system size and radio luminosity 
  from our simulations of time-dependent jets 
  in various environments will provide 
  a tool to constrain the age of jet activity episodes,
  and infer the ambient structure.

For models with an open left boundary,
   the light-curves are dominated by stochastic flashes from jet shocks.
Once such a jet has grown long enough to have at least one pinch shock,
   there isn't any tight relation between size and luminosity.
(For {\sc fast1o}, {\sc fast2o} and {\sc fast1ro}
   the light-curve at late times brightens and dims significantly
   on diverse timescales.)
However for the models with a closed boundary
   (forming a foggy cocoon/lobe)
   relatively simple power-law relations occur.

The integrated intensity light curves $I(t)$
   of the {\sc fast1} and {\sc fast2} jets are quite similar  
   except that the latter drops at later times, 
   when the jet encounters a significant density drop.
During the initial rising stage of
{\sc fast1}
($6.9~\kpc<x_{\rm s}<15~\kpc$)
the total intensity grows like the plasma volume
$I\propto x_{\rm s}^{3.02}\propto t^{3.82}$.
The instantaneous peak intensity
   varies rapidly within a broad envelope that tends to run like
   $I_{\rm max}\propto x_{\rm s}^{0.78}\propto t^{-0.22}$.
The growth curve steepens over time:
   $x_{\rm s}\propto t^{0.36}$ early on
   ($9~\kpc<x_{\rm s}<15~\kpc$).
At later times ($19.5~\kpc<x_{\rm s}<120~\kpc$)
   the growth is $x_{\rm s}\propto t^{0.72}$,
   while the luminosity
   $I\propto x_{\rm s}^{0.61}\propto t^{0.44}$
   and the peak
   $I_{\rm max}\propto x_{\rm s}^{-0.58}\propto t^{-0.42}$.

For {\sc fast2},
   the $x_{\rm s}=x_{\rm s}(t)$ curve steepens as the jet advances,
   but the index is $>1$ for most of the simulation.
When $x_{\rm s}>30~\kpc$ the mean log-slope is $1.25$;
   when $x_{\rm s}>75~\kpc$ this rises to $1.53$.
During the initial rise ($x_{\rm s}\la12~\kpc$)
   the luminosity evolves as
   $I\propto x_{\rm s}^{3.02}\propto t^{2.45}$
   and the peak
   $I_{\rm max}\propto x_{\rm s}^{0.69}\propto t^{0.33}$.
After $x_{\rm s}\ga12~\kpc$,
   the total intensity curve is nearly flat;
   the instantaneous peak values drop roughly like
   $I_{\rm max}\propto x_{\rm s}^{-0.66}$.

The {\sc fast3} jet inflates a globular cocoon/fireball
   with plasma accumulating at a constant rate,
   but a decelerating radial expansion.
At later stages
  ($x_{\rm s}\ga27~\kpc$),
   the growth curve is
   $x_{\rm s}\propto t^{0.44}$.
As the cocoon is opaque for $\beta=1, 10$ or $100$,
   the luminosity rises with the surface area,
   $I\propto x^{2.09} \propto t^{0.92}$,
   (Figure~\ref{fig.temporal.fast3}). 
The (noisy) envelope around the peak intensities
   rises slower,
   $I_{\rm max}\propto x^{1.56} \propto t^{0.69}$.

The {\sc fast4} jets (Figure~\ref{fig.temporal.fast4}), 
   propagating in a ripply background,
   have lightcurves similar to {\sc fast1}, 
   but with undulations, 
   as the jet is retarded by overdense shells,
   and then surges faster through underdense shells.

The light curves of the {\sc slow} jets 
   resemble the {\sc fast1} and {\sc fast2} jets initially
   and {\sc fast4} jets in the later stage.   
Despite the {\sc slow} jet's low Mach number,
   it advances with a power-law behaviour
   resembling that of {\sc fast1}:
   $x_{\rm s}\propto t^{0.77}$ at late stages.
The intensity laws are steeper than for {\sc fast1},
   $I\propto x_{\rm s}^{0.77}$ and $I_{\rm max}\propto x_{\rm s}^{-0.79}$.
During the early rise, $x_{\rm s}\propto t^{0.57}$,
  $I\propto x_{\rm s}^{2.26} \propto t^{1.28}$
  and $I_{\rm max}\propto x_{\rm s}^{0.60}$.
}

\subsubsection{hotspot variability}

\begin{figure*}
\begin{center} 
\begin{tabular}{ccc}
 \includegraphics[width=5cm]{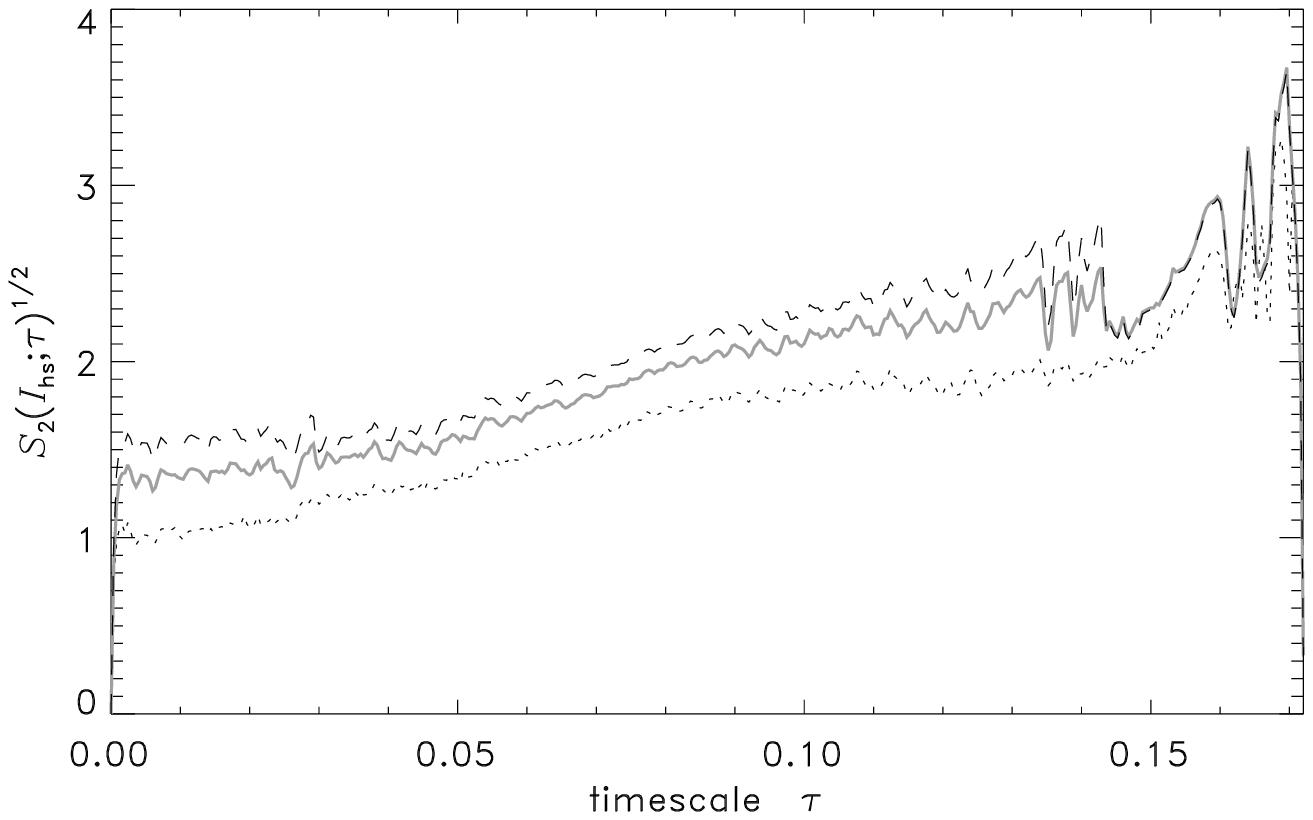}
&\includegraphics[width=5cm]{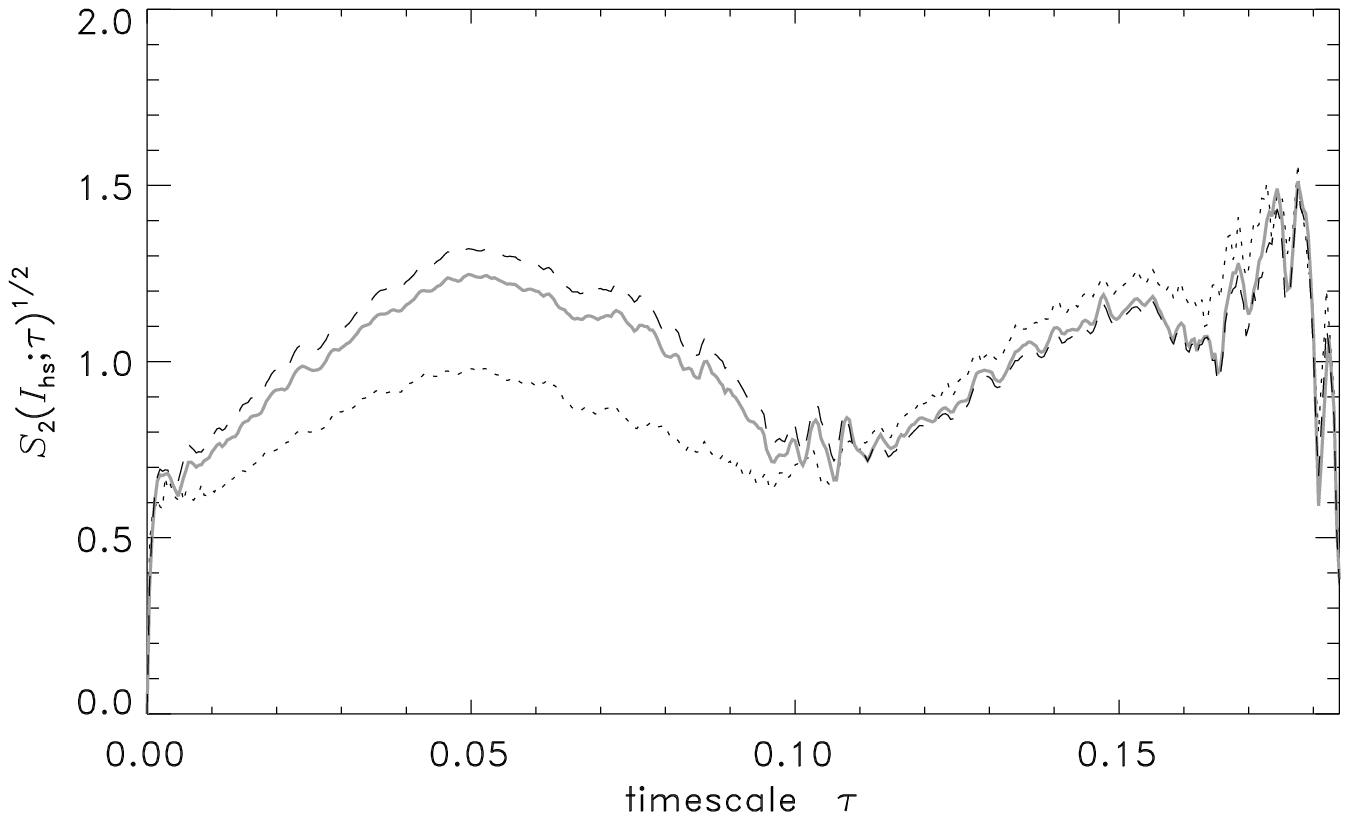}
&\includegraphics[width=5cm]{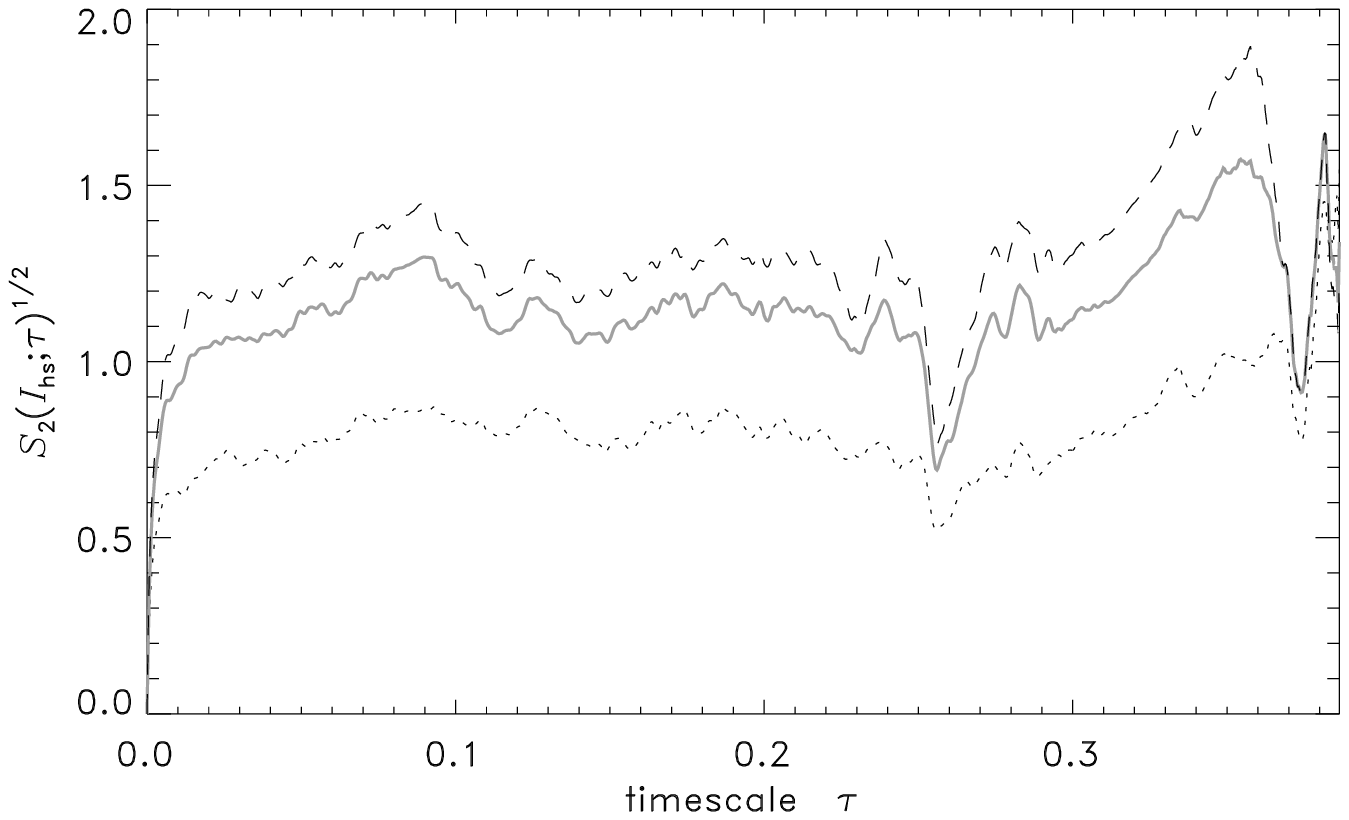}
\\
 \includegraphics[width=5cm]{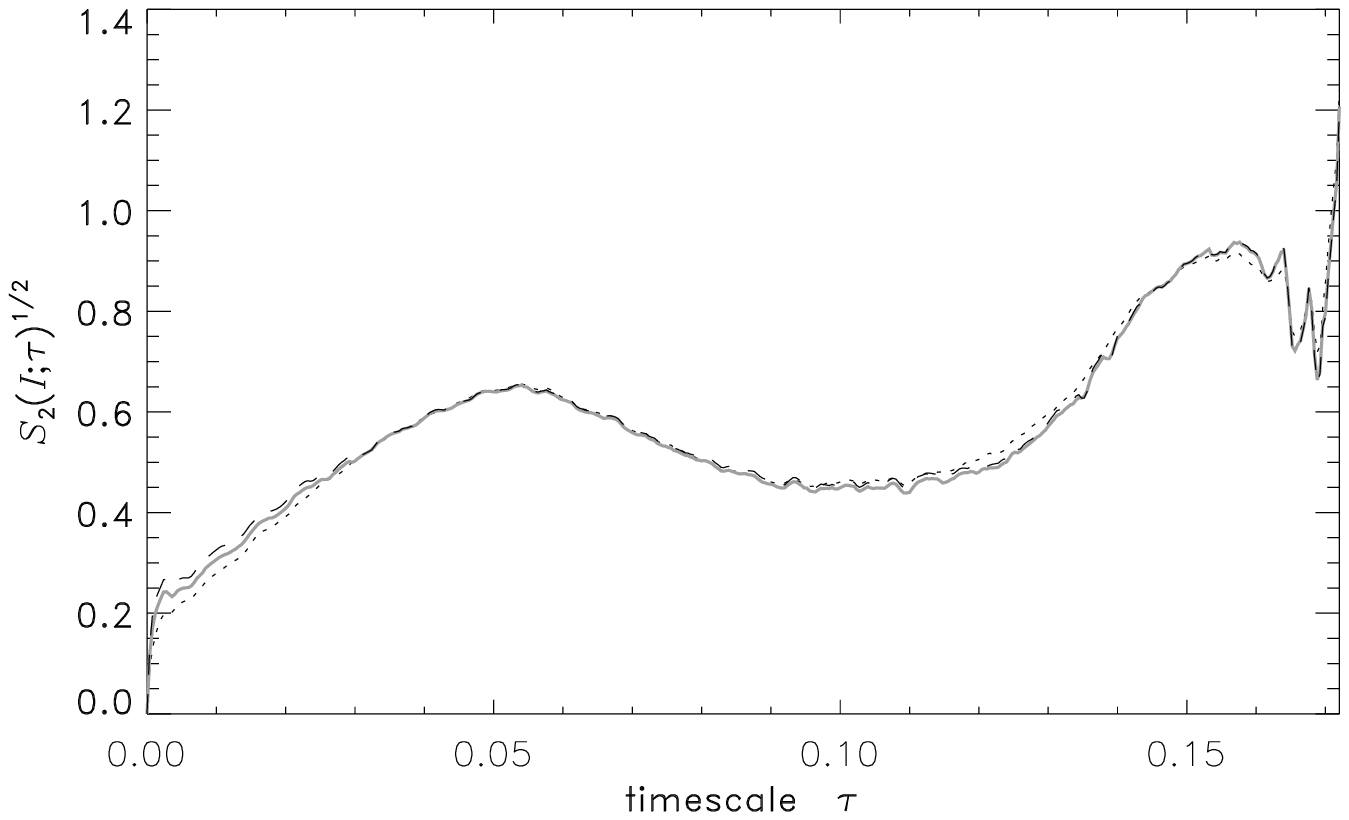}
&\includegraphics[width=5cm]{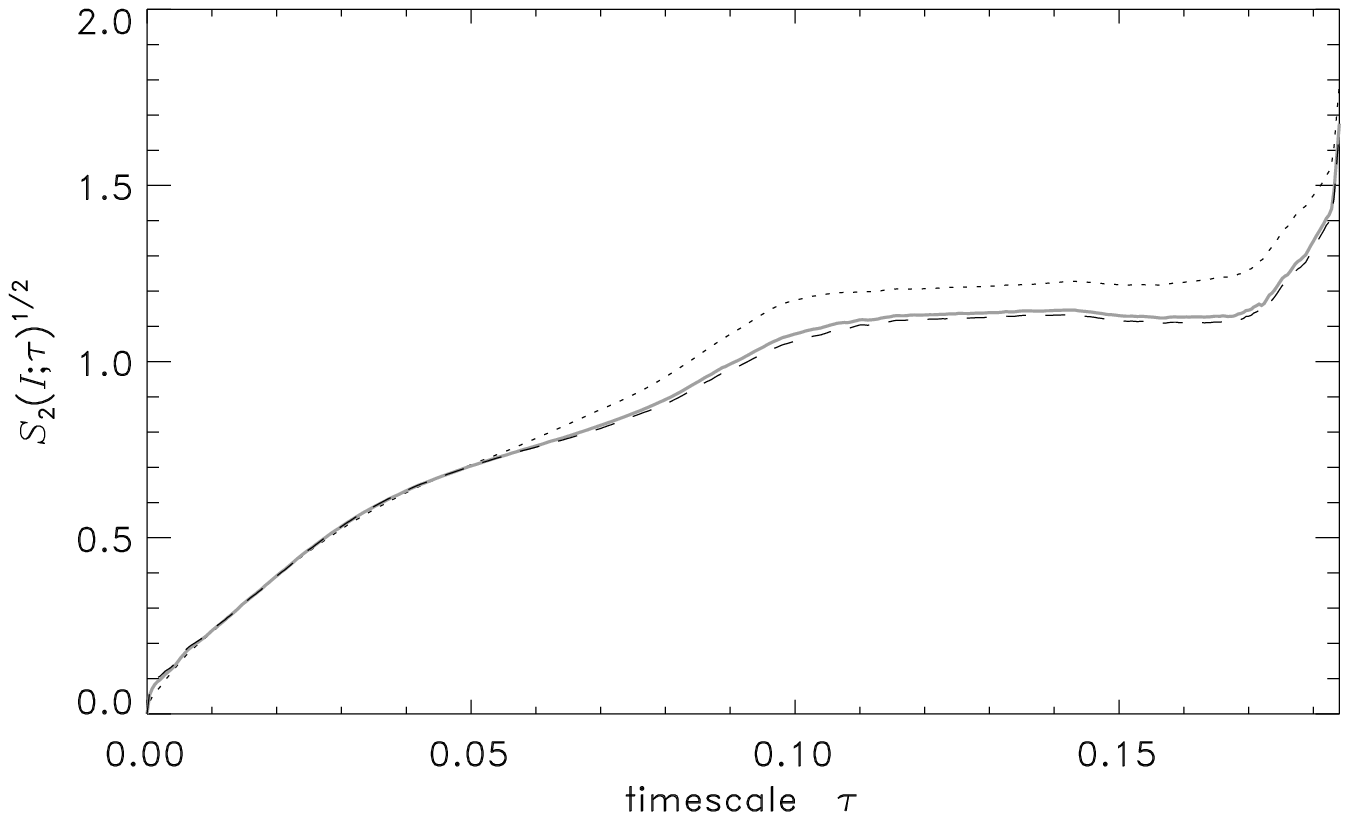}
&\includegraphics[width=5cm]{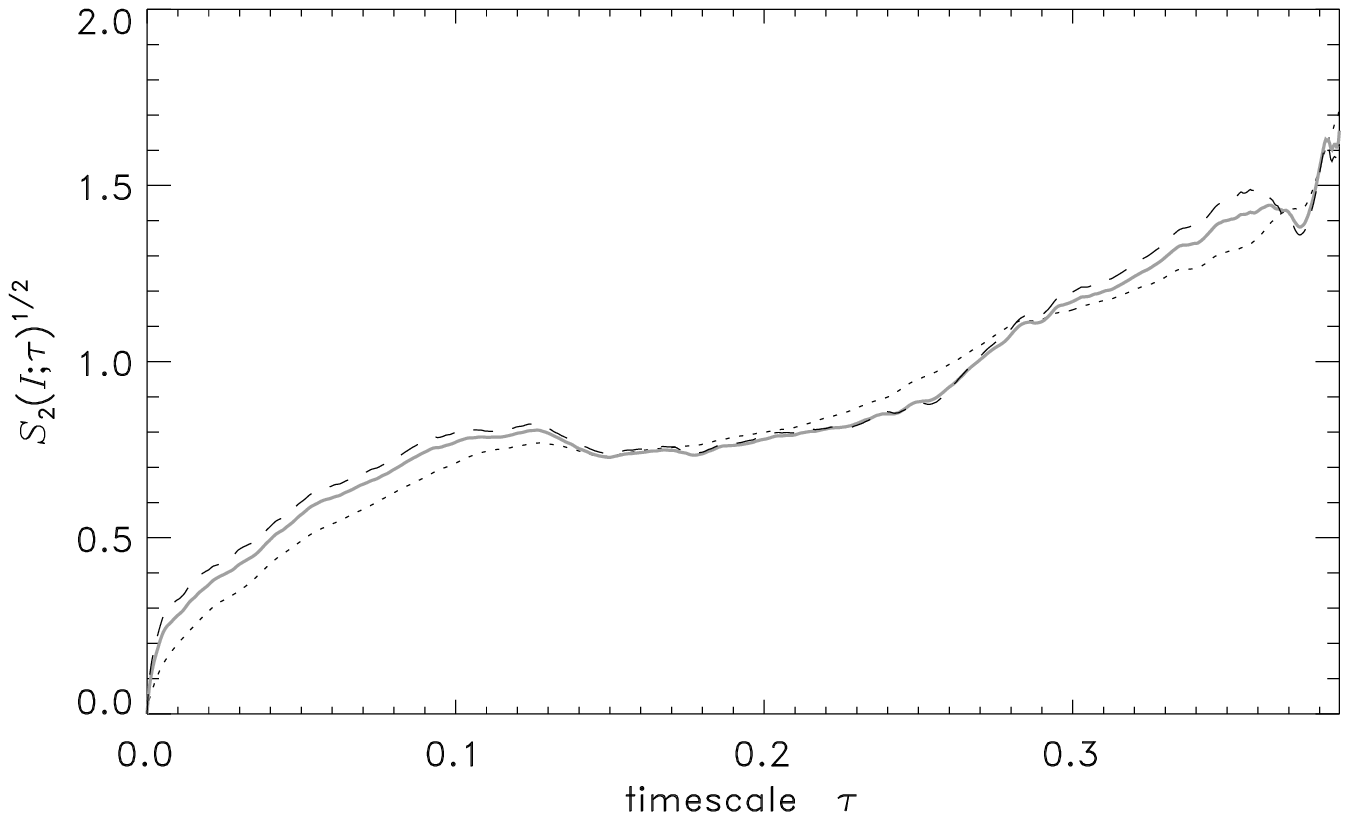}
\end{tabular}
\end{center} 
\caption{ 
\color{Black}
Structure functions of temporal variability
   in several quantities,
   for three open-boundary simulations
   ({\sc fast1o}, {\sc fast2o} and {\sc fast1ro}).
Here we take $\beta=10$ and the magnetic fields are quasi-poloidal
   ($\Bvec\parallel\vvec$).
The top row show structure functions of the hotspot intensity.
The bottom row shows structure functions of the total intensity.
Input variables to the structure functions
   are normalised to their respective means.
Black-dotted, grey and black-dashed curves depict $\beta=1, 10, 100$
   radiative transfer calculations respectively.
}
\label{fig.structure.fn}
\end{figure*}

For models with an open boundary,
   the cocoon is mostly optically thin
   and the hotspot and other bright knots are relatively bare,
   so it is feasible to characterise their variabilities.
In the top row of Figure~\ref{fig.structure.fn}
   we plot the rms variability of the hotspot intensity,
   calculated from the temporal structure function of $I_{\rm hs}(t)$.
\begin{equation}
	S_2(I_{\rm hs};\tau)={1\over{t_{\rm end}-\tau}}\int_0^{t_{\rm end}-\tau}
		\left[{
			I_{\rm hs}(t+\tau) - I_{\rm hs}(t)
		}\right]^2\ dt
	\ .
\end{equation}
The bottom row shows corresponding structure functions
   calculated from the total integrated intensity, $I(t)$.

\begin{figure*}
\begin{center}
\begin{tabular}{ccc}
 \includegraphics[width=5cm]{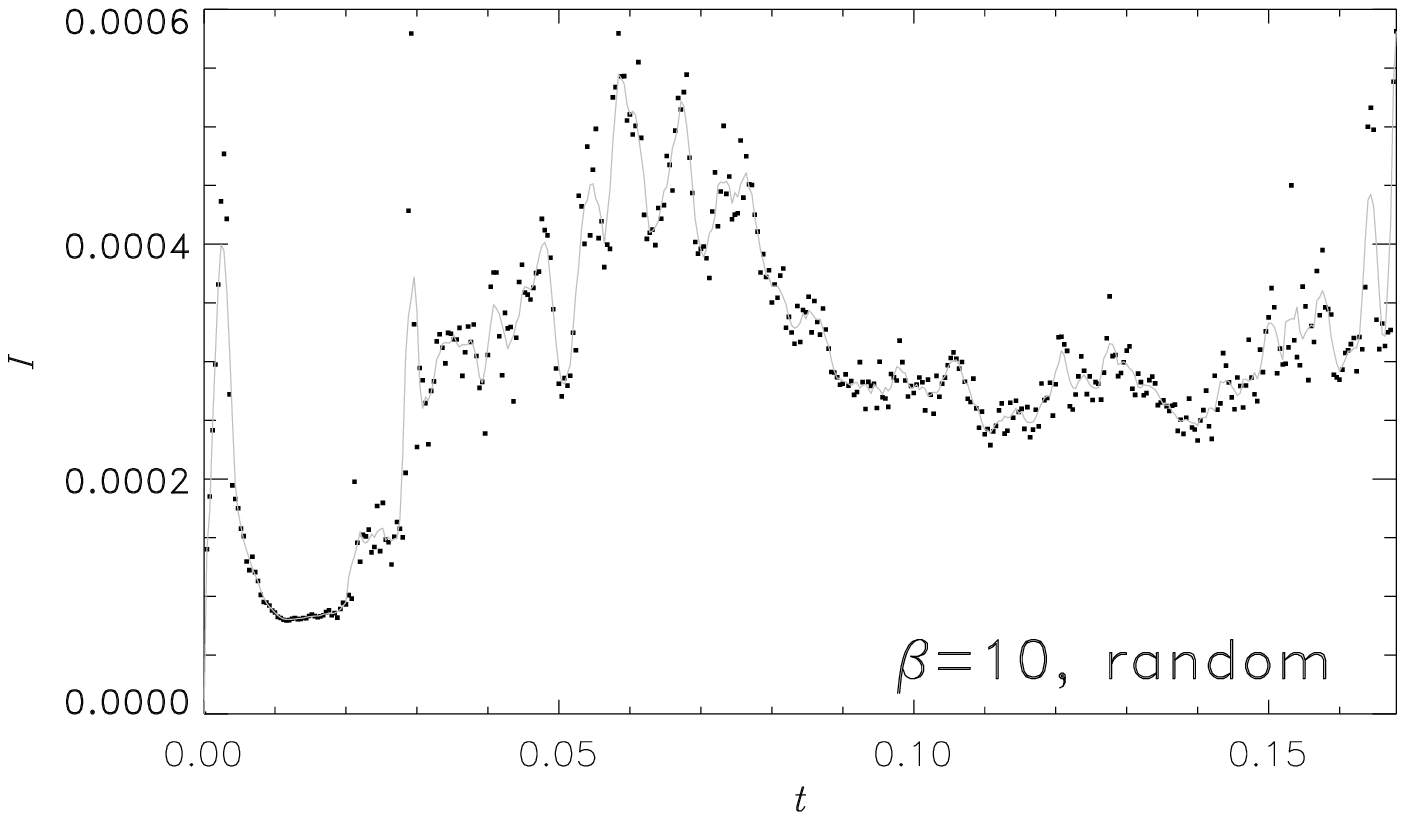}
&\includegraphics[width=5cm]{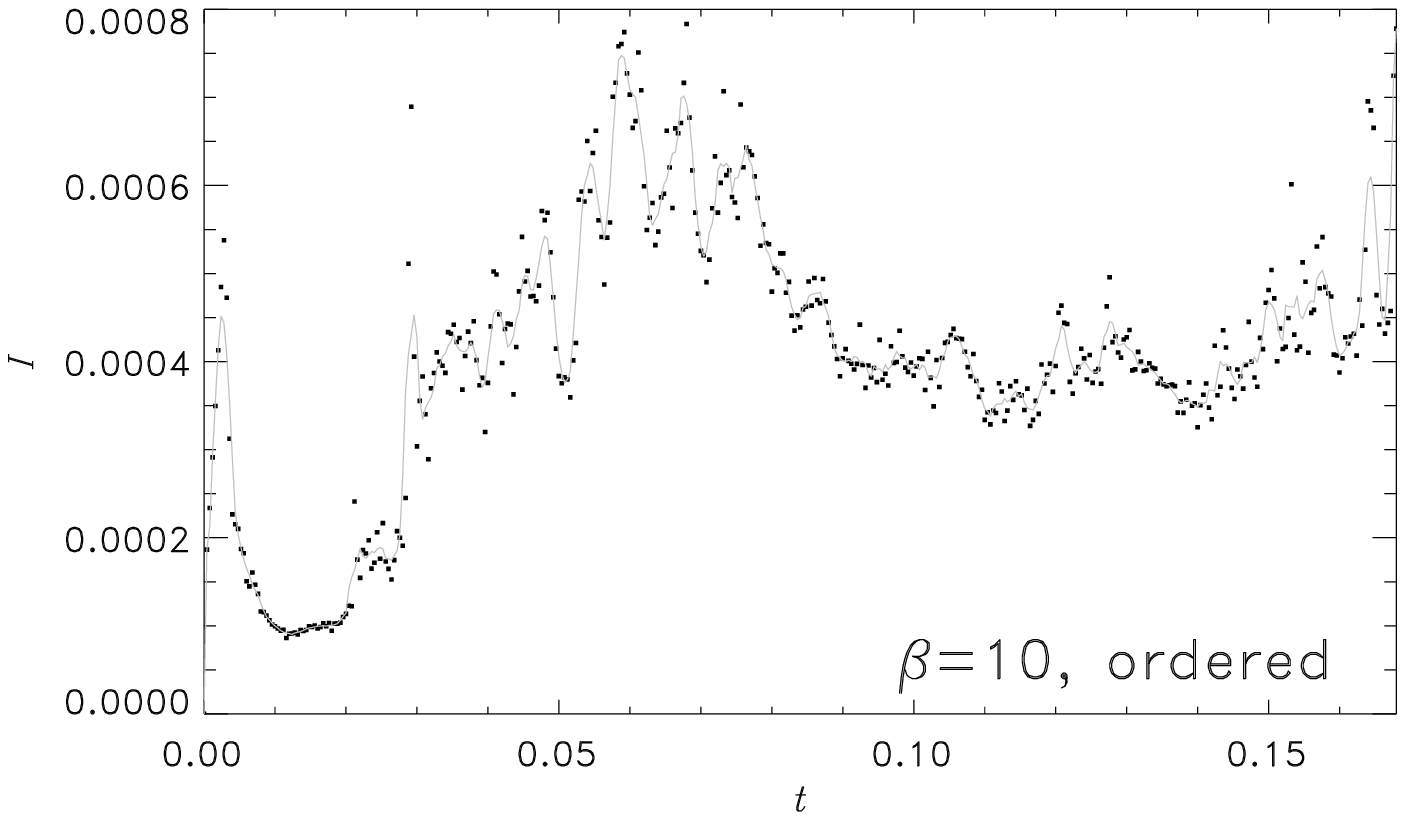}
&\includegraphics[width=5cm]{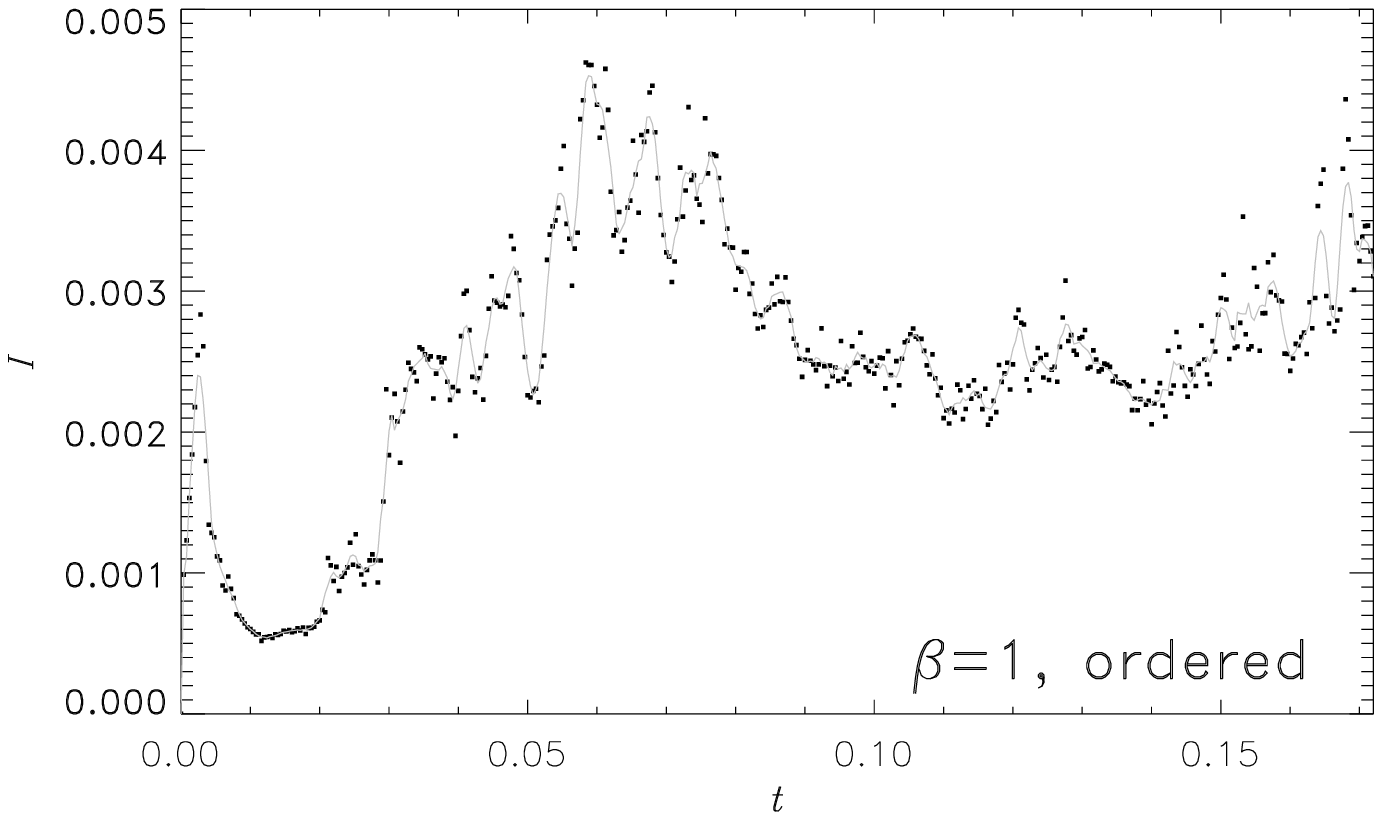}
\\
 \includegraphics[width=5cm]{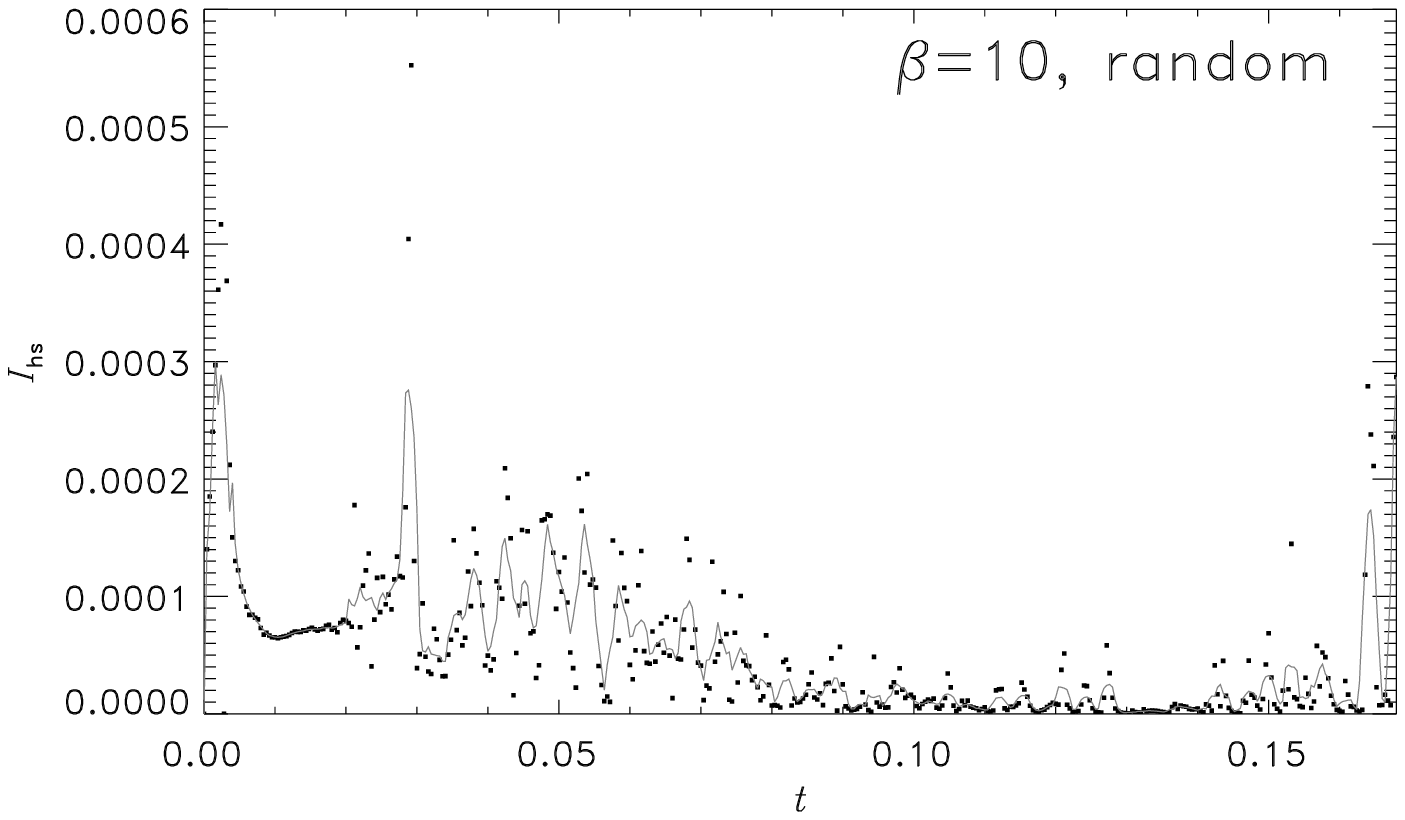}
&\includegraphics[width=5cm]{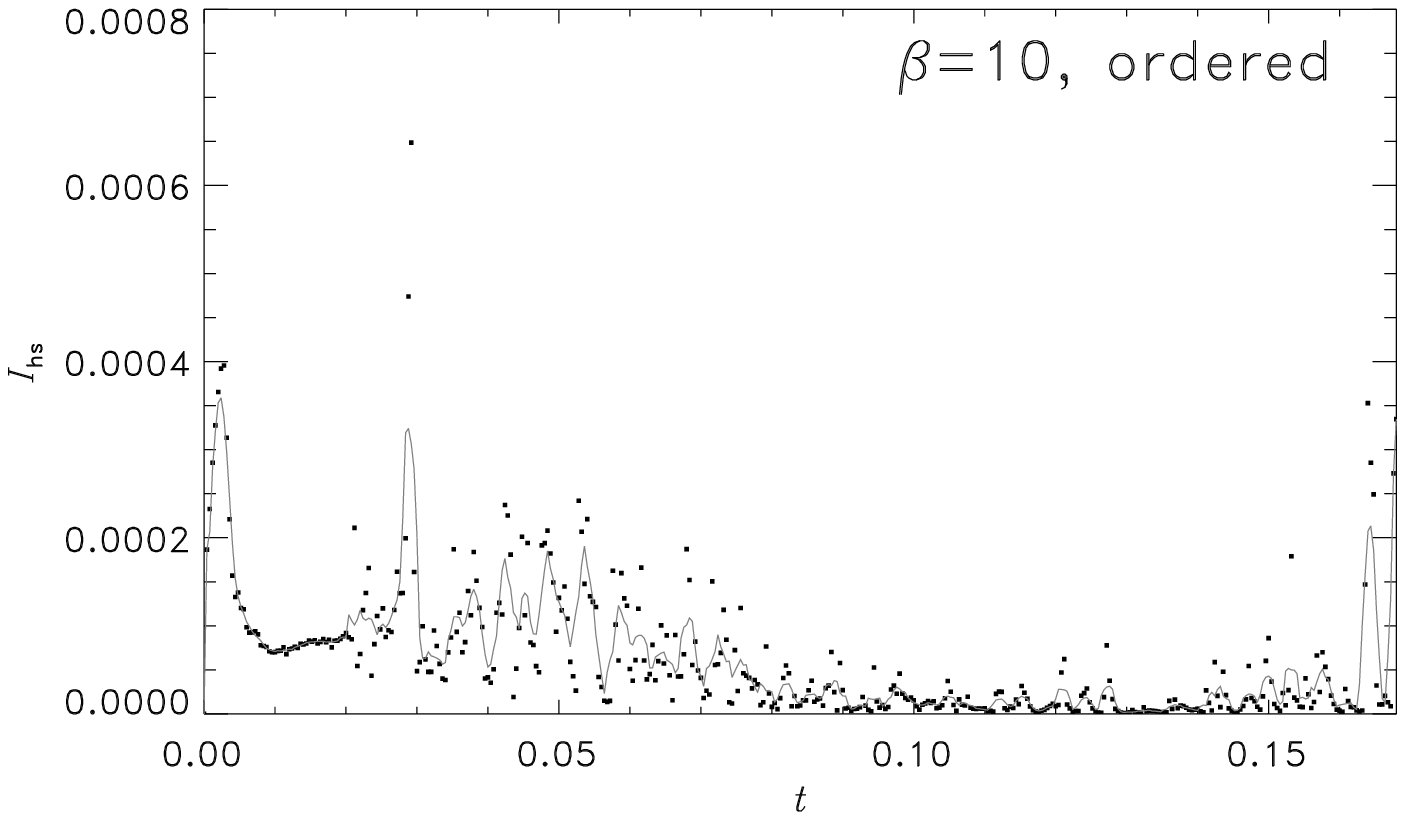}
&\includegraphics[width=5cm]{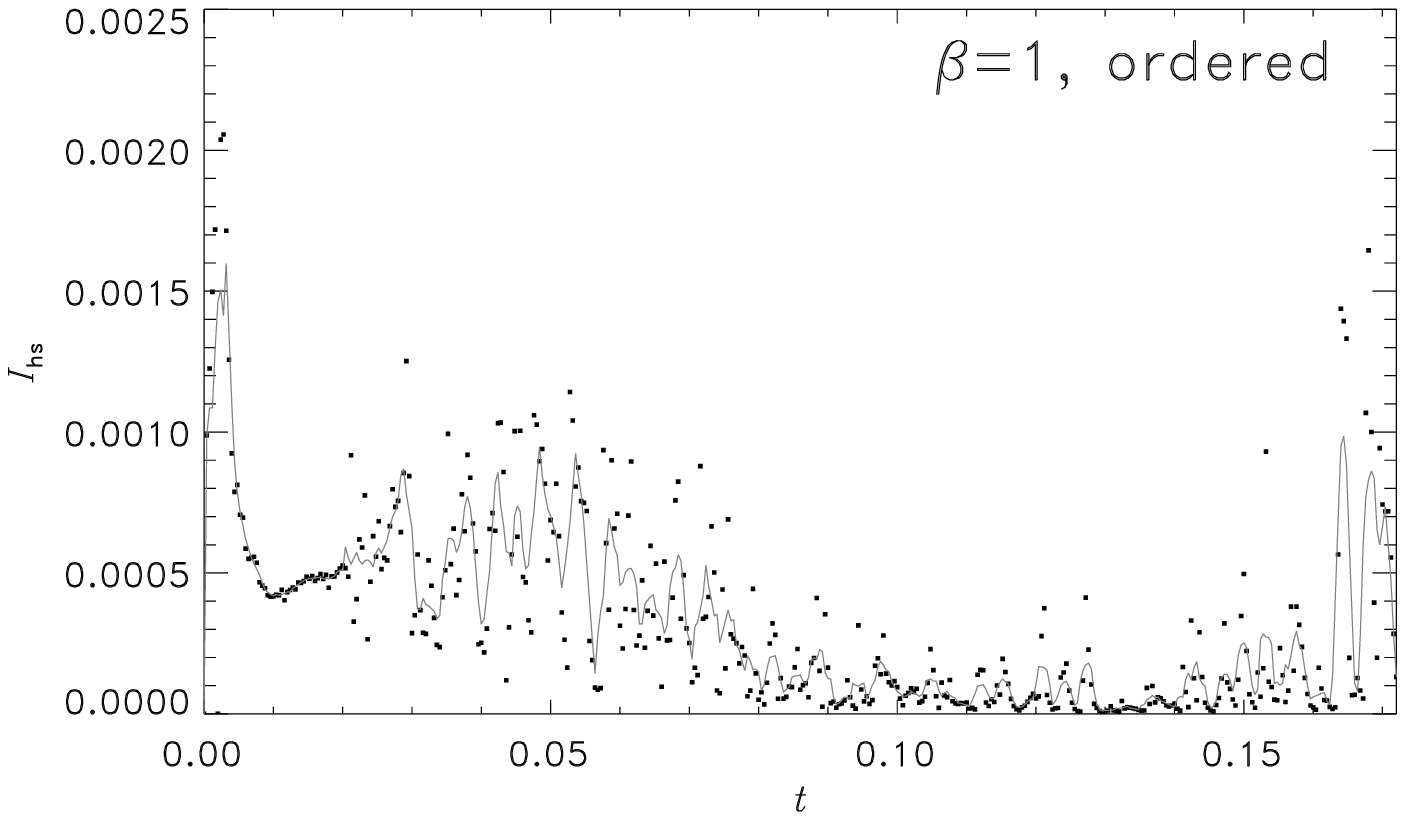}
\\
 \includegraphics[width=5cm]{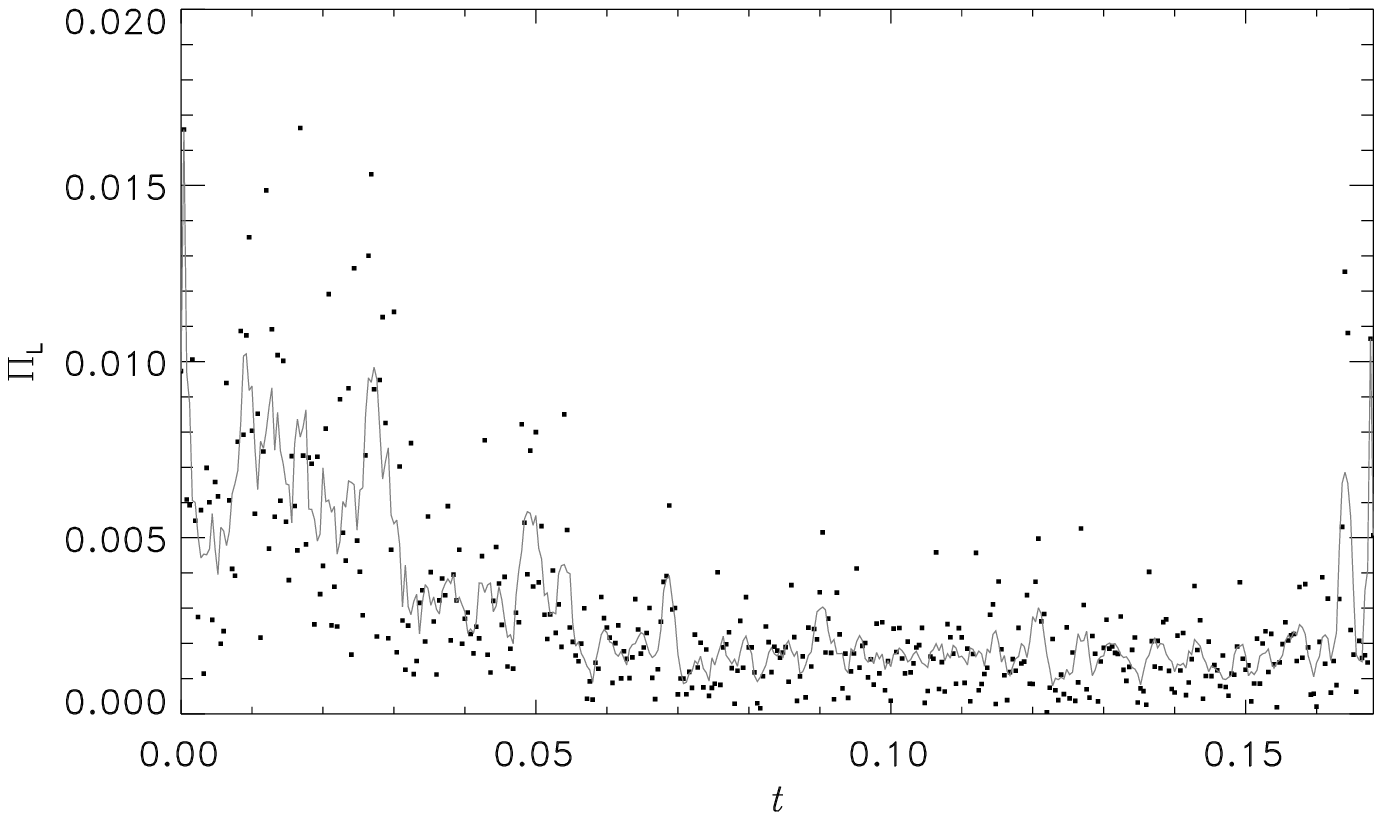}
&\includegraphics[width=5cm]{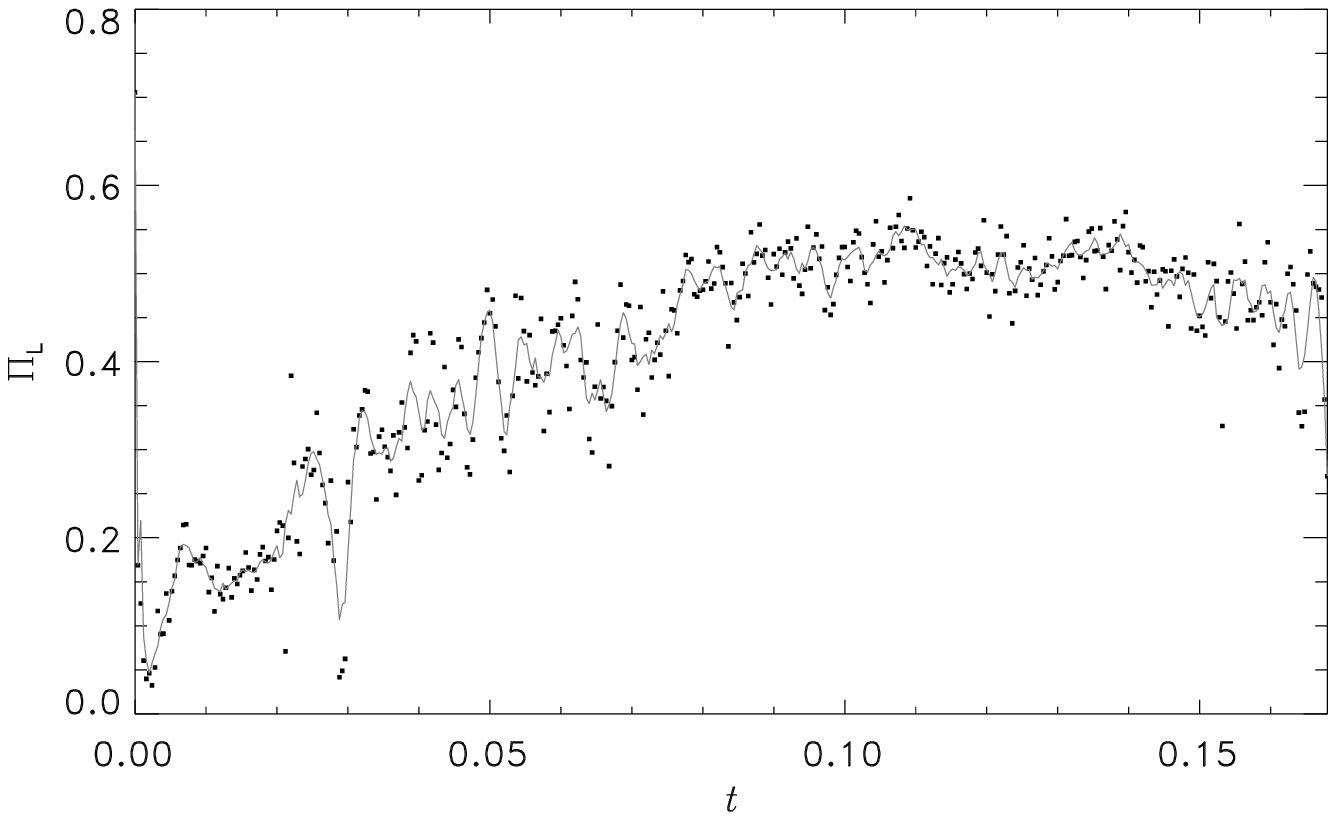}
&\includegraphics[width=5cm]{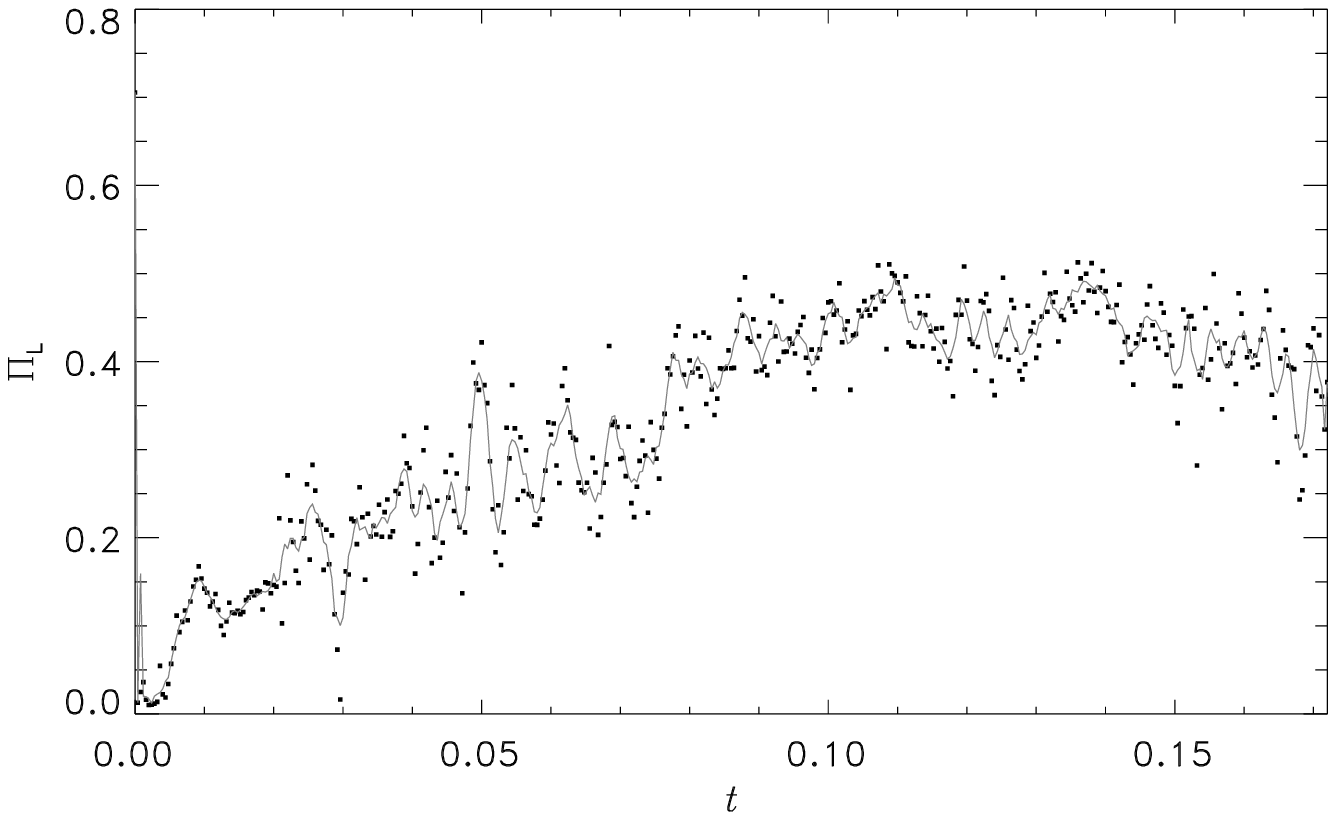}
\\
 \includegraphics[width=5cm]{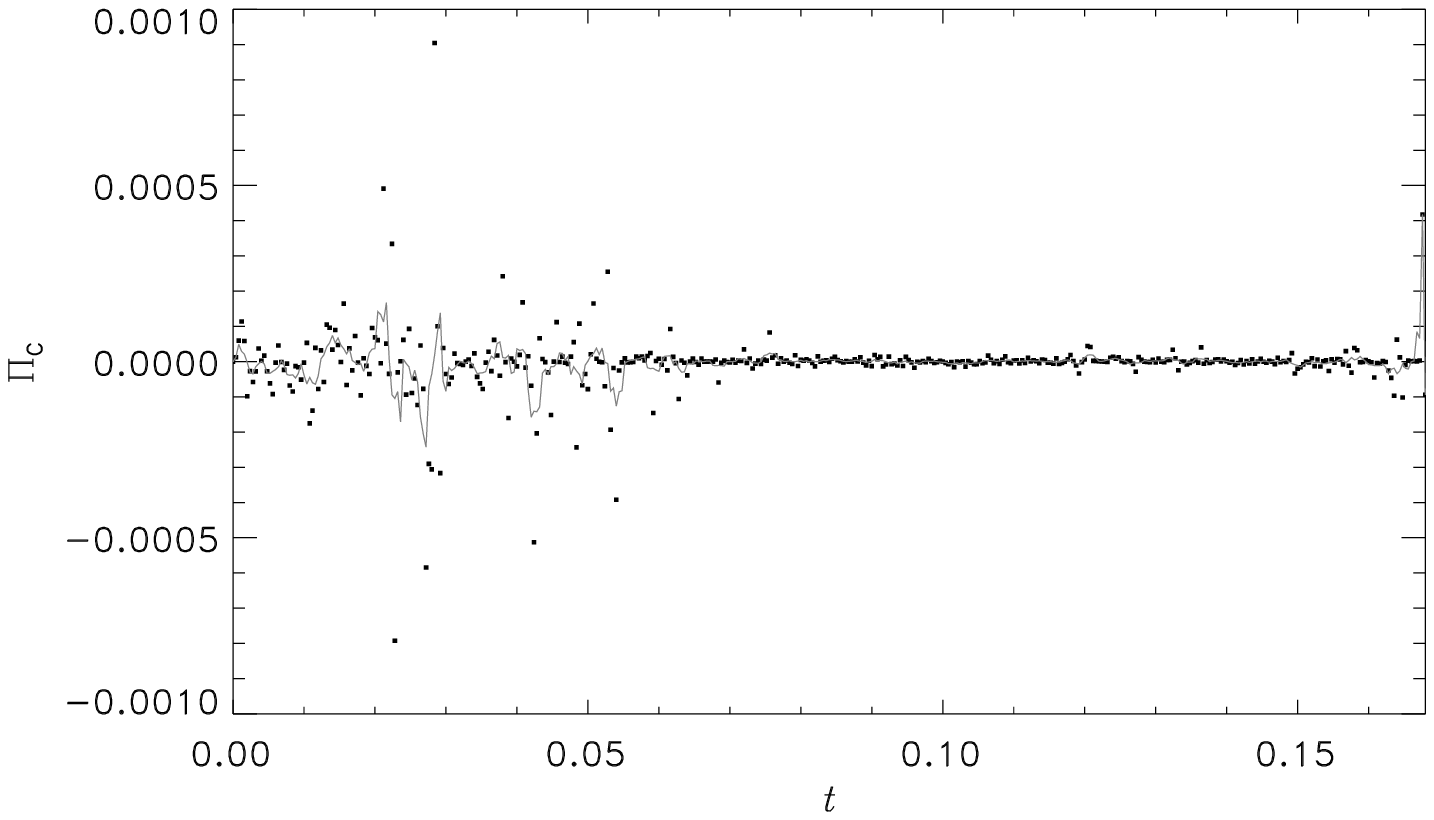}
&\includegraphics[width=5cm]{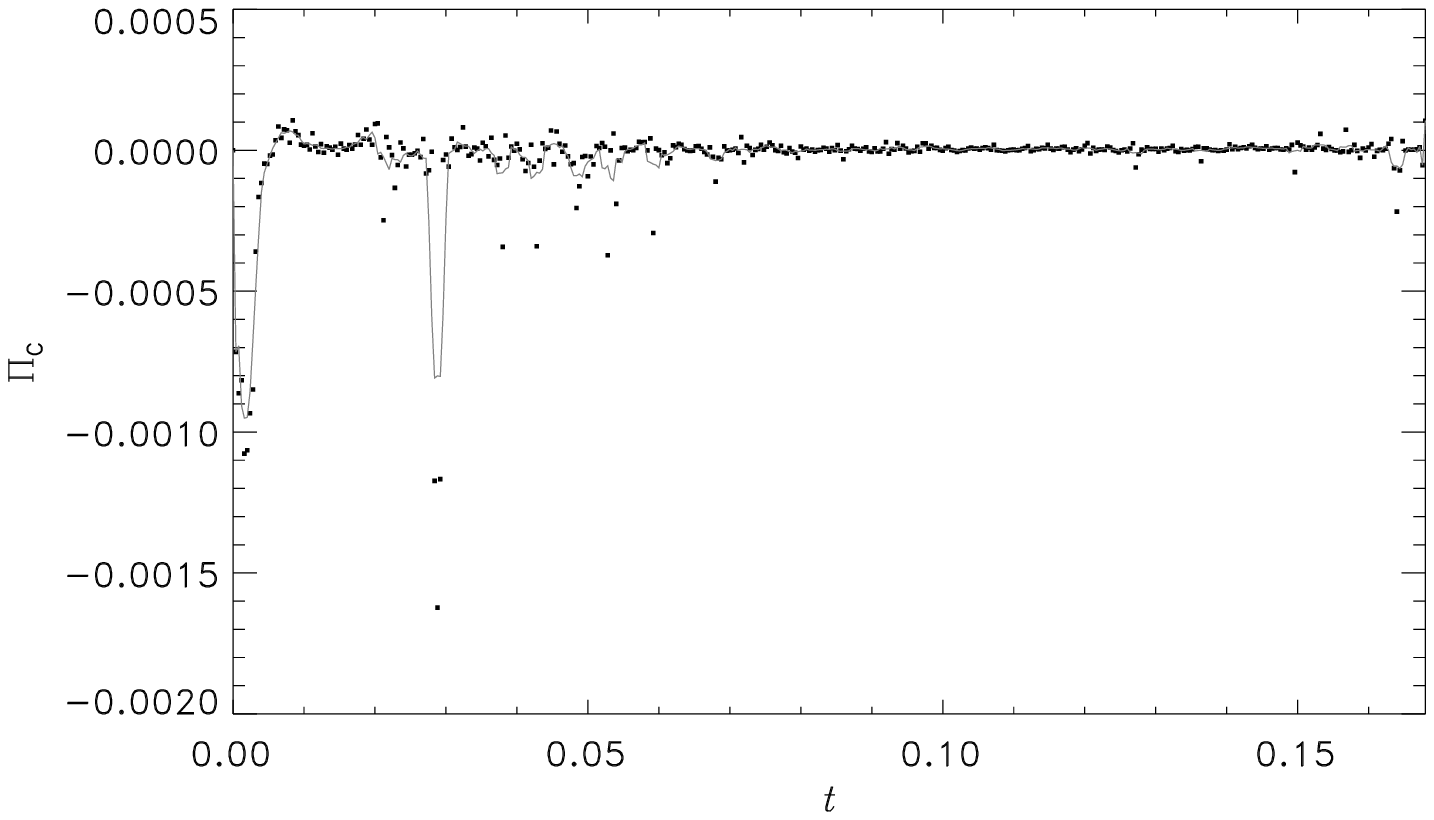}
&\includegraphics[width=5cm]{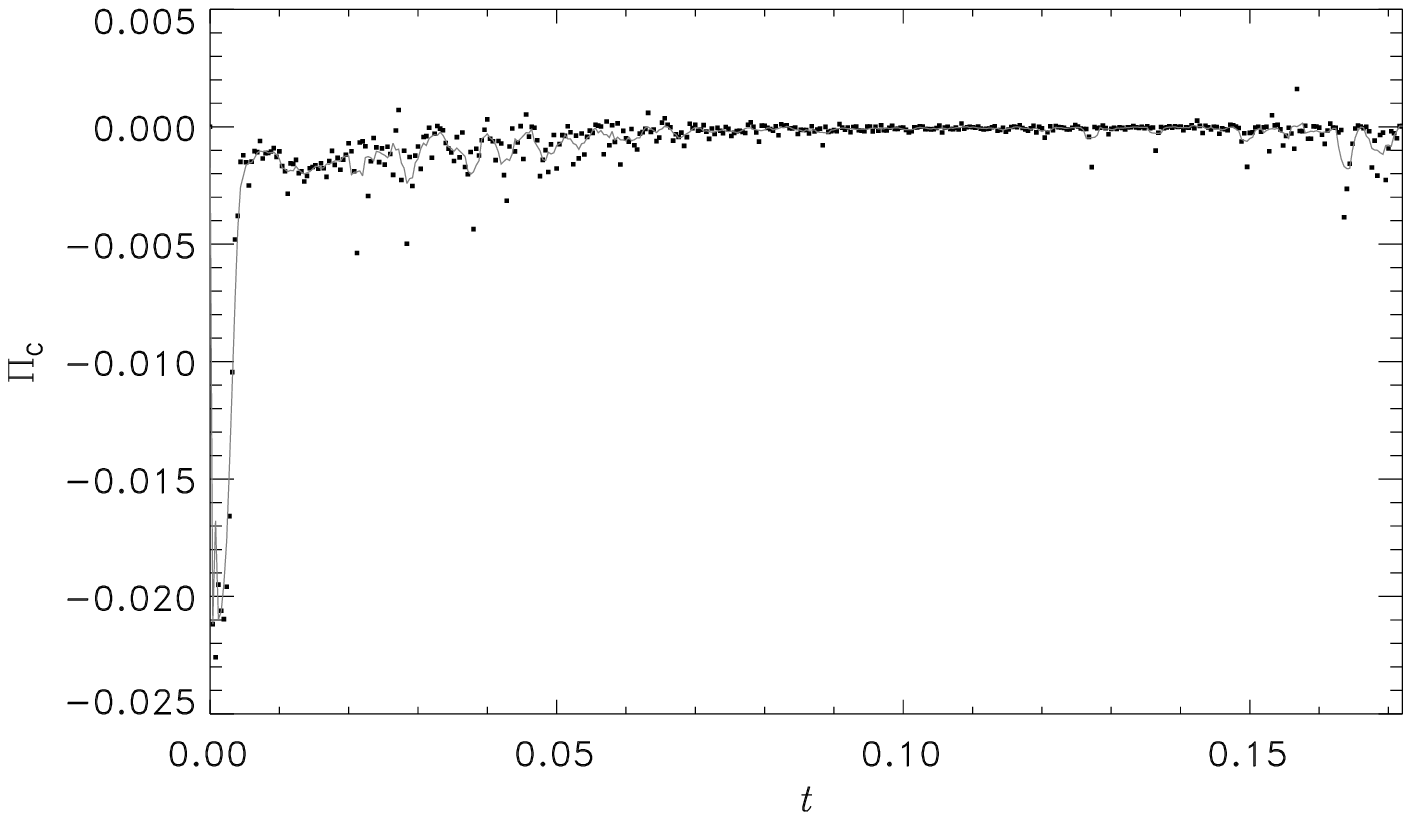}
\end{tabular}
\end{center}
\caption{
\color{Black}
Intensity and polarization light-curves from radiative transfer calculations based on
   the {\sc fast1o} jet simulations, 
   where a fast jet propagates in a uniform background medium.
An open left boundary is adopted.
The jet emission is dominated by distinct and coherent features, 
   such as the hotspot and jet shocks.
Dots mark exact data; grey curves show a 5-point boxcar average.
The spatially integrated intensity light curve (first row)
   is variable on many time scales.
Hotspot intensity ($I_{\rm hs}$, within a 20-cell radius)
   light curves are in the second row.
The hotspot is identified as the rightmost local maximum of intensity
   projected along the jet axis.
The  $I_{\rm hs}$ light curves fluctuate considerably more
   than those of the {\sc slow} jet
   and the {\sc fast1} jet with closed boundary, 
   as the hotspot of the {\sc fast1o} jet is unobscured by the cocoon.
The third and fourth rows show
   the integrated fractional linear and circular polarisation
   curves respectively.
}
\label{fig.temporal.fast1o}
\end{figure*}

{\color{Black}
The basic jet in a uniform background ({\sc fast1o})
   has a structure function that rises towards longer timescales.
Perhaps this reflects the trend for later flashes of the hotspots
   to be dimmer than earlier flashes
   (row two of Figure~\ref{fig.temporal.fast1o}).
The structure function is nonzero down to the smallest $\tau$ scales;
   implying that the time steps of our hydrodynamic snapshots
   are too coarse to resolve the vacillations of the brightest shocks.
Changing the field geometry from random to ordered
   (e.g. with $\beta=10$ fixed)
   has no significant effect on the structure functions.
Greater $\beta$ (weaker fields) raises the size of the $I_{\rm hs}$ fluctuations
   relative to the mean value,
   but the structure function retains the same shape in $\tau$.
}

\begin{figure*}
\begin{center}
\begin{tabular}{ccc}
 \includegraphics[width=5cm]{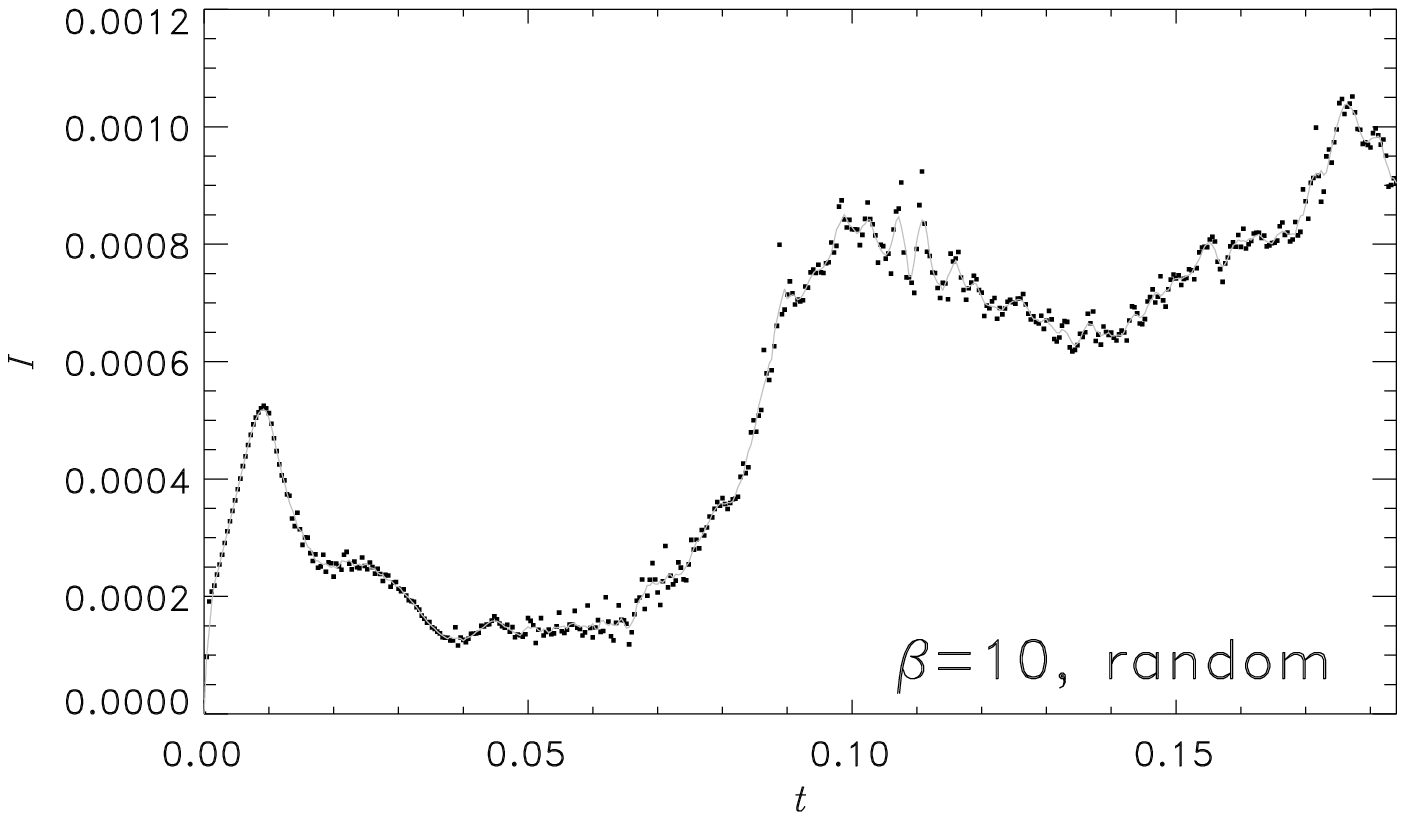}
&\includegraphics[width=5cm]{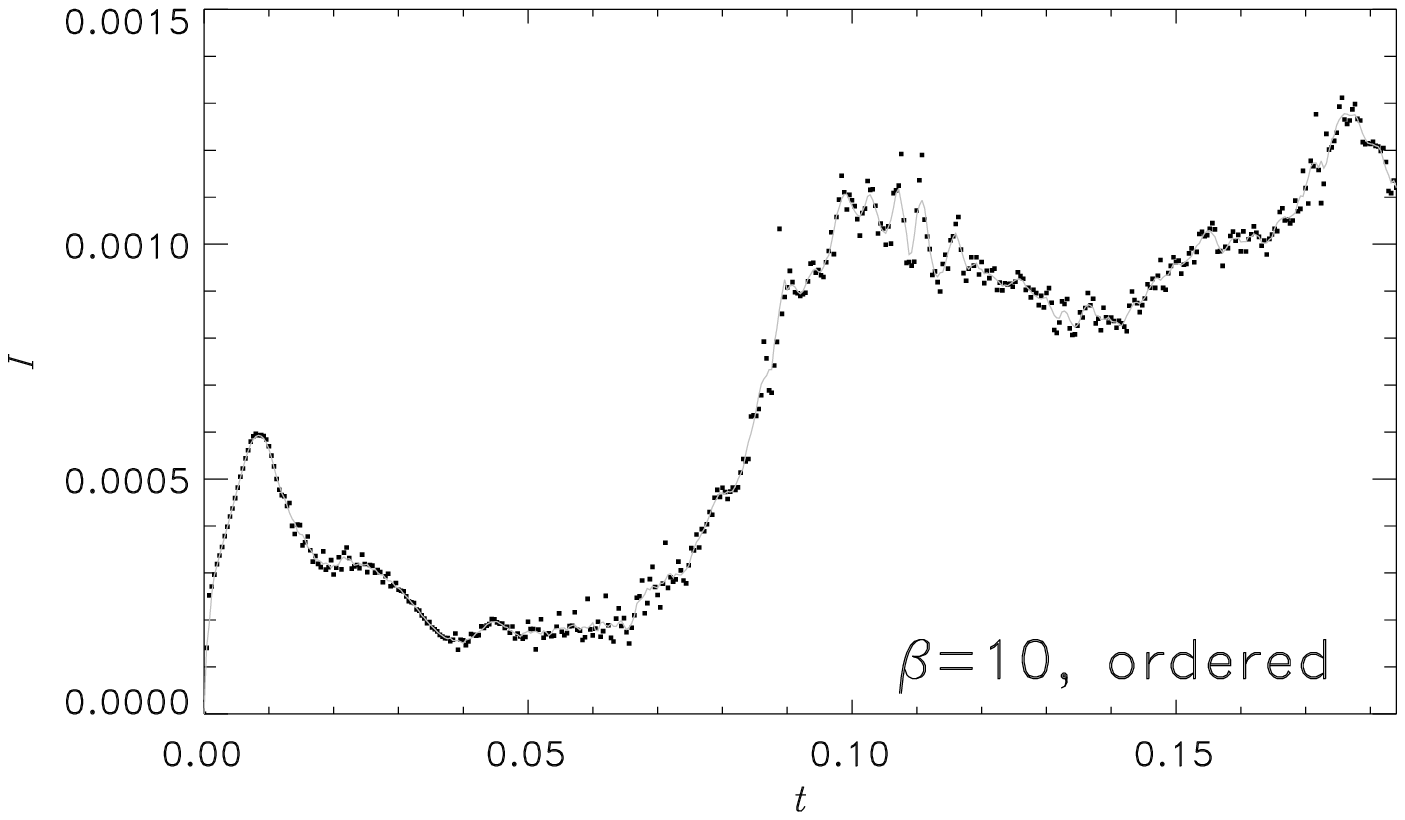}
&\includegraphics[width=5cm]{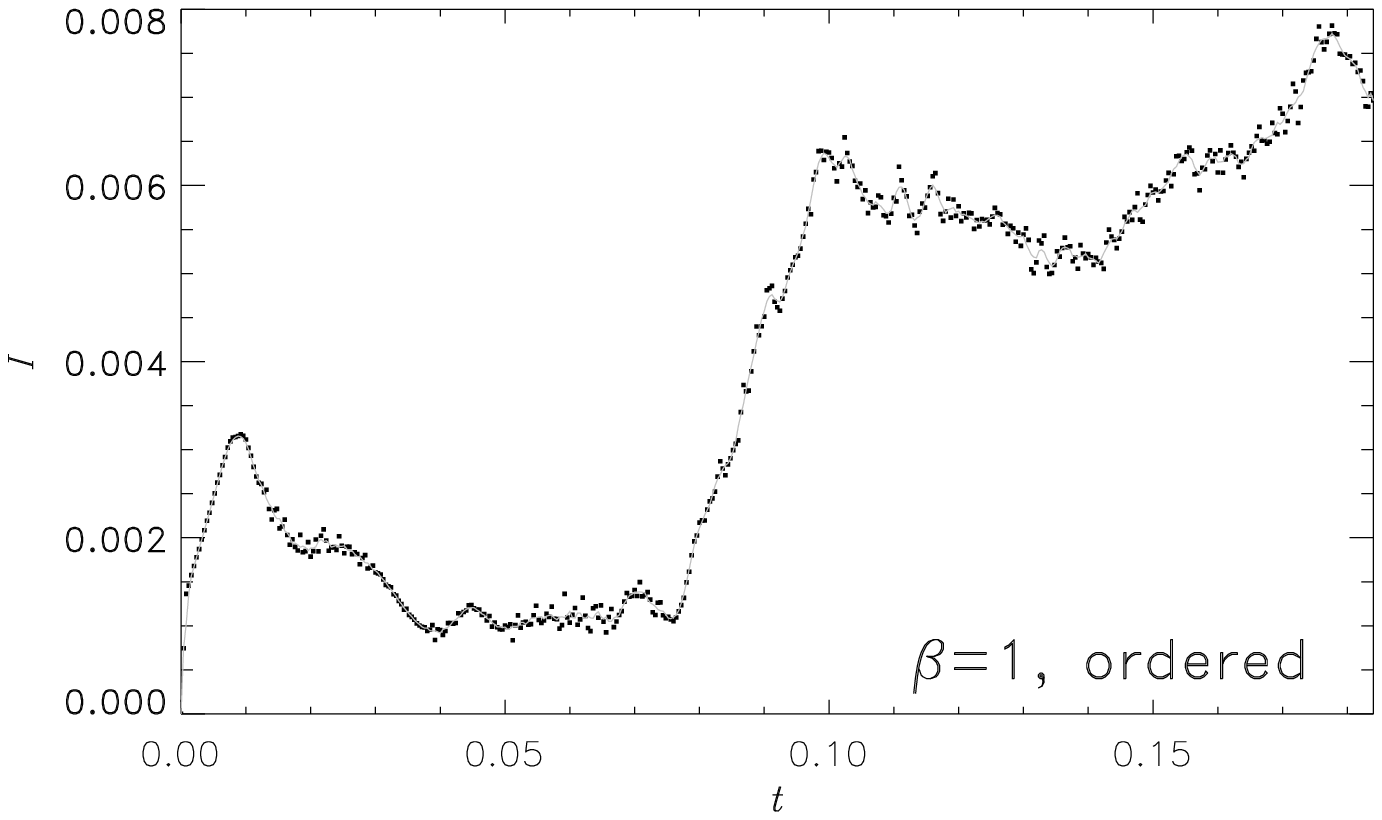}
\\
 \includegraphics[width=5cm]{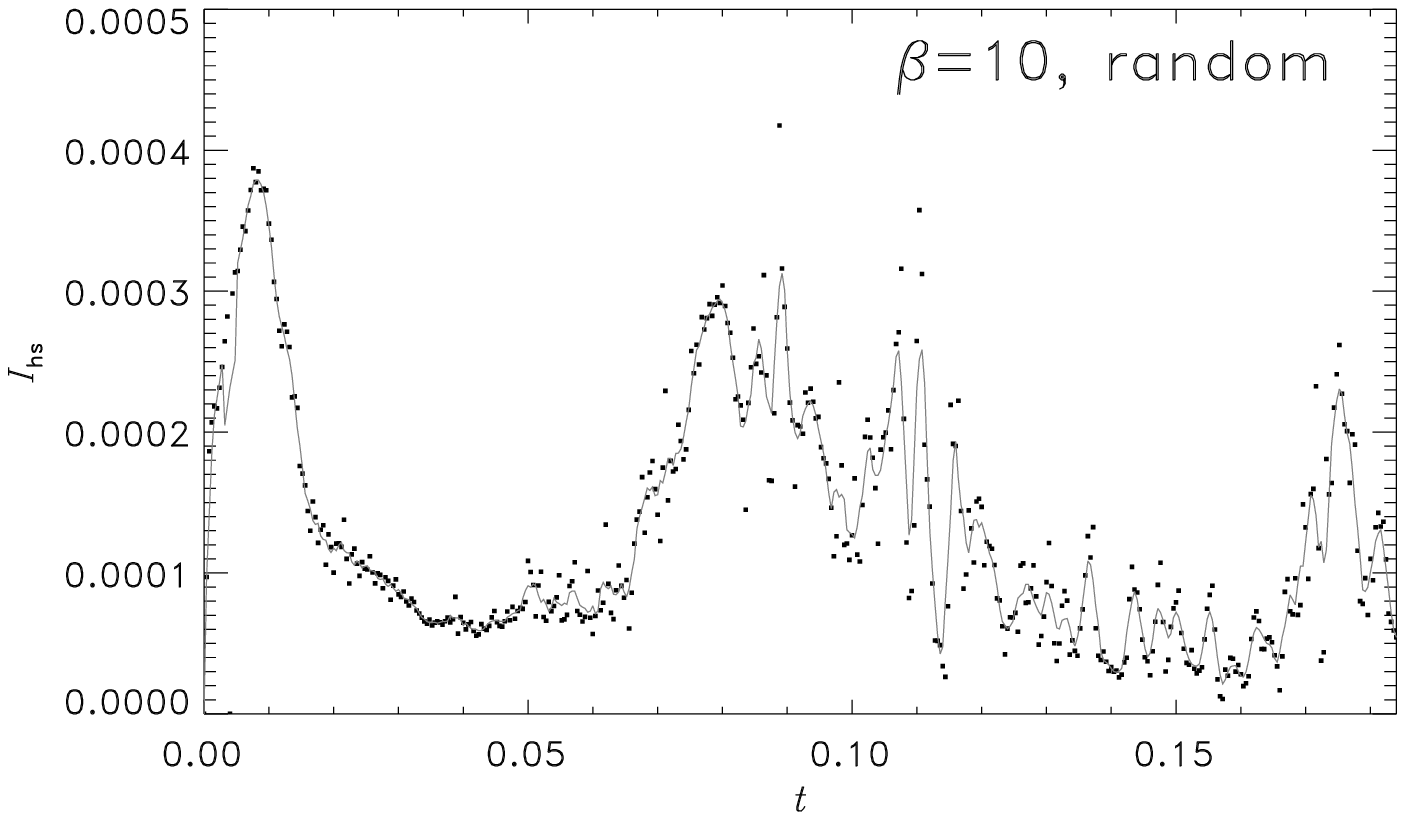}
&\includegraphics[width=5cm]{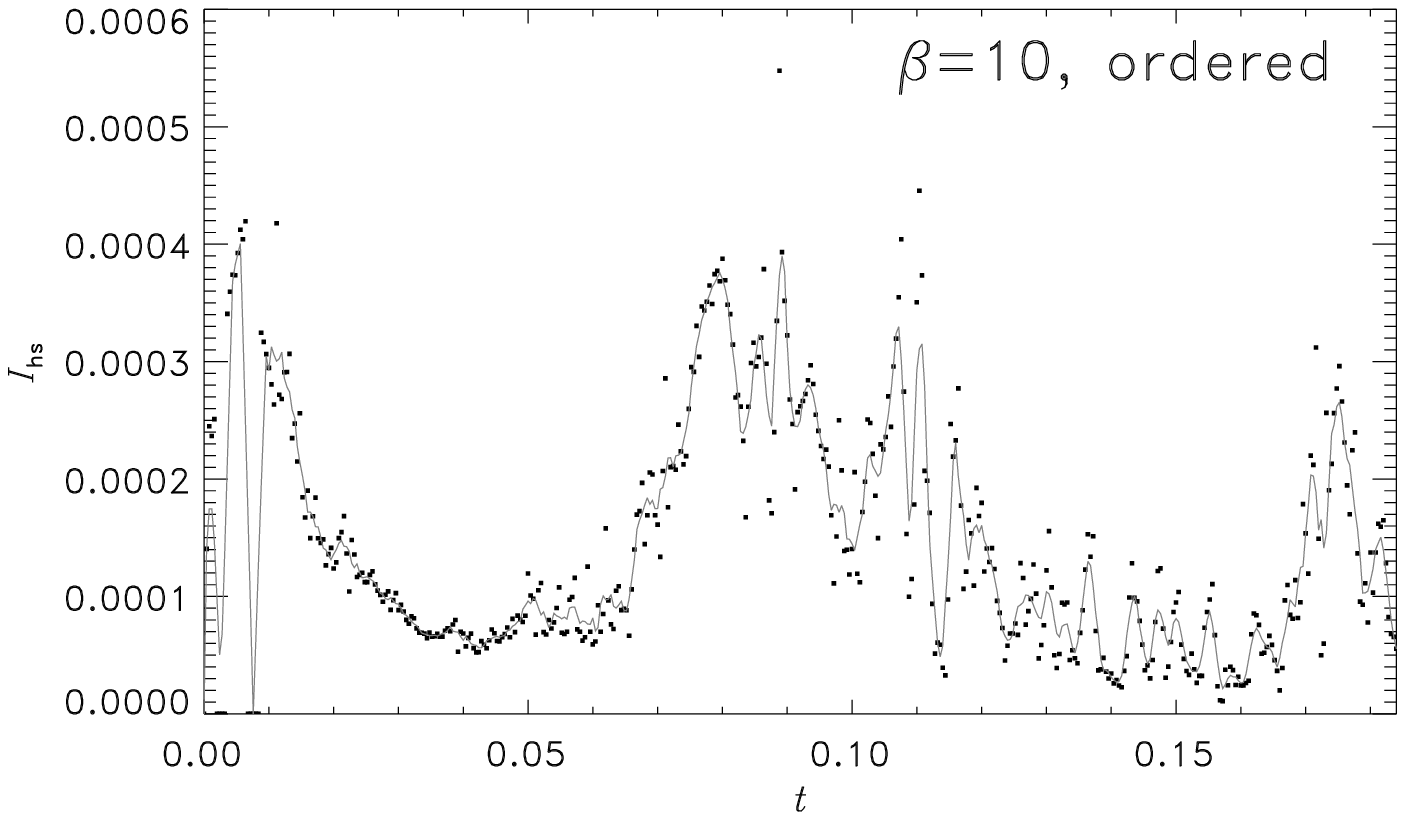}
&\includegraphics[width=5cm]{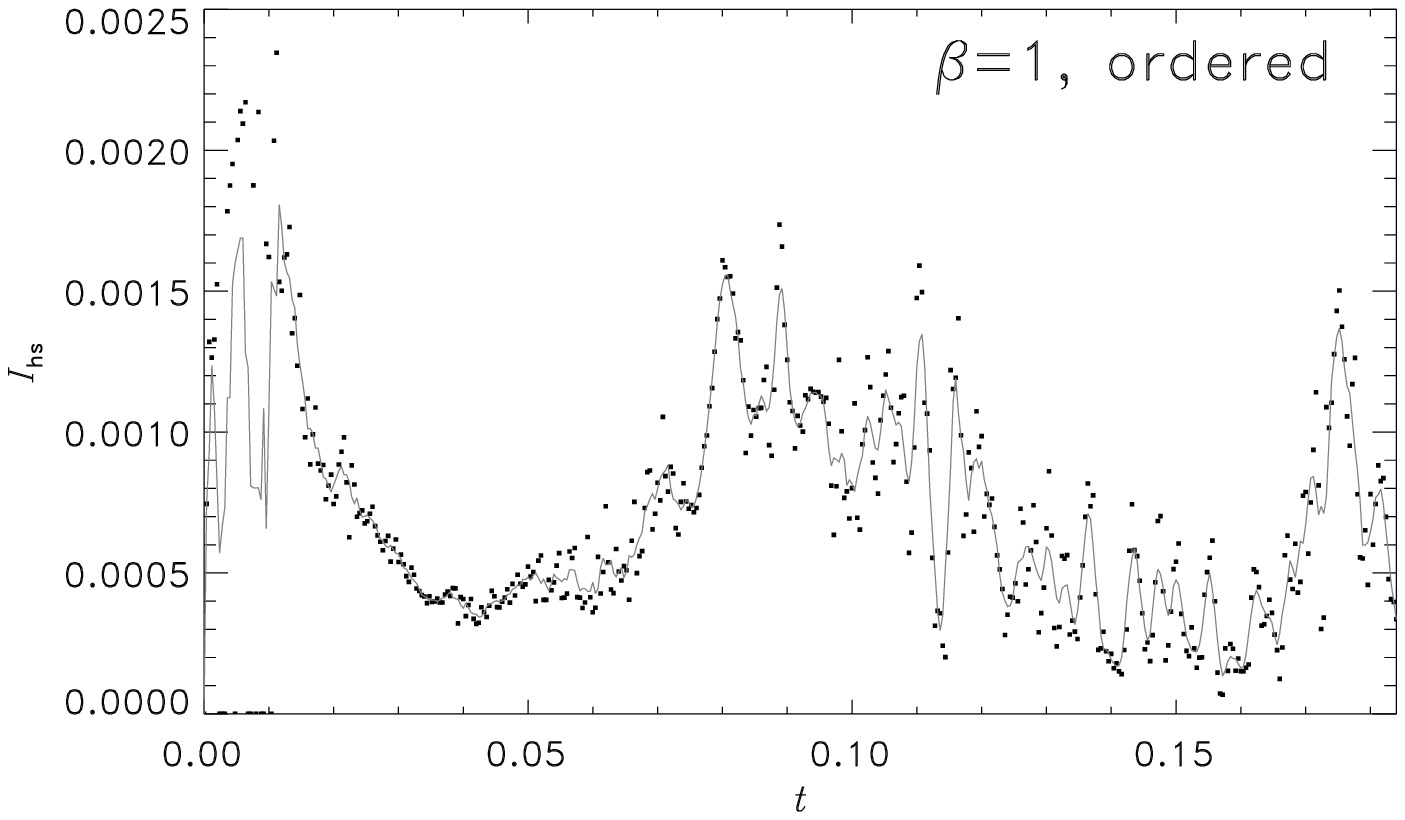}
\end{tabular}
\end{center}
\caption{
Synchrotron radio lightcurves (at 6~cm) of the {\sc fast2o} jets. 
In the simulations,
   the jet traverses a background medium with a declining density. 
The left boundary is open.  
The magnetic field is random or ordered ($\Bvec\parallel\vvec$)
   with $\beta$ values as indicated in each panel.
The open left boundary reduces accumulation of the backflow, 
   and he jet shocks outshine the cocoon emission, 
   (cf.\  the {\sc fast1o} jet simulations).   
The fluctuations of the emission from the hotspot
   are visible in the $I_{\rm hs}$ lightcurves.
}
\label{fig.temporal.fast2o}
\end{figure*}

\begin{figure*}
\begin{center}
\begin{tabular}{ccc}
 \includegraphics[width=5cm]{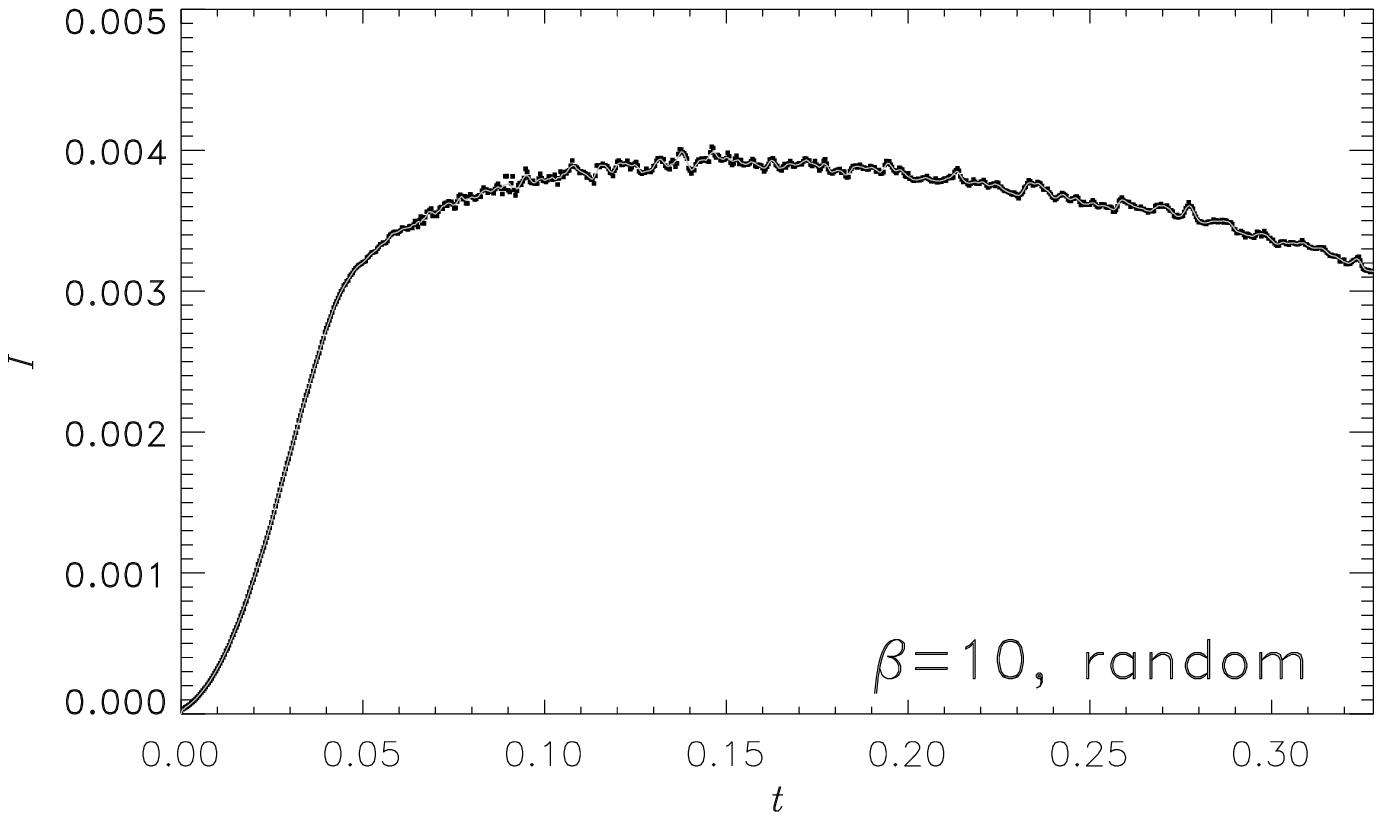}
&\includegraphics[width=5cm]{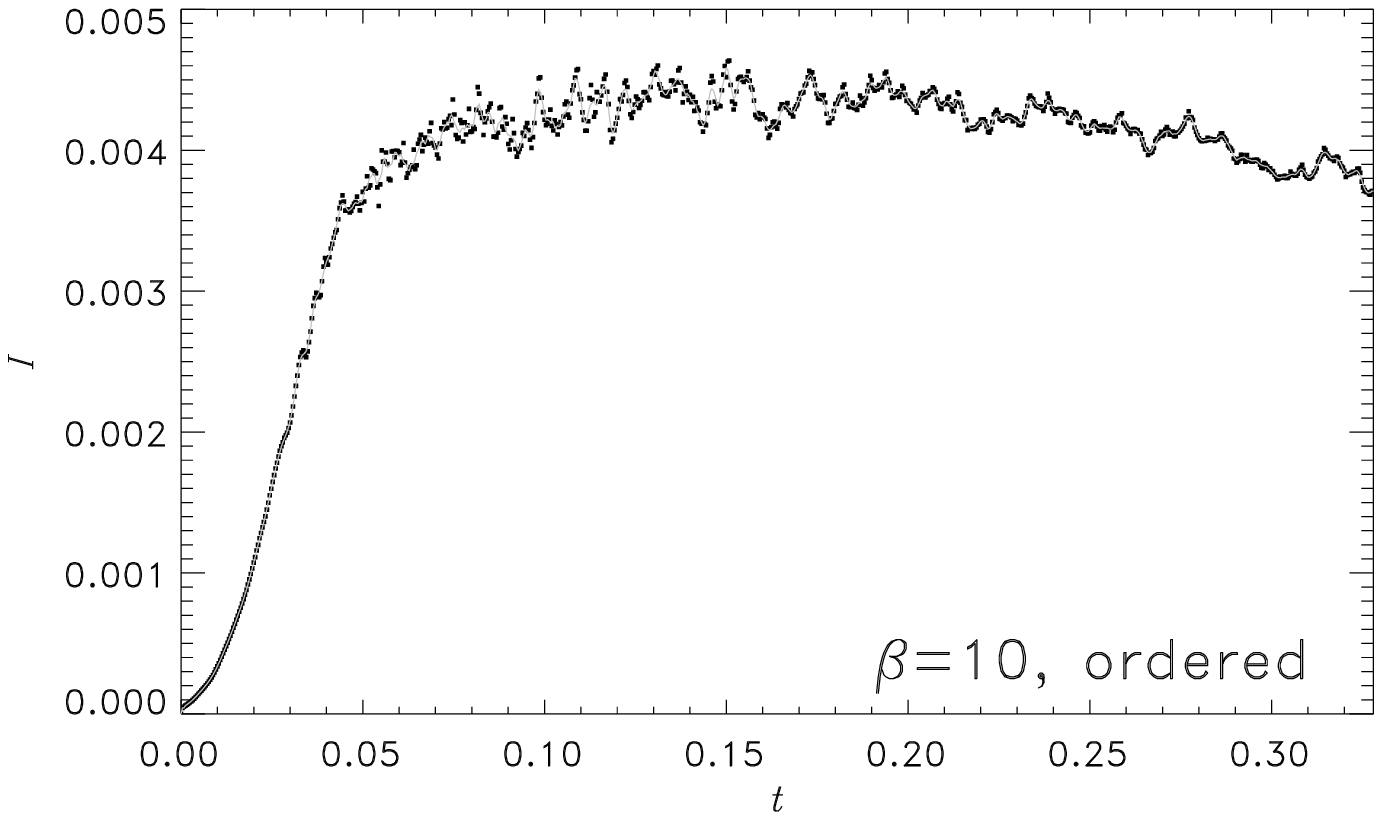}
&\includegraphics[width=5cm]{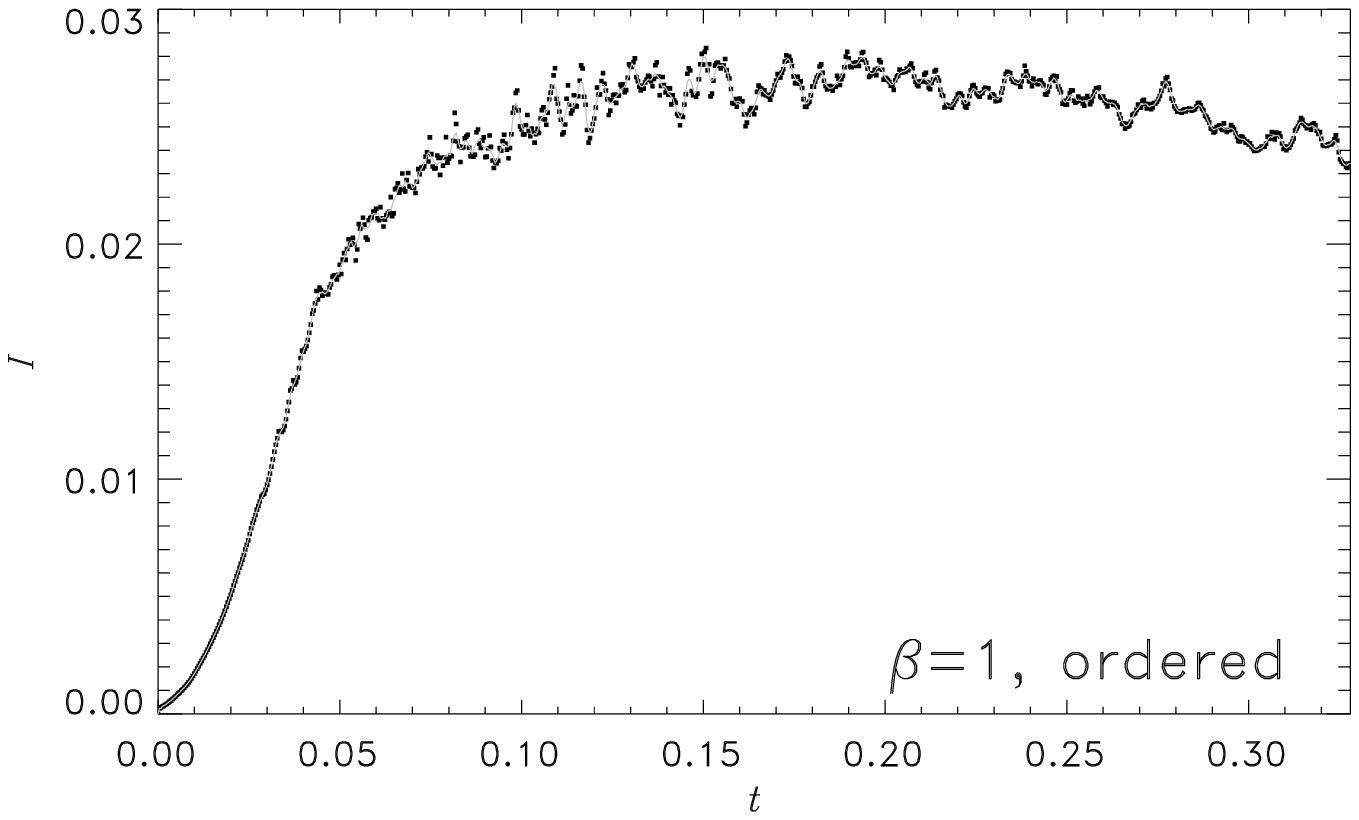}
\end{tabular}
\end{center}
\caption{
Synchrotron radio lightcurves of the {\sc fast2} jets,
   as in Figure~\ref{fig.temporal.fast2o}. 
The simulations adopt same conditions as those in
   the {\sc fast2o} jet simulations, 
   except that the left boundary is reflective.
}
\label{fig.temporal.fast2}
\end{figure*}

\begin{figure*}
\begin{center} 
\begin{tabular}{ccc}
 \includegraphics[width=5cm]{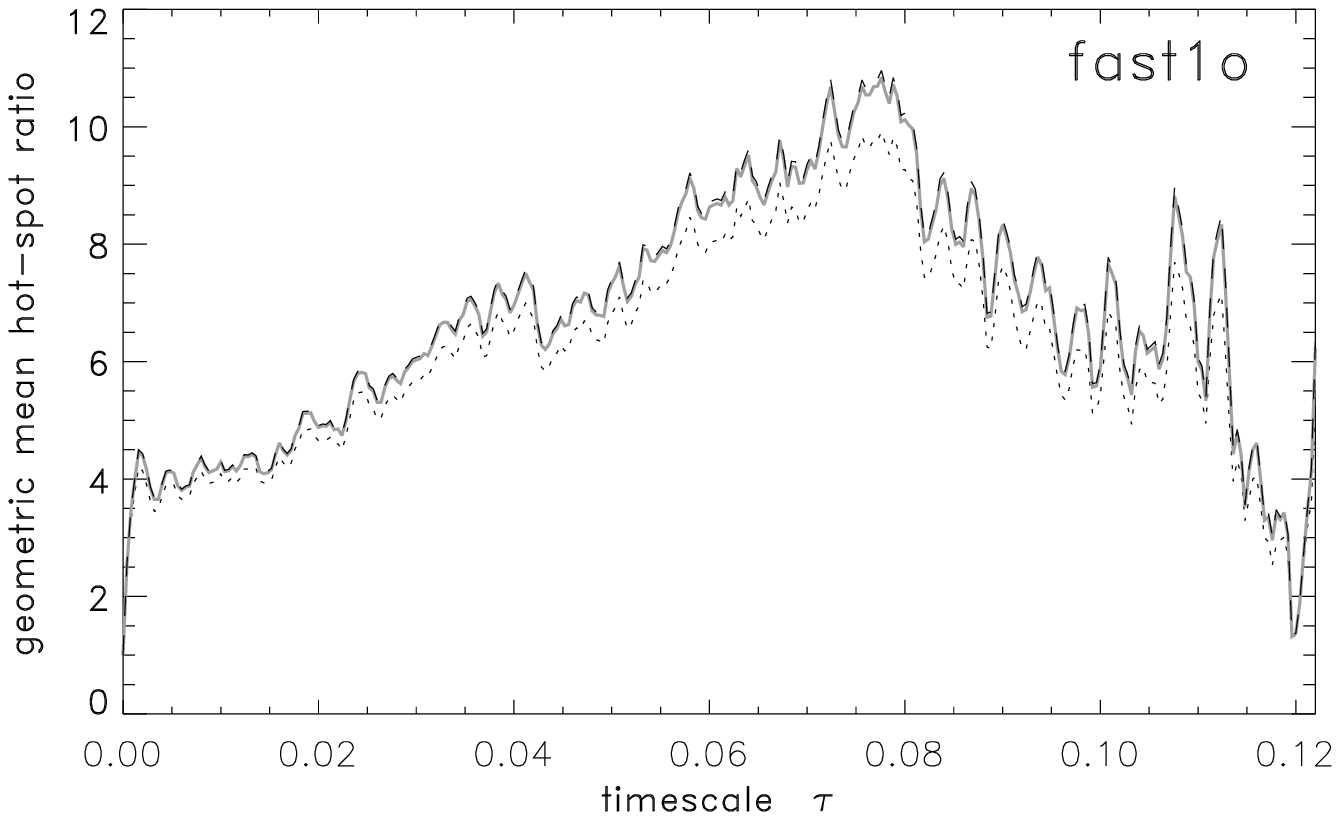}
&\includegraphics[width=5cm]{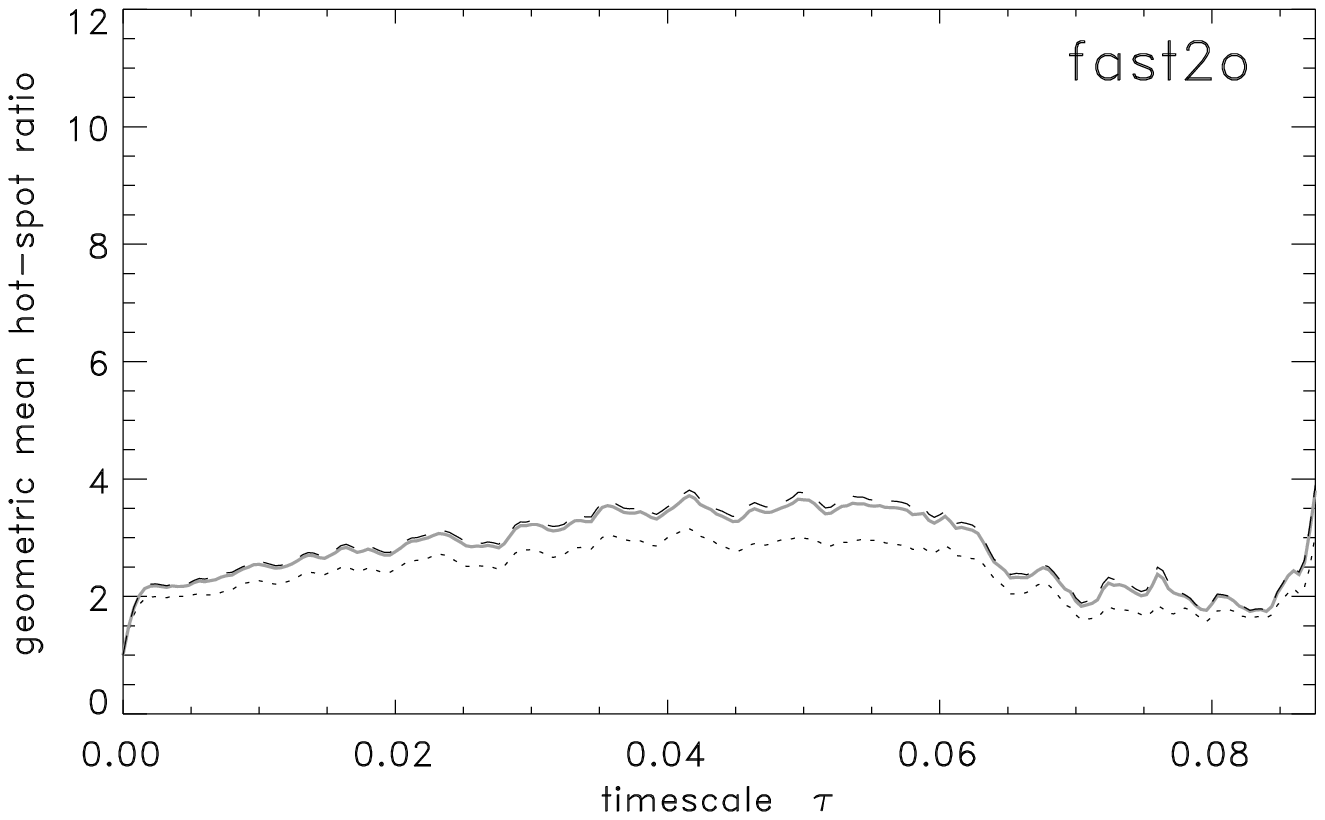}
&\includegraphics[width=5cm]{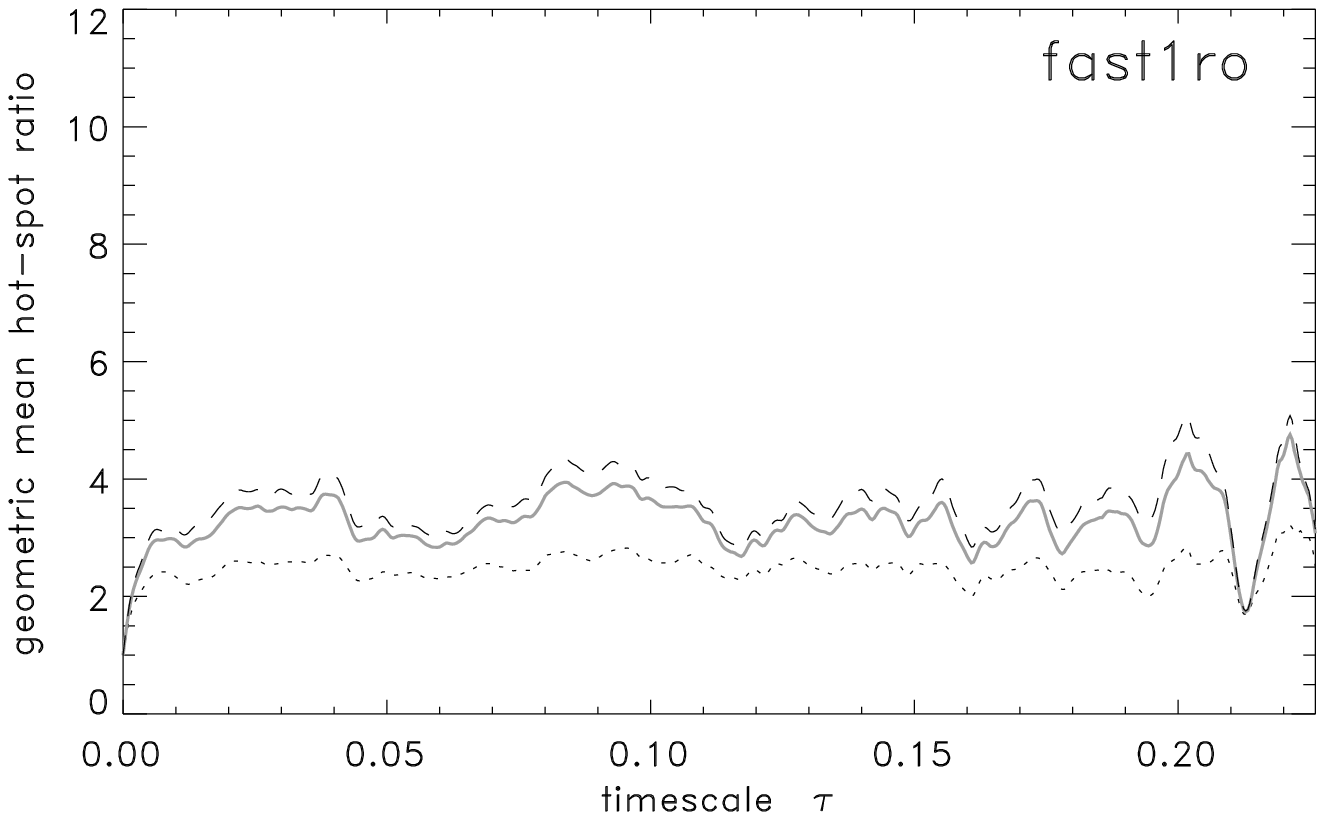}
\\
 \includegraphics[width=5cm]{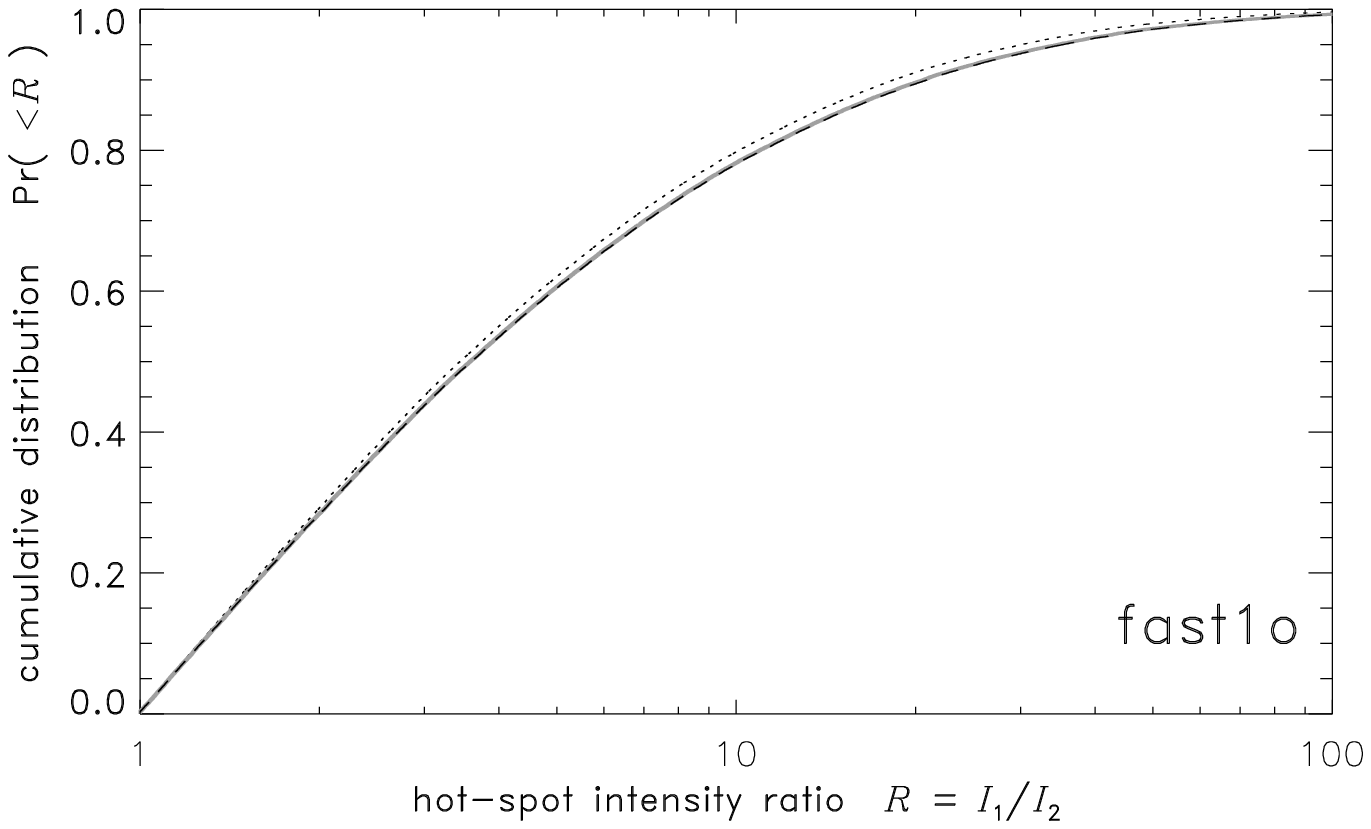}
&\includegraphics[width=5cm]{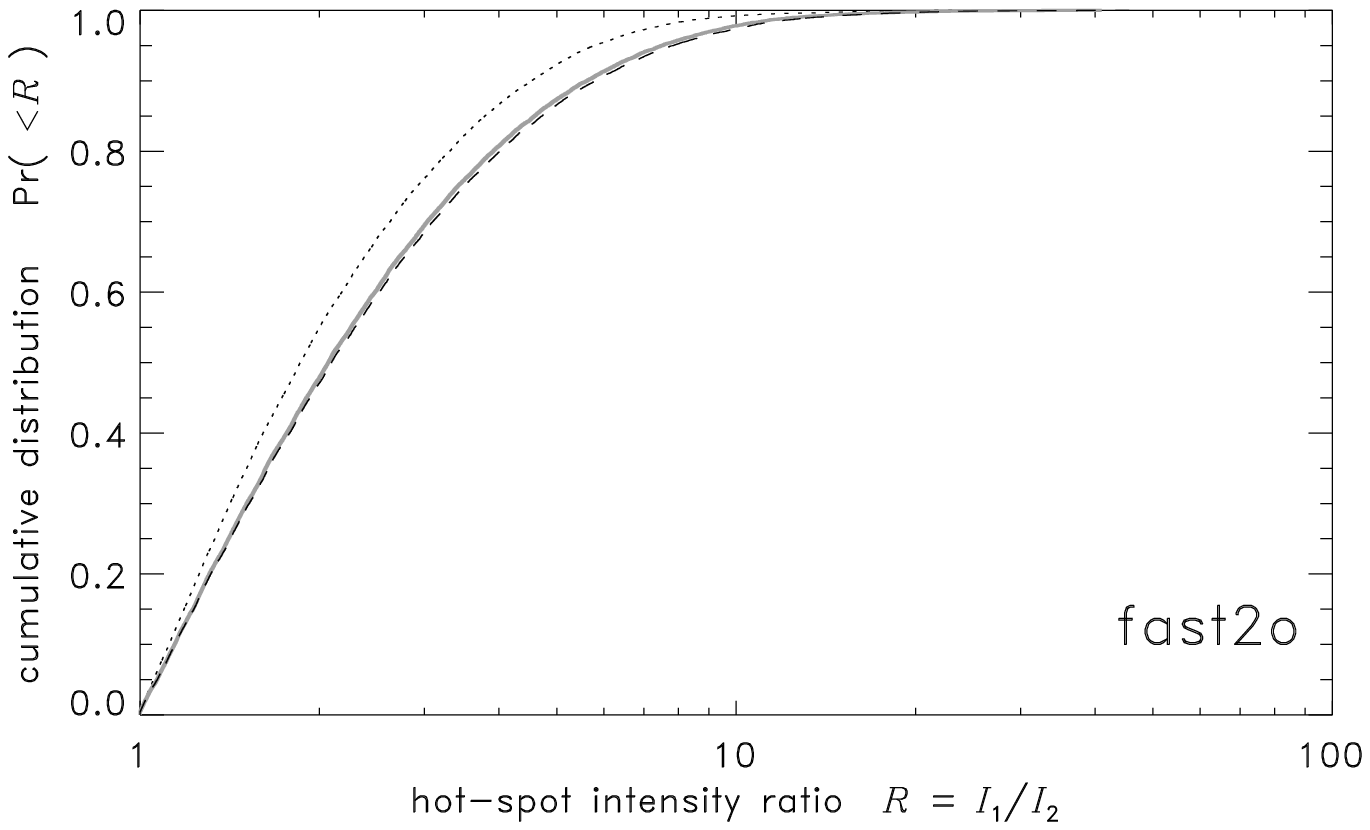}
&\includegraphics[width=5cm]{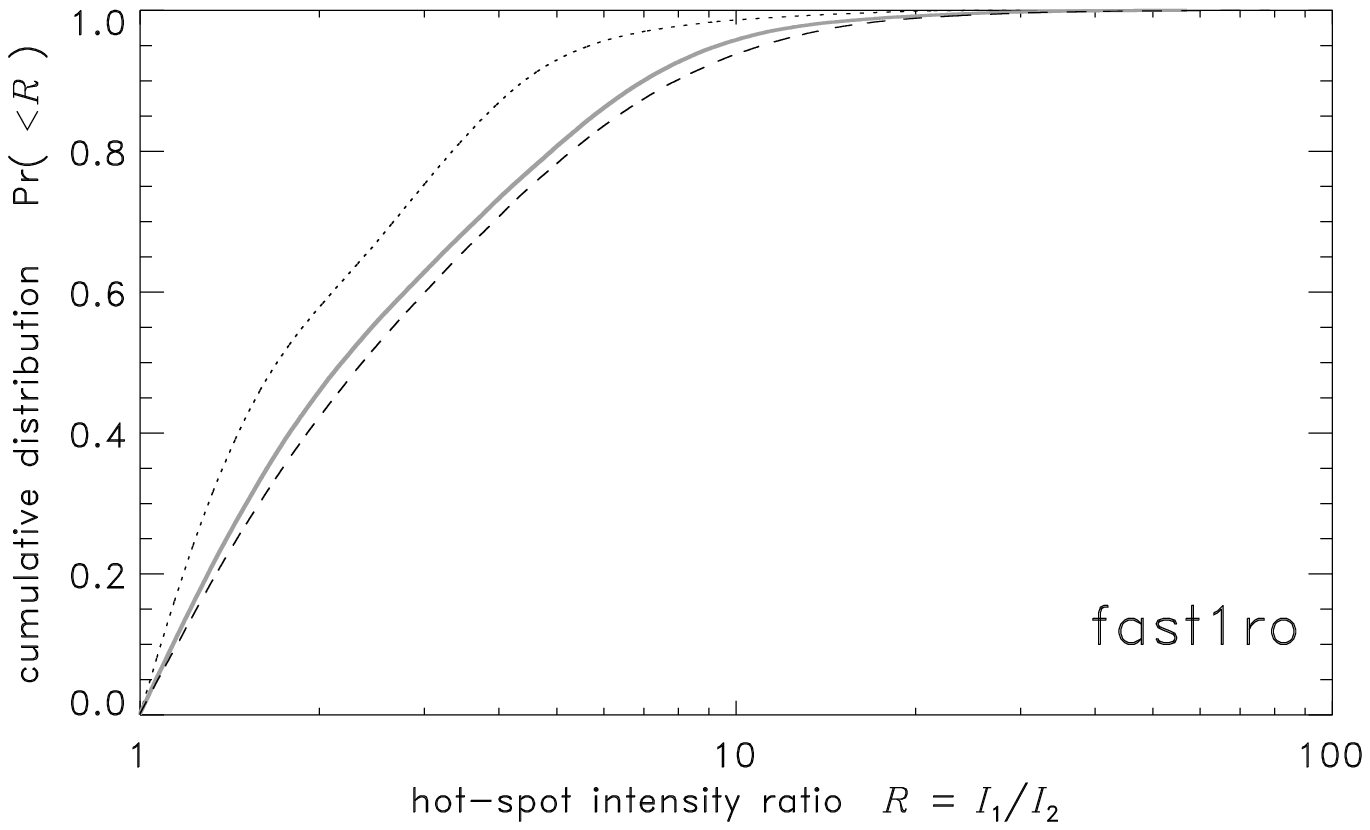}
\end{tabular}
\end{center} 
\caption{ 
\color{Black}
Illustration of the variability of the luminosity ratio of two hotspots
   in open-boundary simulations.
   ({\sc fast1o}, {\sc fast2o} and {\sc fast1ro}
   from left to right respectively).
The magnetic fields are quasi-poloidal,
   $\Bvec\parallel\vvec$.
The top row shows geometric mean hotspot ratios
   obtained from the structure function of $\ln(I_{\rm hs})$,
   for frames when the bow shock has advanced to $x_{\rm s}\ge60~\kpc$.
The bottom row shows cumulative distribution functions
   of the hotspot ratios at independent times.
The dotted, grey and dashed curves
   depict cases with $\beta=1, 10$ and $100$.
}
\label{fig.hotspot.cdf}
\end{figure*}

{\color{Black}
With a declining density profile ({\sc fast2o})
   the structure function of $I_{\rm hs}$
   reveals peaks around timescales
   $\tau\sim0.05U_t$ and $0.16U_t$.
The $0.05U_t$ bump may reflect
   the long interval between prolonged bright flares
   (seen in the initial episode and two later events
   in Figure~\ref{fig.temporal.fast2o}).
The peaks become sharper and more distinct for greater $\beta$:
   given the same flow evolution, a less magnetised hotspot
   flashes relatively more intensely.
The structure function of total intensity $I(t)$
   increases at longer timescales in the domain $\tau\la0.10U_t$,
   perhaps describing with the brightenning trend,
   or the jump in $I$ around the time ($\approx0.08U_t$)
   when the hotspot first separates from the pseudocore shock.
The separation of subsequent diamond shocks at $\approx0.02U_t$ intervals
   do not produce a clear feature in the structure function of $I$.

The jet with a flat background 
   and relativistic equation of state ({\sc fast1ro})
   has a hotspot that fluctuates with similar power at all time scales.
The structure function of $I_{\rm hs}(t)$ is similarly flat
   for $\beta=1, 10$ and $100$.
There is a slight hump at nearly the duration of the simulation;
    this may be due to the early, dim and steady phase at $t\la0.9U_t$.
The general flatness of the temporal structure functions
   suggests that the stochastic variation of the hotspot is like white noise
   (at least for the ``hotspot'' as we have presently defined it).

A radio galaxy of FR2 type has two opposite hotspots.
Even if both the jet and the counter-jet are identical
   and unvarying at the nucleus,
   their hotspots vary independently:
   differently at any given epoch,
   but stochastically in essentially the same manner.
We assume that the envelope or distribution of this innate variability
   is well characterised by the recorded variability
   during sufficiently late stages of the simulations.
Frames occuring after the formation of the pseudocore shock
   should be an adequate selection.

The upper panels of Figure~\ref{fig.hotspot.cdf}
   show what is effectively the goemetric mean
   of the hotspot ratios at a time separation $\tau$,
   as calculated from the structure function of the log-intensities,
	$\exp\sqrt{ S_2(\ln(I_{\rm hs});\tau) }$.
Typical values are $\approx 3$ for {\sc fast2o} and {\sc fast1ro},
   but exceed ten at medium timescales for {\sc fast1o}.
In the lower panels of Figure~\ref{fig.hotspot.cdf}
   we draw random values of $I_{\rm hs}$ from times when $x_{\rm s}\ge 60~\kpc$,
   and plot the cumulative distribution function
   of the hotspot luminosity ratio.
Our radiative transfer calculations imply
   that ratios of factors of a few
   can occur frequently by chance
   (even though the paired jets are equal and steady).
Ratios in excess of ten are not rare for {\sc fast1o}.
For each simulation, the $\beta=1$ calculations
   give the least variable hotspot ratios.
Perhaps this is because stronger magnetism
   raises the opacity of the brightest features,
   so that the foreground emitting surfaces
   effectively hide some of the variability farther from the virtual camera.
The $\gamma=4/3$ simulation {\sc fast1ro}
   shows this effect most strongly. 
The model with a radially declining density profile for the ambient medium
   ({\sc fast2o})
   shows less variable hotspot ratios
   than the standard uniform background ({\sc fast1o}).

The general lesson to be drawn from these plots
   is that the apparent differences between opposite hotspots
   (during a single observational epoch)
   may be largely stochastic and transitory
   (even while the jet nozzles remain steady)
   and that caution is warranted when using these comparisons
   alone to constrain jet orientation, power
   or the behaviour of the nucleus.
Doppler and orientation effects complicate matters further,
   calling for more arduous parameter surveys in future.
To make precise observational inferrences,
   it may be best to involve complementary evidence from other bands,
   or to take a statistical view of large ensembles of AGN.
}

\begin{figure*}
\begin{center}
\begin{tabular}{ccc}
 \includegraphics[width=5cm]{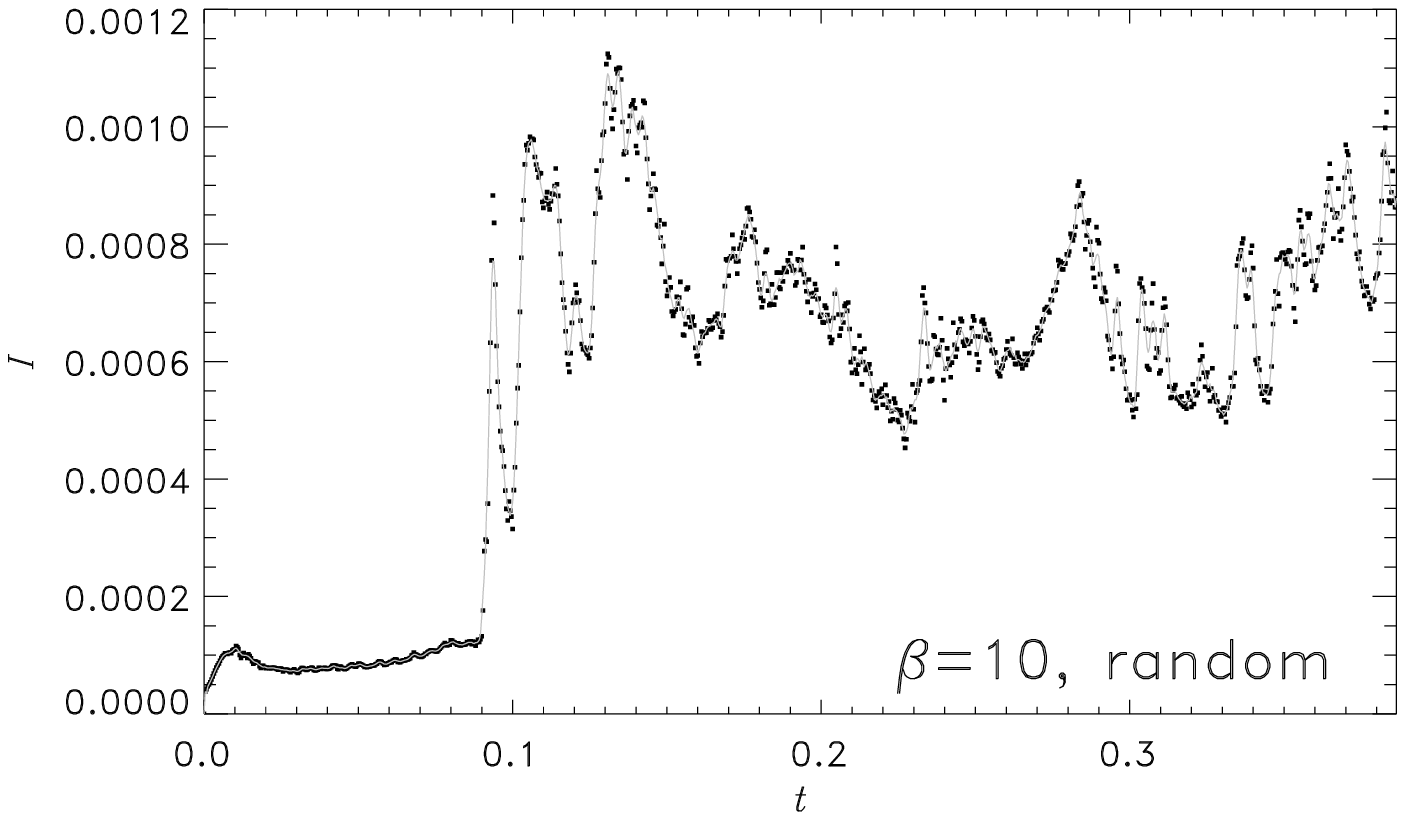}
&\includegraphics[width=5cm]{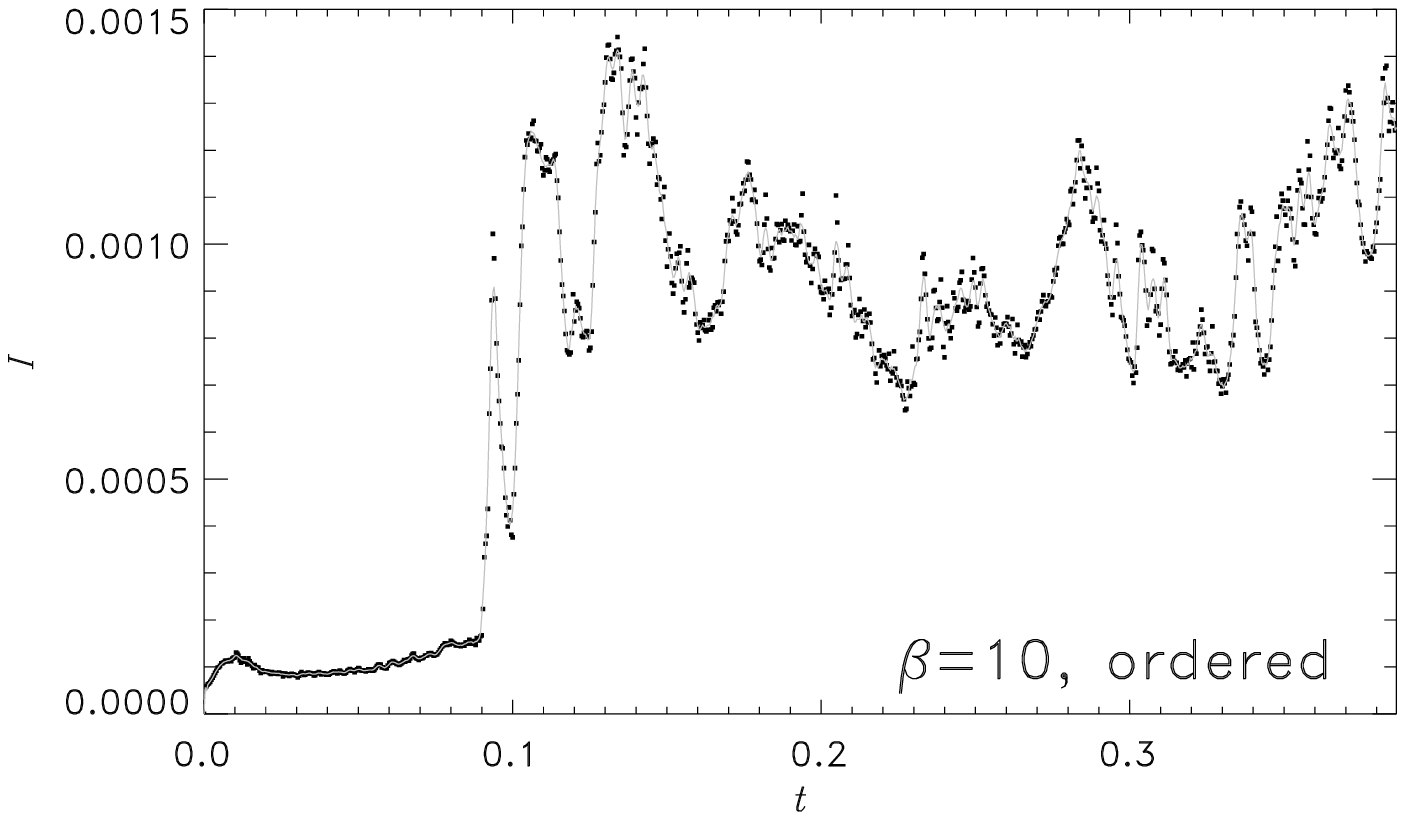}
&\includegraphics[width=5cm]{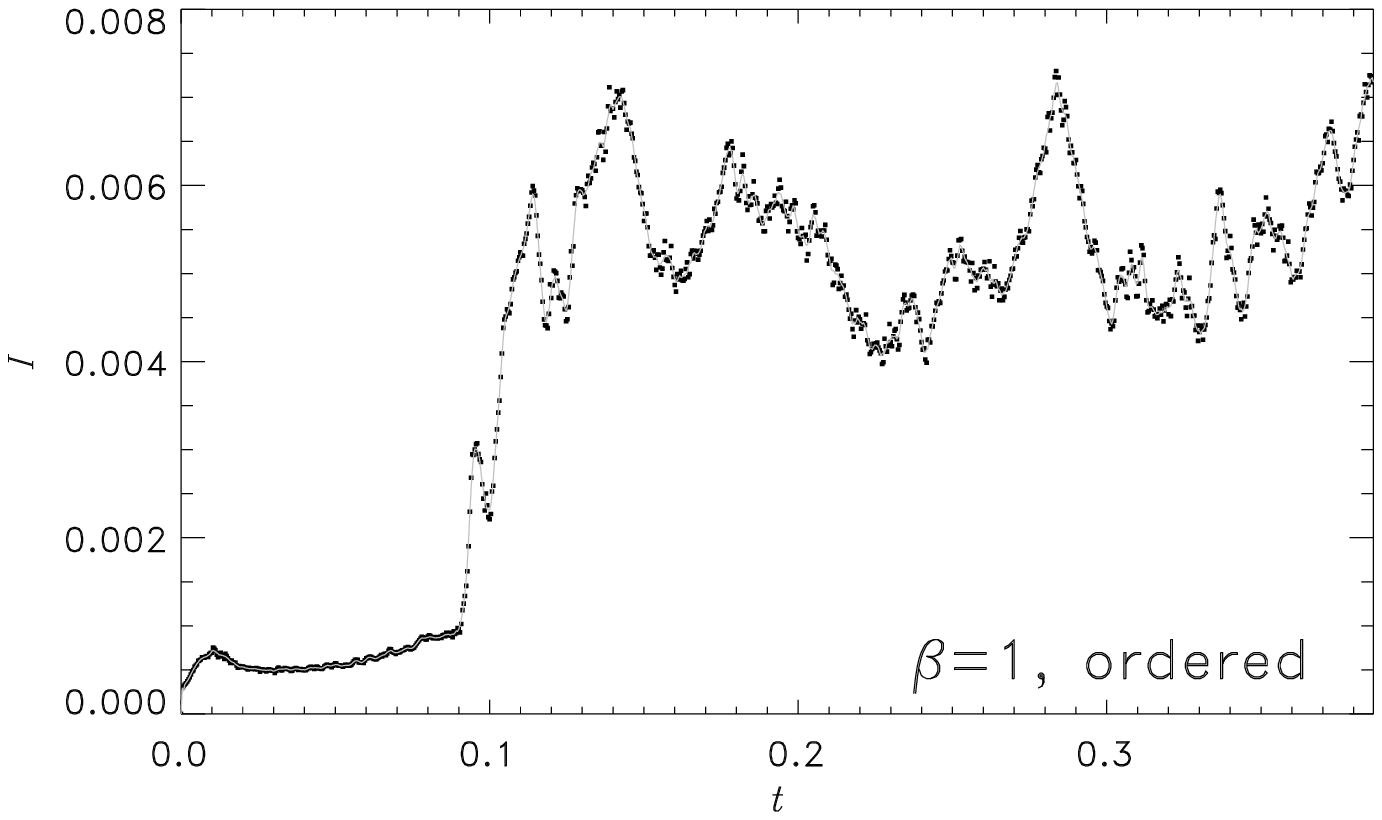}
\\
 \includegraphics[width=5cm]{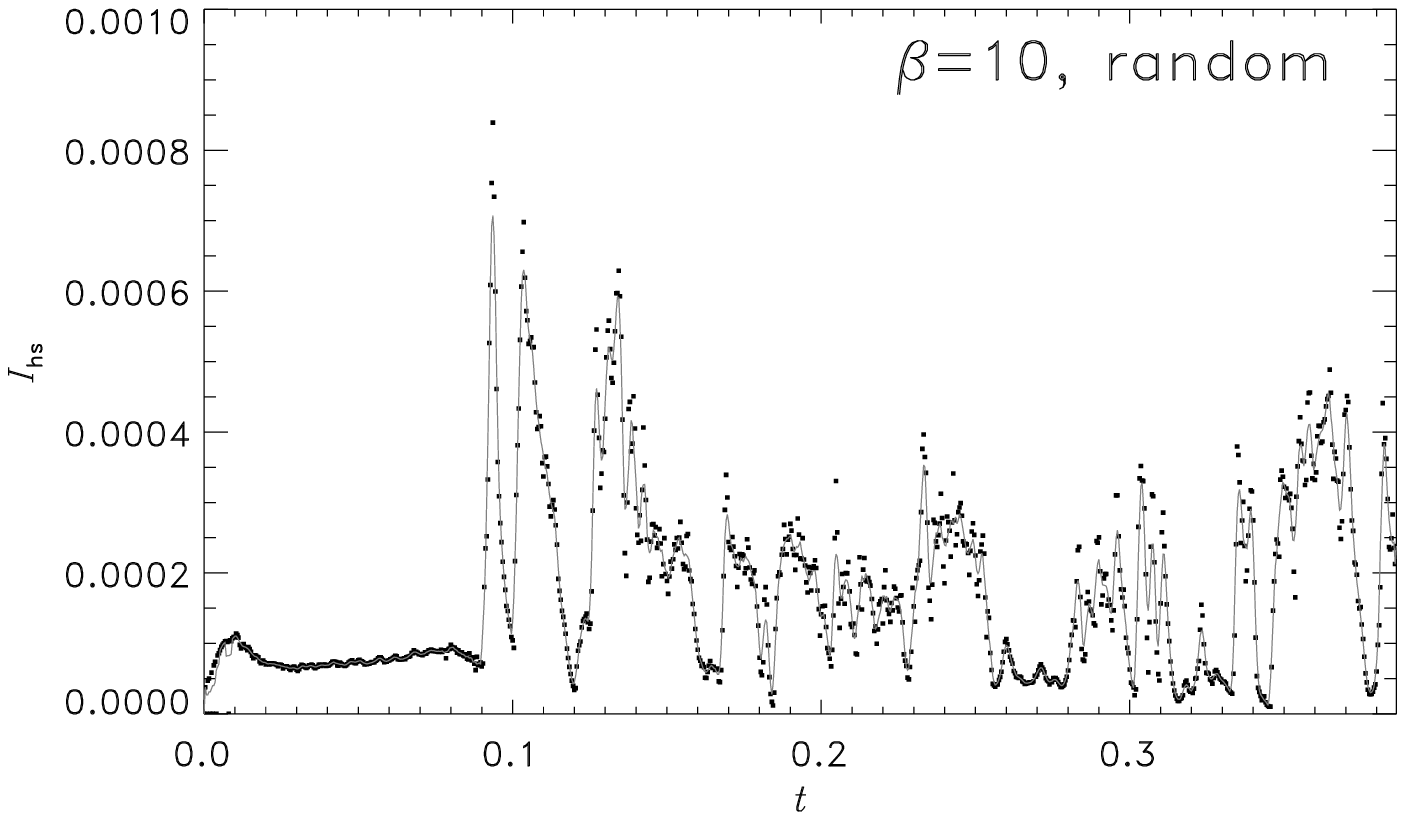}
&\includegraphics[width=5cm]{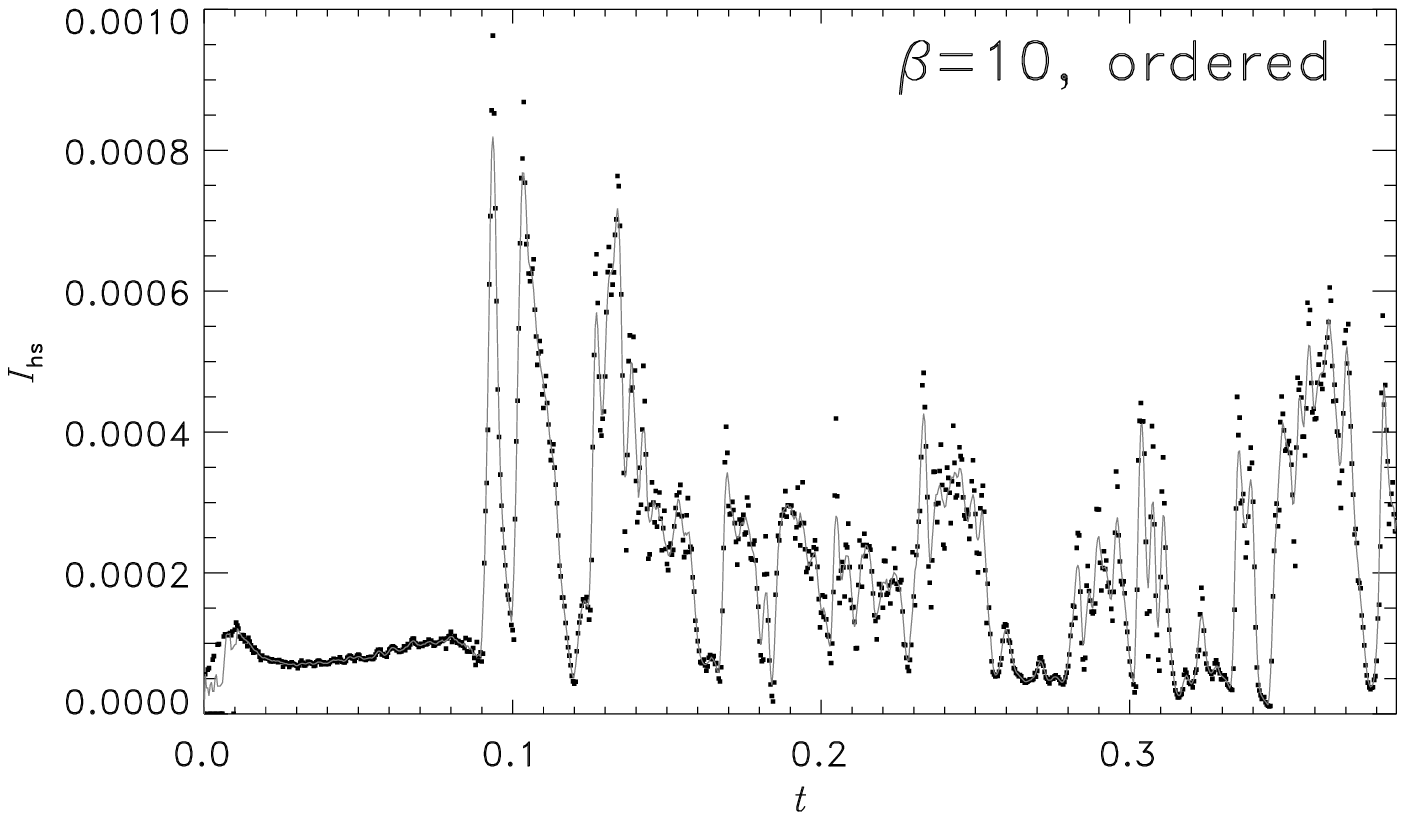}
&\includegraphics[width=5cm]{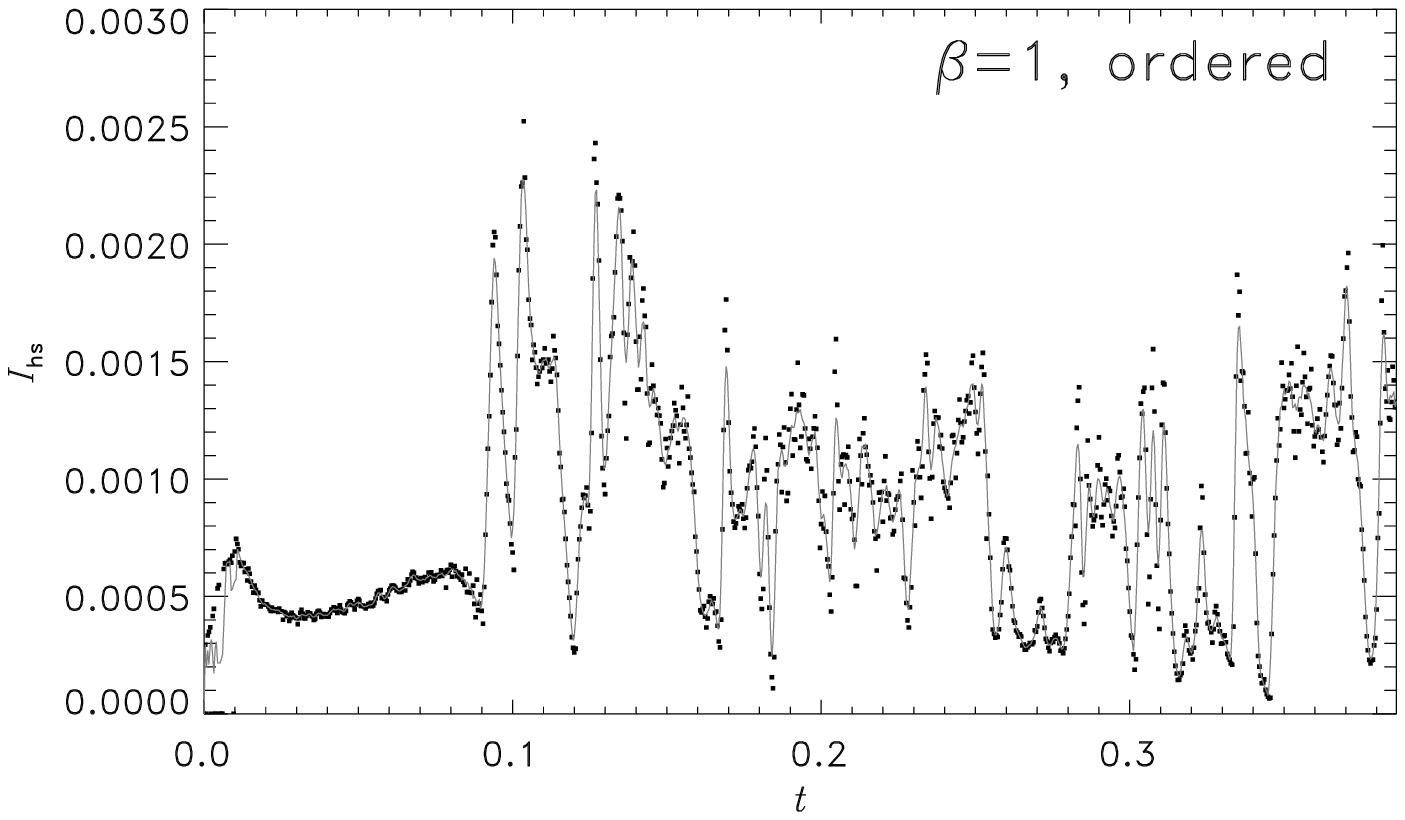}
\end{tabular}
\end{center}
\caption{Same as  Figure~\ref{fig.temporal.slow}, 
  but for the {\sc fast1ro} jet simulations. 
In the simulations, a fast jet propagate in a uniform background medium, 
  and an open boundary is adopted. 
The jet plasma is relativistic, with $\gamma=4/3$.
}
\label{fig.temporal.fast1ro}
\end{figure*}

\subsection{stages of jet activity}
\label{s.stages}

{\color{Black}
Adiabatic hydrodynamic simulations are rescalable
   in the three physical units: length, time and density. 
The implementation of radiative transfer calculations,
   however, limits this freedom, 
   because of the constraint that homologous models have the same optical depths
   (at corresponding scaled radiation frequencies)
   after rescaling the physical structure. 
Nonetheless, some freedom to rescale the models remains. 
Our models are therefore qualitatively relevant to analogous systems
   at smaller sizes, such as microquasar jets
   (although till this section we have interpreted our simulations
   with physical scales typical of large radio galaxies). 
For instance, a parsec-scale dwarf version of the confined fireball {\sc fast3}
   is conceivable.
} 

The earliest stage of jet expansion
   involves a compact ball of jet plasma,
   with at most a single vortex cell
   (a much simpler backflow than at late times).
A prominent termination shock shines brightly,
   but the jet is too short to fit any series of internal shocks.
This is probably what occurs during flares by AGN
   and analogous microquasars
   \citep[e.g.][]{fender2000,fender2002,macquart2002,aller2003}. 
As Figure~\ref{fig.wedge} shows, 
   for the fast jets with an open boundary
   the terminal shock dominates the luminosity during this stage
   ($t\la0.02U_t$ for {\sc fast1o};
   $t\la0.07U_t$ for {\sc fast2o}).
In corresponding models with a closed boundary
   ({\sc fast1}; {\sc fast2})
   the fireball is more uniformly luminous
   due to the accumulated mass of plasma
   surrounding the sides of the jet.

The closed-boundary simulations represent jets
   still propagating near the nucleus.
The accumulated plasma cocoon contributes a fuzz of radio emission
   on lateral scales a considerable fraction of the jet's length.
After the initial fireball stage,
   the locally brightest feature is usually the pseudo-core
   (first jet internal shock)
   but the extended fuzz contributes most of the integrated emission.
In {\sc fast4} the ripply density background
   modulates the cocoon expansion dramatically
   (and causes a long-period throbbing in the $I(t)$ light curve),
   yet the position and variability of the pseudo-core 
   are indistinguishable from the simpler {\sc fast1} case.
In integrated intensity observations of closed systems,
   the fuzzy cocoon may wash out the intrinsic variability of the jet
   (as seen in the slowly varying $I(t)$ light curves).
Spectral aging and dimming of the cocoon plasma
   (not included in our present simulations)
   would leave the jet relatively prominent at later times.

Simulations with an open left boundary represent
   jets that have penetrated far from the nucleus.
Jet internal shocks and the hotspot then dominate the intensity maps
   at all times after the initial fireball.
A comparison of {\sc fast1} with {\sc fast1o}
   or {\sc fast2} with {\sc fast2o}
   shows a greater separation between jet internal shocks
   in cases with an open boundary.
When the background density profile decreases radially
   ({\sc fast2o})
   the pseudo-core is farther from the nucleus
   but the subsequent jet-shocks are closer together
   (compared to {\sc fast1o}).

What about a jet injected into a pre-existing cavity,
   such as the ghost cocoon from a previous episode of nuclear activity?
If the jet meets a smoothly rising density profile of the interstellar medium 
   (like the case of {\sc fast3})
   a very different morphology emerges:
   a prolonged, highly opaque, frustrated fireball.
This spheroid is edge-dim,
   unlike an early-stage fireball with bright termination shock.
The frustrated fireball contains a more complex system of eddies,
   with dark creases visible at eddy interfaces nearer the surface.
If these flows carry locally ordered magnetic fields,
   then these affect the radiative transfer
   differently to the simple fields in an ``early fireball''.
The ``frustrated old fireball'' and ``early fireball''
   may have distinct polarization signatures.
(We study these in detail in a forthcoming paper.)

As shown in Figure~\ref{fig.wedge}, 
   although the internal jet shocks may hover about preferred points, 
   they eventually drift upstream or downstream at apparently high velocities.
Our integrated light profiles and hotspot light-curves
   for the open-boundary fast jets
   (upper two rows of
   Figures~\ref{fig.temporal.fast1o},
   \ref{fig.temporal.fast1ro}
   and \ref{fig.temporal.fast2o})
   show that internal shocks and hotspots
   can brighten or fade by factors of a few
   on timescales of millennia or less.
In smaller analogues of AGN, with shorter timescales,
   it is conceivable that an internal shock
   that fades, moves upstream and rebrightens
   might be mistaken for a new ejection there.

{\color{Black} 
We noted that a pulse of circular polarization ($\Pi_{\rm C}$)
   can occur in the very initial stage 
   of the simulations (bottom row of Figure~\ref{fig.temporal.fast1o}), 
   when the jet has a relatively coherent structure.  
The occurrence of a circular pulse is robust 
   and seems insensitive to the density profile of the external medium.  
It occurs whether the magnetic configuration is set to be quasi-poloidal
   (dragged parallel to the jet flow)
   or quasi-toroidal (circulating around the flow),
   but not if the fields are randomly directed.
Later, the circular polarisation is diluted rapidly
   as the jet evolves
   more numerous, complex, mutually incoherent emitting substructures.
Thus, the circular polarization observed in AGN cores 
  \citep[e.g.][]{homan1999,rayner2000}
  probably indicates certain degrees of coherence in the emission region.
Parametric models with turbulent or ordered,
   poloidal, toroidal or (intermediately) helical magnetic configurations
   have previously been applied to explain polarised emissions of AGN
    \citep[e.g.][and references therein]{beckert2002,ruszkowski2002,ensslin2003,gabuzda2008,vitrishchak2008,homan2009}.
The essential ingredient in these models
   and in our time-dependent models
   is that magnetic fields vary along sightlines through the jet,
   so that linearly polarised radiation from distant locations
   undergoes Faraday conversion to circular modes in foreground plasma.
As expected for this mechanism,
   we see that each spike in $|\Pi_{\rm C}|$
   coincides with a dip in linear polarisation $|\Pi_{\rm L}|$
   (e.g. third row, Figure~\ref{fig.temporal.fast1o}).
Circular polarization at a few percentage level
  was also seen in the initial stage 
  of radio outburst in the microquasar GRO~J1655$-$40 \citep{macquart2002}.  
Our calculations for AGN jets show similar levels.
A detailed discussion of time-dependent circular polarisation in microquasars
  is presented in \cite{saxton2010b}
  and a time-depedent polarimetry of AGN jets will be presented elsewhere 
  (Saxton et al., in preparation).       
}

{\color{Black}   
\section{CONCLUSIONS} 

We performed hydrodynamic simulations 
   for jets encountering external media with conditions and density structures  
   appropriate for active galaxies,
   and carried out time-dependent radiative transfer calculations 
   to determine their polarization and emission properties.  
Our polarised radiative transfer formulation
   takes account of emission, absorption, re-emission, 
   Faraday rotation and Faraday conversion. 
The radiative transfer equations were solved explicitly
   following the jet evolution. 
We applied this method to model the temporal evolution of emission 
   from AGN jets interacting with a variety of structured ambient media.   
Our calculations showed that jet emissions vary considerably 
   even though the launching conditions remain steady at the nucleus.
Their variations are affected by the external density profile;
  thus environmental factors play an important role 
  in determining the observed morphology and temporal properties of radio jets.

Our simulations show that the ambient media influence
   the distribution of jet knots
   and the relation between emission intensity and advance of the jet.
Determining the jet knot distributions 
  and the relation between intensity emission intensity and growth of jets   
  can constrain the properties of the background gas. 
For instance, in an effectively closed cocoon
   (that contributes much of the radio luminosity)
   asymptotic power-laws appear to relate
   the elapsed time, jet length and total intensity. 
The indices of these relations depend on
   the density gradient of the ambient medium. 
For models with an open left boundary  (with the jet terminus far from the nucleus)
   the radio intensity appears to flash independently of the jet length.
The total intensity can fluctuate by factors of a few,
   and the hotspot can fluctuate by tens,
   within dynamically brief intervals
   (though longer than a human lifetime).
Caution is needed when inferring the jet power from a single-epoch observations.
In any particular epoch, the two hotspots
   may differ by three or more times in intrinsic brightness,
   even given identical and constant jet fluxes,
   and neglecting orientation and beaming effects.
}

\section*{Acknowledgments}

The authors acknowledge the use of UCL Research Computing
   facilities and services in the completion of this work.
Part of the hydrodynamic simulations
  and all the numerical radiative transfer calculations 
  were performed using the UCL {\em Legion} supercomputer. 
We thank Chris Wegg for contributions in developing some of the algorithms 
  used in the numerical polarized radiative transfer calculations.


\input{bbl.tex}
\appendix

{\color{Black}
\section{An remark on radiative transfer} 

In our calculations,
   radiative transfer is computed in eulerian cells that
  do not co-move with jet plasma.
}
Thus, the transfer calculation has not included
   relative Doppler shift explicitly.    
We now assess whether or not the omission
   affects the calculated results significantly.
The quantity $\nu^{-3}I_\nu$ is Lorentz invariant  
  (where $\nu$ is the frequency and $I_\nu$ is the specific intensity).  
The radiative transfer equation in a co-moving frame thus takes the form: 
\begin{equation} 
    \left[{ \left( \frac{\nu}{\nu_0} \right)^3  \hat{\mathbsf D}  + 
      \left( \frac{\nu}{\nu_0} \right)^2  {\mathbsf K}     }\right]
	[[{\Ivec}]] 
       = \left( \frac{\nu}{\nu_0} \right)^2 [[{\Jvec}]]    
\end{equation}   
   \citep[see e.g.][]{mihalas1970,peraiah2002}.
Here, variables evaluated at the local rest frame
   are denoted by the subscript $0$.     
The co-moving transfer equation implies that 
   Doppler effects are of the first order in
   $[1- ({\nvec} \cdot {\vvec})/c]$ 
   \citep[see also][]{caster1972},
   where ${\vvec}$ is the medium local velocity.      
If ignoring terms of ${\cal O} [(v/c)^2]$
   or higher and performing a  series expansion,  
   then the inclusion of Doppler shift
   simply modifies the propagation operator:  
\begin{equation}  
	\partial_{s} \rightarrow  \left({
		1 - \frac{{\nvec}\cdot {\vvec}}{c}
	}\right) \partial_{s} + 
     {\cal G}  \ ,   
\end{equation}   
  where  
\begin{equation} 
 {\cal G} \approx  
  3 \left\{{
	({\nvec} \cdot {\GRADvec})
	\left({
		\frac{{\nvec}\cdot {\vvec}}{c} 
	}\right) 
	}\right\}  \  . 
\end{equation}      
The $[1- ({\nvec} \cdot {\vvec})/c]$ term
   in front of the operator $\partial_{s}$ 
   re-scales time and length along the ray 
   and is unrelated to the absorption and re-emission of the radiation, 
   and the Faraday effects on the Stokes parameters.    
The term ${\cal G}$, which is proportional to
   $\Delta \nu/\nu_0~ (\equiv (\nu-\nu_0)/\nu_0)$,  
   induces a  ``diffusive drift'' of power across the frequency space.    
It can be omitted, provided that 
  there is no sharp relativistic velocity gradient
  across the computational cells   
  (e.g.\ in the presence of relativistic turbulence).    
Thus, omission of Doppler effects might 
  distort the perceived time scale and length of the systems 
  but would not affect the qualitative results obtained 
  for the global morphology and polarization level of the jets.  
Simulations with explicit inclusion of Doppler effects 
  require substantial modifications in the numerical algorithm 
  and additional constraints set by the relativistic physics.  
We will leave this for a separate study.

\end{document}

%% file: aas_macros.tex
\def\jref@jnl#1{{\rm#1}}

\def\aj{\jref@jnl{AJ}}                   
\def\araa{\jref@jnl{ARA\&A}}             
\def\apj{\jref@jnl{ApJ}}                 
\def\apjl{\jref@jnl{ApJ}}                
\def\apjs{\jref@jnl{ApJS}}               
\def\ao{\jref@jnl{Appl.~Opt.}}           
\def\apss{\jref@jnl{Ap\&SS}}             
\def\aap{\jref@jnl{A\&A}}                
\def\aapr{\jref@jnl{A\&A~Rev.}}          
\def\aaps{\jref@jnl{A\&AS}}              
\def\azh{\jref@jnl{AZh}}                 
\def\baas{\jref@jnl{BAAS}}               
\def\jrasc{\jref@jnl{JRASC}}             
\def\memras{\jref@jnl{MmRAS}}            
\def\mnras{\jref@jnl{MNRAS}}             
\def\pra{\jref@jnl{Phys.~Rev.~A}}        
\def\prb{\jref@jnl{Phys.~Rev.~B}}        
\def\prc{\jref@jnl{Phys.~Rev.~C}}        
\def\prd{\jref@jnl{Phys.~Rev.~D}}        
\def\pre{\jref@jnl{Phys.~Rev.~E}}        
\def\prl{\jref@jnl{Phys.~Rev.~Lett.}}    
\def\pasp{\jref@jnl{PASP}}               
\def\pasj{\jref@jnl{PASJ}}               
\def\qjras{\jref@jnl{QJRAS}}             
\def\skytel{\jref@jnl{S\&T}}             
\def\solphys{\jref@jnl{Sol.~Phys.}}      
\def\sovast{\jref@jnl{Soviet~Ast.}}      
\def\ssr{\jref@jnl{Space~Sci.~Rev.}}     
\def\zap{\jref@jnl{ZAp}}                 
\def\nat{\jref@jnl{Nature}}              
\def\iaucirc{\jref@jnl{IAU~Circ.}}       
\def\aplett{\jref@jnl{Astrophys.~Lett.}} 
\def\apspr{\jref@jnl{Astrophys.~Space~Phys.~Res.}}
\def\bain{\jref@jnl{Bull.~Astron.~Inst.~Netherlands}} 
\def\fcp{\jref@jnl{Fund.~Cosmic~Phys.}}  
\def\gca{\jref@jnl{Geochim.~Cosmochim.~Acta}}   
\def\grl{\jref@jnl{Geophys.~Res.~Lett.}} 
\def\jcp{\jref@jnl{J.~Chem.~Phys.}}      
\def\jgr{\jref@jnl{J.~Geophys.~Res.}}    
\def\jqsrt{\jref@jnl{J.~Quant.~Spec.~Radiat.~Transf.}}
\def\memsai{\jref@jnl{Mem.~Soc.~Astron.~Italiana}}
\def\nphysa{\jref@jnl{Nucl.~Phys.~A}}   
\def\physrep{\jref@jnl{Phys.~Rep.}}   
\def\physscr{\jref@jnl{Phys.~Scr}}   
\def\planss{\jref@jnl{Planet.~Space~Sci.}}   
\def\procspie{\jref@jnl{Proc.~SPIE}}   